\documentclass[fleqn,usenatbib]{mnras}
\usepackage{newtxtext,newtxmath}
\usepackage[T1]{fontenc}
\usepackage[table]{xcolor}

\DeclareRobustCommand{\VAN}[3]{#2}
\let\VANthebibliography\thebibliography
\def\thebibliography{\DeclareRobustCommand{\VAN}[3]{##3}\VANthebibliography}

\usepackage{graphicx}	% Including figure files
\usepackage{amsmath}	% Advanced maths commands

\usepackage{amssymb}	% Extra maths symbols
\usepackage{adjustbox}
\usepackage{natbib}
\usepackage{float}
\usepackage{placeins}
\title[Distances of Galactic Radio Pulsars]{Distances of Galactic Radio Pulsars; First Quadrant: -2$\degr$ $<$ $\ell$  $<$ 90$\degr$ and -2$\degr$ $<$ $b$ $<$ 2$\degr$}

\author[P. Kütükcü et al.]{
Pınar Kütükcü,$^{1}$\thanks{E-mail: pinarkutukcu@gmail.com}
Aşkın Ankay,$^{2}$
Efe Yazgan$^{3}$
and Kutsal Bozkurt$^{1}$
\\
$^{1}$Physics Department, Yıldız Technical University, Davutpaşa Cad., 34220, Esenler/Istanbul, Turkey\\
$^{2}$Physics Department, Boğaziçi University, North Campus, KB Building Floor 3-4, 34342 Bebek/Istanbul, Turkey\\
$^{3}$Physics Department, National Taiwan University, Taipei, Taiwan
}

\date{Accepted XXX. Received YYY; in original form ZZZ}

\pubyear{2021}

\begin{document}
\label{firstpage}
\pagerange{\pageref{firstpage}--\pageref{lastpage}}
\maketitle

\begin{abstract}
Distance versus dispersion measure relations are constructed for Galactic radio pulsars in small solid angle intervals. The calculations are based on some basic criteria as well as using the independent distance measurements of well examined pulsars for the first Galactic quadrant including Galactic central directions. Values of average free electron density for these regions are derived from the fits to distance versus dispersion measure relations and checked for consistency and smoothness. The effects of plasma in the Galactic arms and within the central parts of the Galactic bulge region are also compared and discussed. Our adopted distances for the radio pulsars are compared with the ones given in some other models. Some basic results on distributions of the radio pulsars and the plasma are presented.
\end{abstract}


%%%%%%%%%%%%%%%%%%%%%%%%%%%%%%%%%%%%%%%%%%%%%%%%%%

%%%%%%%%%%%%%%%%% BODY OF PAPER %%%%%%%%%%%%%%%%%%

\section{Introduction}

%This is a simple template for authors to write new MNRAS papers.
%See \texttt{mnras\_sample.tex} for a more complex example, and \texttt{mnras\_guide.tex}
%for a full user guide.

%All papers should start with an Introduction section, which sets the work
%in context, cites relevant earlier studies in the field by \citet{Fournier1901},
%and describes the problem the authors aim to solve \citep[e.g.][]{vanDijk1902}.
%Multiple citations can be joined in a simple way like \citet{deLaguarde1903, delaGuarde1904}.

Since the first observation of a pulsar (PSR) in 1967, more than 2800 pulsars were observed up to date according to the ATNFPC (Australia Telescope National Facility Pulsar Catalog) \citep{manchester2005}. Most of these pulsars are observable at radio frequencies due to their synchrotron emission. It is essential to know the distances of pulsars from the Sun mainly in order to calculate their radiative power and to determine their positions and distributions in the Galaxy.

Most of the known pulsars are isolated objects and none of the directly observable intrinsic properties of pulsars allows for a distance determination of a pulsar unlike some other sources, e.g. cepheids. For a few of the pulsars that are close enough, the parallax method could be utilized. Because of the high speeds of neutron stars gained during asymmetric core-collapse supernova explosions  and the short observable lifetimes (10$^4$-10$^5$ yr) of supernova remnants (SNRs) \citep{guseinov2007}, only young neutron stars can reliably be associated to SNRs. As the number of Galactic SNRs including neutron stars, pulsar wind nebulae and/or bow shock structures is not more than half of the known SNRs \citep{green2019} and the fact that SNR distances are not small in most of the cases (\citealt{guseinov2007}, and references therein, \citealp{ranasinghe2018revised},  \citealt{,green2019}, and references therein, \citealt{lee2019near}), there are only a few neutron star -- SNR pairs with reliable distances. The distances of pulsars located in globular clusters can be determined with a better precision, however the observed number of such pulsars is small.

In this work, we present an improved version of a basic and reliable method to determine distances of all radio pulsars with known dispersion measure (DM) values based on constructing distance -- DM relations \citep{yazgan2007, ankay2016galactic}. It is possible to derive average free electron densities in the line of sight (LOS) of the radio pulsars from these relations to constrain the distributions of the plasma as well as the radio pulsars in the Galaxy. Figures and tables for distance-DM relations in the solid angle interval 2$\degr$ $<$ $\ell$  $<$ 90$\degr$; -2$\degr$ $<$ $b$ $<$ 2$\degr$ are given in Appendix ~\ref{app}.

Most of the pulsars listed in the ATNFPC have precise timing observations that allow distance determinations based on the 3D free electron distributions. Passing through the interstellar plasma, radio waves from pulsars get dispersed depending on the electron density along the line of sight and the frequency of the radio pulse ($\nu$). Thanks to the pulsed emission, the difference in the arrival times between two frequencies, both much greater than the plasma frequency, is measurable. This delay $\Delta{t}$ can be written as \citep{Lipunov}

\begin{equation}
\Delta{t}=t_{2} - t_{1}=\bigg(\frac{1}{\nu_{2}^{2}} - \frac{1}{\nu_{1}^{2}}\bigg)\frac{e^{2}}{2m_{e}c}\int\displaylimits_{0}^{D}n_{e}ds
\end{equation}

The last term in equation (1) is defined as DM in pc/cm$^{3}$:

\begin{equation}
DM \equiv \int\displaylimits_{0}^{D}n_{e}ds
\end{equation}

Here $c$ is the speed of light in vacuum, $m_{e}$ is the mass of electron, $e$ is the electron charge,  $n_{e}$ is the free electron density in the LOS in cm$^{-3}$, and D is the distance of the radio pulsar from the Sun. Distance of the source can be found using this equation if the electron density in the LOS is known.

This method is practically applicable only in the case of pulsed low frequency emission making it the standard method applicable to all radio pulsars as a rule. The DM measurements have in general small inaccuracies and the variations in the DM values due to displacements of filaments in SNRs are rare and mostly insignificant \citep{Petroff}. The main and effectively the only difficulty in using this method is related to the mostly unknown distributions of the plasma regions (mainly SNRs and HII-regions) throughout the Galaxy and their densities. The first mathematical model on DM-based distance determination \citep{taylor1993pulsar} used in ATNFPC led to overestimated and underestimated distance values especially in the Galactic center directions. An improved version of such a model was introduced by \citet{ne2001}, yet this model also gives overestimated/underestimated pulsar distances. The plasma distribution in our galaxy, though it is relatively more uniform compared to the distributions of molecular clouds and HI clouds, can not be adequately described by simple mathematical models using general assumptions. However, the models are being improved, e.g. as introduced in \citet{yao2017new} (YMW17 model) which is the latest model used in ATNFPC as the default model. This model also includes systematic biases which are discussed below.

Alternatively, distance – DM relations can be constructed for small solid angles choosing narrow longitude and latitude intervals including pulsars with measured DM to fit and using the independently measured distance values of calibrator pulsars \citep[e.g.][]{yazgan2007, guseinov2002trustworthy, ankay2016galactic}. The average free electron number density values can be determined from these relations for a large number of solid angle intervals covering the whole galaxy. The criteria mainly introduced by \citet{guseinov2002trustworthy} and \citet{yazgan2007} in adopting distances for some radio pulsars with measured DM values with some improvements will be presented below.

The rotation measure (RM) based on the Faraday rotation is also a precisely measured quantity similar to DM but for a smaller number of pulsars. RM depends on the average magnetic field of the interstellar medium in the LOS as well as the average free electron density. So the ratio RM/DM gives information on the average magnetic field of the interstellar medium which can in principle be used to check and compare the distance values of pulsars located in the same region (Manchester and Taylor (1977)). RM is defined through the following equations: 

\begin{equation}
RM \equiv \frac{e^{3}}{2\pi m_{e}^{2}c^{4}}\int\displaylimits_{0}^{D}n_{e}Bcos\theta ds
\end{equation}

\begin{equation}
\left\langle Bcos\theta\right\rangle =  1.232\frac{RM}{DM}
\end{equation}

Here B is the magnetic field strength and $\theta$ is the angle between the line of sight and the magnetic field. The unit of B is $\mu$G and the unit of RM is rad/m$^{2}$.

Having a precise knowledge of the locations and the sizes of the Galactic arms is essential to determine both the Galactic plasma (in particular the free electron density) distribution and the locations of Galactic young pulsars. Giant molecular clouds, HII regions and masers are used as the basic sources and tracers in most of the Galactic arms structure models \citep[e.g.][and references therein]{georgelin1976spiral, russeil2003star, hou2009spiral, xu2013nature, zhang2013parallaxes, choi2014trigonometric, garcia2014giant, hou2014observed, reid2014trigonometric, sato2014trigonometric, vallee2014spiral, vallee2014catalog, wu2014trigonometric, hachisuka2015parallaxes, vallee2015different}.

Recent observations \citep{ponti2015xmm, schnitzeler2016radio}, found that the plasma at the core of the Galactic bulge is very dense, probably denser than the plasma in the most dense parts of the Galactic arms, including some "holes" with low density. This is discussed below based on the observations of radio pulsars close to or in the direction of the Galactic center especially a magnetar which is probably connected to the dense plasma region Sgr A located at the center.

\section{Basic Criteria in Adopting Distances of Radio Pulsars and Its Relation to the Galactic Plasma Distribution}

We have constructed distance -- DM relations for radio pulsars in the first quadrant of the Galaxy for b = 0$\pm$2 degrees including also the Galactic central directions in the intervals $\ell$ = b = 0$\pm$2 degrees. We have divided the pulsars into 20 groups, each group having 4 degrees by 4 degrees solid angle, up to $\ell$ = 78 degrees. The average free electron density for each group has been calculated and compared with each other for adjacent groups to check the smoothness and continuity of the plasma distribution. The number of radio pulsars in the longitude interval $\Delta$$\ell$ = 78-90 degrees with b = 0$\pm$2 degrees is only three, therefore, we have only checked the possible distances of these pulsars without making a fit.

Some basic criteria based on observational results are listed below to construct distance versus DM relations \citep{ankay2016galactic}. We assume that the main contribution to the measured DM values is mainly due to the plasma in the HII regions, SNRs and probably the Galactic central region. A low density plasma distribution between the Galactic arms and at high latitudes are also taken into consideration.

The first and the fourth quadrants of the Galaxy must include most of the radio pulsars, and in general most of the Galactic sources, as the volume of this region is significantly larger than the anti-center region. This is also the case for young pulsars as they are supposed to be located within or close to the star formation regions (SFRs) that form the Galactic arms. On the other hand, as the Galactic arm structure beyond the Galactic center is not well known in the first and the fourth quadrants, it is much more difficult to determine the locations of distant pulsars. Yet, some of the most luminous radio pulsars are definitely located in the interval -40$\degr$<l<40$\degr$. Based on the available observational data (ATNFPC), the number of such luminous pulsars in this interval is not as large as expected suggesting that a larger portion of the Galactic arms are outside of this longitude interval. The distribution of the space velocities of pulsars and the activity of the star formation regions producing core-collapse supernovae are considered to get a conclusive result.

In order to have an estimate on the radio pulsar luminosity, we take Crab pulsar as a reference to put an upper limit for each radio pulsar distance using the measured fluxes at 1400 MHz and 400 MHz. Crab pulsar is taken as a reference to put an upper limit because it is close by and young. Its well-known luminosity helps prevent large overestimation in adopting pulsar distances. About 96$\%$ of radio pulsars are less luminous than the Crab pulsar at 400 MHz and  about 87$\%$ radio pulsars are less luminous than Crab pulsar at 1400 MHz (ATNFPC). For almost all of the radio pulsars, the 1400 MHz distance limit is much smaller than the one at 400 MHz. Therefore, the luminosity of the Crab pulsar can be used as a lower limit to prevent large over-estimations in adopting pulsar distances. \citet{Guseinovluminosity} constructed luminosity functions for young radio pulsars ($\tau$<10$^{7}$yr) at 400 MHz and 1400 MHz and showed that most of the PSRs have low luminosity values. For a detailed review on pulsar luminosities see \citet{Bagchi}.

In some Galactic longitude intervals, some parts of the arms are offset with respect to the Galactic plane. Some of these deviations are well known by observation, especially at smaller distances from the Sun, and are considered both for the free electron distribution in the Galaxy and when adopting radio pulsar distances.

The distances of pulsars from the Galactic plane are related to their ages. The spin-down age of a pulsar is given by 

\begin{equation}
t = \frac{P}{(n-1)\dot{P}}\bigg[1-\bigg(\frac{P_{0}}{P}\bigg)^{(n-1)}\bigg]
\end{equation}

where n is the braking index, P is the period, P$_{0}$ is the birth period and $\dot{P}$ is the time derivative of the period of the pulsar. Assuming P$_{0}$ $\ll$ P and n $\neq$ 1

\begin{equation}
t = \frac{1}{(n-1)}\frac{P}{\dot{P}}
\end{equation}
 The characteristic age is defined for n=3 (for magnetic dipole braking) as \citep{manchester1977pulsars, lyne2012pulsar}

\begin{equation}
\tau = \frac{P}{2\dot{P}}
\end{equation}

Note that the characteristic age is not a measure of the true age \citep{Jiang}. For example, for the Crab pulsar $\tau$=1256.8yr (ATNFPC) and the real age is equal to the age of SN 1054 which is $\sim$970 yr. \citet{espinoza2011} give a list of young galactic pulsars with reliable measured braking indices in the range 2.14-2.91. So, the real age can be greater than $\tau$ up to a factor of 2 based on these measured values of breaking index. Such small differences in age values do not affect our analysis or change our results in this work. For the pulsar J1734-3333 \citet{espinoza2011} measure a braking index value 0.9, mentioning a recent glitch would most likely disrupt  the $\ddot{P}$ value. \citet{Lyne1996} give a braking index value 1.4 for the Vela pulsar (J0835-4510), where they mention a large uncertainty in the $\dot{P}$ measurement caused by glitch together with the timing noise. For J1640-4631 ($\tau$=3.35$\times$$10^3$ yr (ATNFPC)), \citet{archibaldn} measure a braking index value 3.15 and argue the reason of a high braking index value may occur because of an unseen glitch when the timing noise is ignored. This braking index value for J1640-4631 must be in accordance with energy loss due to pure magnetic dipole radiation within uncertainties. For a similar example for characteristic age and real age values (J1801-2451), see section~\ref{sec:3.2}.  According to \citet{guseinov2004strong} magnetic fields of pulsars decay exponentially and the characteristic time scale for this process is about 3$\times$$10^6$ yr. As shown by \citet{guseinov2004strong}, the correlation between $\tau$ and the kinematic age (depending on the pulsar distance from the plane) no longer exists for $\tau$>$10^7$ years. This is also considered in adopting distances of older pulsars, in particular for the ones in the regions of the Galactic arms that are deviated from the Galactic plane.

 According to \citet{faucher2006birth} average space velocity of pulsars is ${380_{-60}^{+40}}$ km/s. \citet{verbunt2017observed} suggest that the average space velocity of 42$\%$ of pulsars is 120 km/s and 58$\%$ of pulsars is 580 km/s. There are 422 PSRs whose transverse speed  (V$_{trans}$) is given according to the adopted distances by the ATNFPC and the average value of the transverse speeds of these pulsars is 265.7 km/s. There are some very large V$_{trans}$ values given in ATNFPC (e.g. PSR J1327-0755 V$_{trans}$=11705 km/s, b=53.848$\degr$, d$>$25 kpc; PSR J0134-2937 V$_{trans}$=2279.86 km/s, b=-80.250$\degr$, d$>$25 kpc; PSR J0151-06354 V$_{trans}$=1419.49 km/s, b=-65.004$\degr$, d$>$25 kpc; PSR J2116+1414 V$_{trans}$=1612.10 km/s, b=-23.409$\degr$, d$>$25 kpc). This is mainly because of some of the PSRs with $\lvert$b$\rvert$>1 having overestimated distances as adopted in ATNFPC. About 75$\%$ of 422 PSRs in ATNFPC have V$_{trans}$ values below the average 265.7 km/s value. Although the number of PSRs with overestimated distances is relatively small, they still cause the average speed value to be overestimated. Average space velocity of PSRs depends on the kick (asymmetry of the explosion), regardless of where the PSR was born \citep{yazgan2007,Katsuda2018}. According to \citet{allakhverdiev1997} and  \citet{hansen1997pulsar} the average space velocity of pulsars is 250-300 km/s. Considering that young pulsars ($\tau<$ 8$\times$$10^5$ yr) must be located in or close to the star formation regions (molecular clouds, HI clouds, giant HII regions, SNRs, OB associations and open clusters) within the Galactic arms and the references given above, the average space velocity of pulsars must be about 300-350 km/s.
 
 While constructing distance-DM relations, both characteristic age and average space velocity of pulsars (300-350 km/s) are taken into account as discussed in previous two paragraphs. Young pulsars whose $\tau<$ 8$\times$$10^5$ yr are considered to be inside (or close to) the star formation regions (Galactic arms). The distance and the thickness of the Galactic arms are taken into account as criteria while determining distance-DM relations. Older pulsars ($\tau$$>$10$^7$ years) can be located within Galactic arms, in between them or far away from them unless their space velocity is too small, so that the characteristic age is not a valid criterion for such pulsars in general.

There must be a correlation between the distribution of the HII regions, SNRs, OB associations and the Galactic distribution of free electrons in the vicinity of the Galactic plane. Their densities must also be correlated which is harder to show due to the lack of necessary observational data. Such correlations must be stronger at smaller distances and when considering only the Galactic arm regions. At large distances, the correlations must become weaker because of the presence of low density plasma between the arms. At larger distances from the Galactic plane, the free electron distribution must become more uniform with a decrease in its average value. Because of this fact, pulsars at larger Galactic latitudes must be located at relatively larger distances as compared to the pulsars at smaller latitudes with similar DM values at the same longitudes. The pulsars with large $\tau$ and/or the ones in the longitude intervals of the deviated arm regions must be considered separately.

The plasma regions in the Galactic bulge is not well known as compared to the plasma in arms nearer to the Sun, but their contribution to the values of DM can be more important than the plasma in the arms as seen in the case of the largest DM pulsar (SGR J1745-29).

The location, thickness and scale height of each Galactic arm are taken into account in adopting distances for the pulsars located at or close to the Galactic plane. For this, we assumed that the average value of the free electron number density increases as the line of sight passes through each arm as a function of distance for the pulsars in the same longitude and latitude intervals.

When adopting distances for pulsars in each solid angle interval (these groups of pulsars are given in the next section) independent distance measurements for them (like trigonometric parallax and associations with SNRs or stellar clusters) are also taken into consideration. Using the results of such measurements and making comparisons between pulsars in each group data lead to improved values of both distance and uncertainty in distance. The data used are mainly $\ell$, b, DM, $\tau$ and also some of the other data displayed in the text of the next section and in Tables ~\ref{tab:1}--\ref{tab:21}). In these tables, data from ATNFPC and the ones adopted and calculated in this study are shown in separate columns.

\section{Radio Pulsar Groups in Different Solid Angle Intervals in the First Quadrant}

Below, some basic information and observational results for groups of Galactic radio pulsars in different solid angle intervals, covering
the first quadrant of the Galaxy, are presented. Luminosity comparisons between radio pulsars and the Crab pulsar which is taken as a reference, are done using the measured flux at 1400 MHz to put a limit on the distance unless otherwise is stated. The measured transverse speeds (V$_{trans}$) of pulsars displayed in the ATNFPC are also presented.

\subsection{358$\degr$ $<$ $\ell$  $<$ 2$\degr$; -2$\degr$ $<$ $b$ $<$ 2$\degr$}
\label{sec:3.1} 

There are 29 radio pulsars in these Galactic central directions including some of the largest DM pulsars. Nine of the pulsars are young ($\tau<$ 8$\times$$10^5$ yr) (see Table~\ref{tab:1}) that each of them must be located at or close to one of the Galactic arms or at the Galactic bulge.

One of the young pulsars in this group is the magnetar SGR J1745-29 with the largest DM (1778 pc/cm$^3$) among all the observed pulsars up to date. The magnetar's LOS passes through the HII region G359.964-00.101 with unknown distance \citep{anderson2014wise} and possibly through low-density parts (i.e. unobserved) of several other HII regions. SGR J1745-29 is probably located close to the Galactic center with its direction passing through Sgr A. It has also the largest RM measured up to date suggesting a location close to the Galactic center. Since SGR J1745-29 is only 3400 years old, it is also possible that this pulsar is located at a Galactic arm beyond the center considering its extremely large DM and the small number of RM measurements for pulsars to make conclusive comparisons. The magnetar has V$_{trans}$=251.023 km/s at d=8.3 kpc (ATNFPC), therefore if it is located at a distant arm beyond the center, its speed can be about about 2 times larger than the average speed of pulsars.

Eight of the 29 pulsars in this group have DM $>$ 800 pc/cm$^3$ (see Table~\ref{tab:1}) strongly suggesting the presence of dense plasma at the Galactic central regions. There is no observed pulsar with DM in the range 600 pc/cm$^3$ $<$ DM $<$ 800 pc/cm$^3$ possibly corresponding to the interval between the Norma arm and the Galactic central region.

Young pulsar J1745-3040 ($\ell$=358.553$\degr$, b= -0.963$\degr$, DM=88.373 pc/cm$^3$, d=2.343 kpc (YMW17), $\tau$ = 5.46$\times$10$^5$ yr) is located at Sagittarius arm. It has V$_{trans}$=46.78 km/s at d=0.2 kpc (0-1.3 kpc, kinematic) according to ATNFPC. The speed of this pulsar is comparable to the average value at our adopted distance. Very young pulsar 1740-3015 ($\ell$=358.294$\degr$, b= 0.238$\degr$, DM=151.96 pc/cm$^3$, d=2.944 kpc (YMW17), $\tau$ = 2.06$\times$10$^4$ yr) is located at Scutum arm. For this pulsar, 0.4 kpc (0.1-2.1 kpc, kinematic) is adopted in ATNFPC. J1752-2806 ($\ell$=1.54$\degr$, b= -0.961$\degr$, DM=50.372 pc/cm$^3$, d=1.335 kpc (YMW17), $\tau$ = 1.1$\times$10$^6$ yr) is probably at or close to Sagittarius arm, which has the adopted distance of 0.2 kpc (0.1-1.3 kpc, kinematic) in ATNFPC. There is no Local arm source in or close to the directions of these pulsars. If such very small distances (0.2-0.4 kpc) were correct, the average free electron density in the nearby region of the Sun would have been very large. herefore, the distances adopted in ATNFPC for these three pulsars based on the kinematic measurements listed in YMW17 model, must have an order of magnitude underestimated values.

RM values of J1752-2806 and J1745-3040 are both positive and similar in magnitude (see Table~\ref{tab:1}). Considering also their DM values, these two pulsars must be located in or close to the Sagittarius arm.

Pulsar J1752-2806 (d=1.1-1.5 kpc) has a larger upper limit on the distance (taking the Crab pulsar as a reference) at 1400 MHz as compared to 400 MHz which is an exceptional case. This pulsar may have a luminosity similar to that of Crab.

Young pulsar J1747-2958 ($\ell$  = 359.305$\degr$, b = -0.841$\degr$, DM = 101.5 pc/cm$^3$, d = 2.5 (ATNFPC) kpc, $\tau$ = 2.55$\times$10$^4$ yr) is possibly associated with PWN G359.23-0.82 \citep[d$<$3 kpc,][]{uchida1992hi}; \citep[d = 2 kpc][]{camilo2002heartbeat}; \citep[d$<$5.5 kpc (HI)][]{camilo2002heartbeat}; \citep[d = 4.8$\pm$0.8 kpc (X-ray)][]{abdo2013} and located in the direction of the HII region G359.50-0.60 (d = 1.5$\pm$0.2 kpc, \citet{hou2009spiral} with a lateral separation of only 8 pc.

Pulsars J1739-2903, J1740-3015, J1739-3023 and J1738-2955 have negative RM values ranging from -74$\pm$18 to -236$\pm$18 (see Table~\ref{tab:1}). Together with their DM values and considering the fact that all of them are young, they must be located within the Scutum arm. Pulsar J1741-2945 has a direction close to the directions of J1738-2955, J1739-3023 and J1740-3015 with a significantly larger DM. Therefore, the line of sight of this pulsar must pass through at least some part of the Norma arm as well and it may even be located within this arm.

Pulsars J1736-2843 and J1741-2719 have the largest latitudes in this group (see Table~\ref{tab:1}). Based on their DM values, the lines of sight of these pulsars are possibly not passing through the Scutum and the Norma arms. Their $\tau$ values are not small either, making the uncertainties in their distances larger.

Pulsars J1749-3002 (b = -1.244$\degr$) and J1752-2821 (b = -0.976$\degr$) have DM values comparable to each other (see Table~\ref{tab:1}). Model dependent and adopted distances (which are the same for each of them) for these two pulsars given in ATNFPC are 12.7 kpc and 5.4 kpc, respectively, with a distance ratio of about 2.4. If the latitude of J1752-2821 were even slightly smaller making its b $<$ -1$\degr$, the distance of this pulsar would be at least twice the value given in ATNFPC according to YMW17 model. These two pulsars must probably be located at similar distances.

There is the giant molecular cloud Sagittarius (Sgr) B2 \citep[$\ell$= 0.67, b = -0.03, d = 7-8.6 kpc][]{reid2009,reid2014trigonometric} possibly located close to the Galactic center. There are 451 HII regions in $\Delta$$\ell$= $\Delta$b = 0$\degr$$\pm$2$\degr$ intervals around the Galactic center direction \citep{anderson2014wise}. Distance measurements are avaliable only for 17 of the HII regions locating all of them at 8 kpc in the intervals $\Delta$$\ell$= 358.53$\degr$ -- 358.95$\degr$ and $\Delta$b = -0.116$\degr$ -- 0.085$\degr$. None of the pulsars in this group has a LOS passing through any one of these HII regions.

\begin{figure}
	
	\includegraphics[width=\columnwidth]{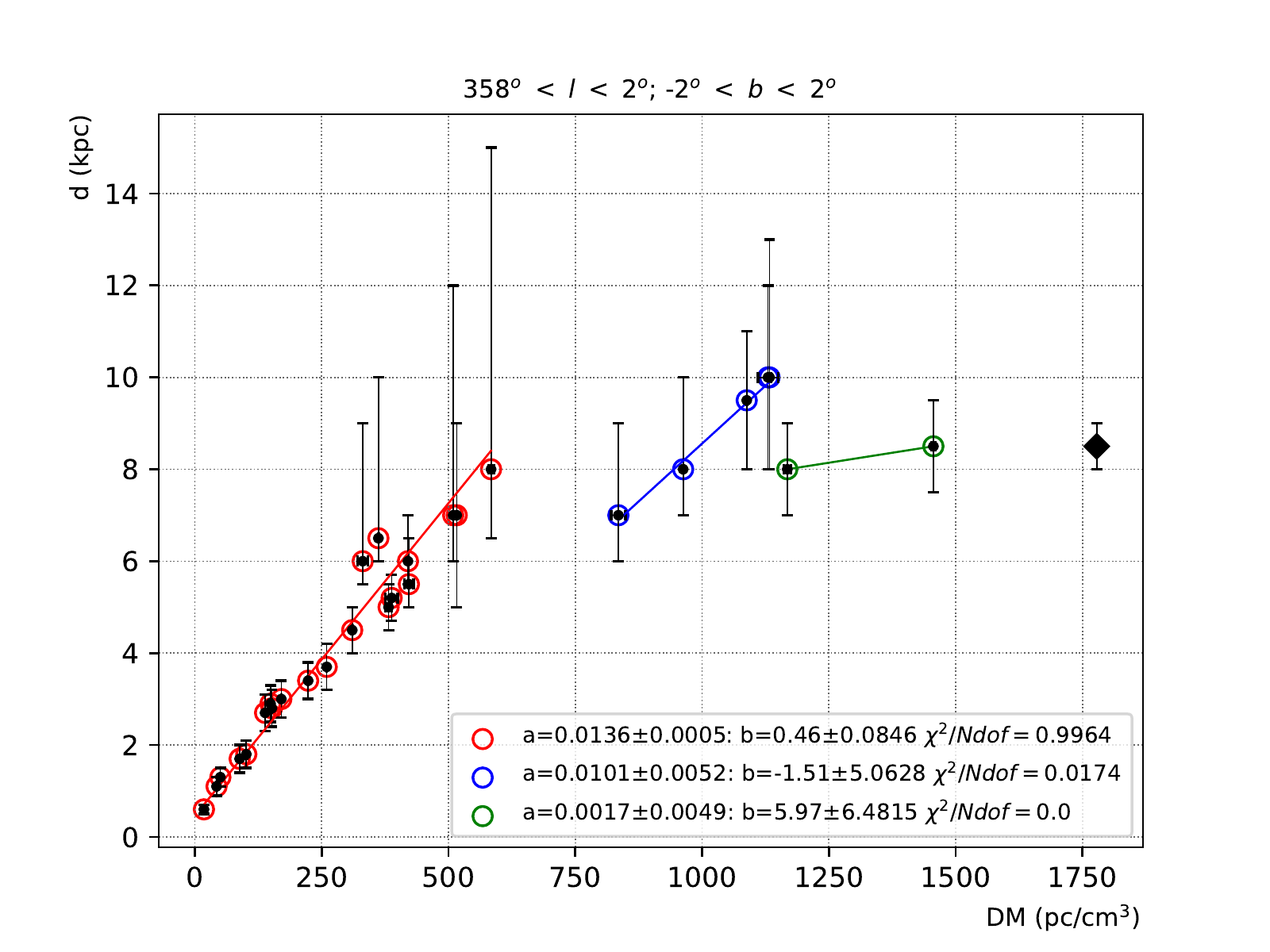}
	\caption{Distance-DM relations for radio pulsars in the solid angle interval 358$\degr$ $<$ $\ell$  $<$ 2$\degr$; -2$\degr$ $<$ $b$ $<$ 2$\degr$. The data are fit to linear functions (shown in red, blue and green) with slopes a, and intercepts b.}
	\label{fig:1}
\end{figure}

Based on our adopted distances, J1749-3002 must be more luminous than the Crab pulsar by a factor of 2.8-4.8, while J1741-3016 and J1746-2850 must be more luminous or comparable to the Crab. J1747-2802, J1740-3015 and J1748-3009 may be comparable to and J1745-3040 must be comparable to or less luminous than the Crab pulsar (see Table~\ref{tab:1} and Figure~\ref{fig:1}).

\begin{table}
	\centering
	\caption{Radio pulsars in the solid angle interval 358$\degr$ $<$ $\ell$  $<$ 2$\degr$; -2$\degr$ $<$ $b$ $<$ 2$\degr$ with measured DM. In the first column J--Name of the pulsar is given. The second and third columns are galactic latitude ($\ell$) and longitude (b) of the pulsar in degrees. The fourth and fifth columns are dispersion measure (DM) values and their error in pc/cm$^3$. The sixth and seventh columns are rotation measure (RM) values and their error in rad/m$^2$. The eighth column is the characteristic ages ($\tau$) of the pulsars in Myr. The ninth column is the ratio of rotation measure to dispersion measure (RM/DM). The tenth and eleventh columns are the adopted distances (d) and their error in kpc.  The twelfth column is the Crab limit for the distance (d$_{Crab}$) in kpc. The thirteenth column is the distance to the Galactic plane (z) calculated from the adopted distance in pc. The symbol $^{\dagger}$ indicates the Crab (S400) luminosity limit.}

	\label{tab:1}
	\begin{adjustbox}{width=0.48\textwidth}

	\begin{tabular}{lccccccc|ccccr} 

		\hline

		\multicolumn{8}{c|}{Data from ATNFPC} & \multicolumn{5}{c}{Adopted and Calculated Data in this Study} \\ \hline

		PSR & $\ell$ & b &  DM & err & RM & err & $\tau$ & RM/DM & d & err & d$_{Crab}$ & z \\

		& degree & degree & pc/cm$^3$ & pc/cm$^3$ & rad/m$^2$ & rad/m$^2$ & Myr & & kpc & kpc   & kpc & pc\\

		\hline

		J1753-28	&	0.887	&	-1.377	&	18	&	$\pm$0.9	&	-	&	-	&	-	&	-	&	0.6	&	$\pm$0.1	&	-	&	-14	\\

		J1751-2857	&	0.646	&	-1.124	&	42.84	&	$\pm$0.03	&	-	&	-	&	5530	&	-	&	1.1	&	$\pm$0.2	&	31	&	-22	\\

		J1752-2806	&	1.54	&	-0.961	&	50.372	&	$\pm$0.008	&	96	&	$\pm$0.2	&	1.1	&	1.9	&	1.3	&	$\pm$0.2	&	1.4$^{\dagger}$, 1.08	&	-22	\\

		J1745-3040	&	358.553	&	-0.963	&	88.373	&	$\pm$0.004	&	101	&	$\pm$7.0	&	0.546	&	1.1	&	1.7	&	$\pm$0.3	&	5.8$^{\dagger}$, 1.6	&	-29	\\

		J1747-2958	&	359.305	&	-0.841	&	101.5	&	$\pm$1.6	&	-	&	-	&	0.0255	&	-	&	1.8	&	$\pm$0.3	&	15	&	-26	\\

		J1739-2903	&	359.206	&	1.064	&	138.55	&	$\pm$0.02	&	-236	&	$\pm$18	&	0.65	&	-1.7	&	2.7	&	$\pm$0.4	&	4	&	50	\\

		J1741-2733	&	0.636	&	1.582	&	149.2	&	$\pm$1.7	&	-89.6	&	$\pm$5.1	&	95.9	&	-0.6	&	2.9	&	$\pm$0.4	&	6	&	80	\\

		J1740-3015	&	358.294	&	0.238	&	151.96	&	$\pm$0.01	&	-168	&	$\pm$0.7	&	0.0206	&	-1.1	&	2.8	&	$\pm$0.4	&	9.5$^{\dagger}$, 2.5	&	12	\\

		J1739-3023	&	358.085	&	0.336	&	170.5	&	$\pm$0.1	&	-74	&	$\pm$18	&	0.159	&	-0.4	&	3	&	$\pm$0.4	&	7	&	18	\\

		J1738-2955	&	358.378	&	0.724	&	223.4	&	$\pm$0.6	&	-200	&	$\pm$20	&	0.0857	&	-0.9	&	3.4	&	$\pm$0.4	&	15	&	43	\\

		J1751-2737	&	1.761	&	-0.376	&	260	&	$\pm$0.1	&	-	&	-	&	-	&	-	&	3.7	&	$\pm$0.5	&	-	&	-24	\\

		J1741-2945	&	358.798	&	0.38	&	310.3	&	$\pm$1.2	&	-531.1	&	$\pm$9.1	&	5.58	&	-1.7	&	4.5	&	$\pm$0.5	&	10	&	30	\\

		J1736-2843	&	359.142	&	1.764	&	331	&	$\pm$10	&	-491.3	&	$\pm$7.6	&	3.4	&	-1.5	&	6	&	+3.0, -0.5	&	11	&	246	\\

		J1741-2719	&	0.907	&	1.604	&	361.9	&	$\pm$0.4	&	-	&	-	&	68.2	&	-	&	6.5	&	+3.5, -0.5	&	17	&	252	\\

		J1741-3016	&	358.346	&	0.13	&	382	&	$\pm$6	&	-449.9	&	$\pm$3.4	&	3.34	&	-1.2	&	5	&	$\pm$0.5	&	5	&	11	\\

		J1750-28	&	0.663	&	-0.737	&	388	&	$\pm$12	&	-	&	-	&	3.56	&	-	&	5.2	&	$\pm$0.5	&	25	&	-67	\\

		J1748-3009	&	359.272	&	-1.147	&	420.2	&	$\pm$0.01	&	-	&	-	&	-	&	-	&	6	&	+1.0, -0.5	&	6	&	-120	\\

		J1746-27	&	0.971	&	0.494	&	422	&	$\pm$9	&	-	&	-	&	-	&	-	&	5.5	&	+1.0, -0.5	&	15	&	47	\\

		J1749-3002	&	359.459	&	-1.244	&	509.4	&	$\pm$0.3	&	-290	&	$\pm$3.0	&	1.23	&	-0.6	&	7	&	+5.0, -1.0	&	13.2$^{\dagger}$, 3.7	&	-217	\\

		J1752-2821	&	1.267	&	-0.976	&	516.3	&	$\pm$1.3	&	-	&	-	&	2.93	&	-	&	7	&	+5.0, -1.0	&	13	&	-170	\\

		J1748-30	&	359.117	&	-1.141	&	584	&	$\pm$5	&	-	&	-	&	-	&	-	&	8	&	+7.0, -1.5	&	19	&	-239	\\

		J1747-2802	&	0.971	&	0.121	&	835	&	$\pm$14	&	421	&	$\pm$26	&	18.6	&	0.5	&	7	&	+2.0, -1.0	&	11	&	15	\\

		J1746-2850	&	0.134	&	-0.044	&	962.7	&	$\pm$0.7	&	-	&	-	&	0.0127	&	-	&	8	&	+2.0, -1.0	&	8	&	-6	\\

		J1745-2910	&	359.763	&	-0.054	&	1088	&	$\pm$0.1	&	-	&	-	&	-	&	-	&	9.5	&	$\pm$1.5	&	-	&	-9	\\

		J1745-2912	&	359.788	&	-0.175	&	1130	&	$\pm$20	&	-535	&	$\pm$110	&	-	&	-0.5	&	10	&	$\pm$2.0	&	-	&	-31	\\

		J1747-2809	&	0.869	&	0.076	&	1133	&	$pm$3	&	-	&	-	&	0.00531	&	-	&	10	&	+3.0, -2.0	&	-	&	13	\\

		J1746-2856	&	0.126	&	-0.233	&	1168	&	$\pm$7	&	13253	&	$\pm$53	&	1.2	&	11.3	&	8	&	$\pm$1.0	&	-	&	-33	\\

		J1746-2849	&	0.134	&	-0.03	&	1456	&	$\pm$3	&	10104	&	$\pm$100	&	1.84	&	6.9	&	8.5	&	$\pm$1.0	&	12	&	-4	\\

		J1745-2900	&	359.944	&	-0.047	&	1778	&	$\pm$3	&	-66080	&	$\pm$24	&	0.0034	&	-37.2	&	8.5	&	$\pm$0.5	&	-	&	-7	\\

		\hline

	\end{tabular}

\end{adjustbox}
\end{table}	

\subsection{2$\degr$ $<$ $\ell$  $<$ 6$\degr$; -2$\degr$ $<$ $b$ $<$ 2$\degr$}
\label{sec:3.2}
Out of the 30 pulsars in this group, 9 are young ($\tau$$<$ 8$\times$$10^5$ yr), though it must be noted that 4 of the pulsars have unknown $\tau$ (see Table~\ref{tab:2}). The solid angle intersects the Sagittarius, Scutum and Norma arms as well as the Galactic bulge and unknown possible arms beyond the bulge.

There is a Scutum arm source (G5.88-0.39) located at d = 2.82 -- 3.18 kpc from parallax measurements \citep{sato2014trigonometric,reid2014trigonometric,reid2019}.

The young pulsar J1801-2451 ($\ell$=5.254$\degr$, b= -0.882$\degr$, DM=291.55 pc/cm$^3$, d=4.5 kpc, $\tau$ = 1.55$\times$10$^4$ yr, see Table~\ref{tab:2}) has V$_{trans}$=199.15 km/s  at d=3.803 kpc (ATNFPC) and z=-69 pc at d=4.5 kpc. There may be a deviation of the SFR from the Galactic plane in the direction of this pulsar at about d=4-5 kpc as J1801-2451 is possibly connected to SNR G5.27-0.9 (d=5 kpc (proper motion) \citealt{thorsett2002decapitating,guseinov2003observational}). PSR -- SNR G5.4-1.2 connection is established in \citet{frail1991unusual} (d$>$4.3 kpc \citealt{frail1994radio}, d $\sim$ 5 kpc \citealt{caswell1987galactic}, d=4.5 kpc \citealt{guseinov2003observational}, d>4.3 kpc (HI absorption suggests) \citealt{green2019,thorsett2002decapitating}).

Pulsar J1748-2446A ($\ell$=3.836$\degr$, b= 1.696$\degr$, DM=242.15 pc/cm$^3$, d=5 kpc, $\tau$ is unknown, (see Table~\ref{tab:2})) is a member of the globular cluster Ter 5 including 42 pulsars none of which has a measured $\tau$ and this globular cluster has a large uncertainty in its distance measurement \citep[d=4.6--10.3 kpc][]{harris1996catalog,cohn2002hubble,Ortolani1996,Ortolani2001,Ortolani2007,ferraro2009two,kong2010fermi,baumgardt2019mean,valenti2007near}.

Five of the pulsars in this group may be more luminous than the Crab pulsar. See Figure~\ref{fig:2} for distance-DM relation for this solid angle interval.

\subsection{6$\degr$ $<$ $\ell$ $<$ 10$\degr$; -2$\degr$ $<$ $b$ $<$ 2$\degr$}

There are 26 pulsars, 7 of which are young ($\tau$$<$ 8$\times$10$^5$ yr, see Table~\ref{tab:3}). The solid angle intersects the Sagittarius, Scutum and Norma arms, possibly also intersecting the Perseus arm and the distant parts of the Sagittarius and Scutum arms beyond the Galactic central region.

There is a Norma arm source \citep{reid2014trigonometric,reid2019} in this interval: G9.62+0.19 (d = 4.6 -- 5.8 kpc \citealt{sanna2009}).

PSR J1756-2251 (d=2.1 kpc) has V$_{trans}$=8.38 km/s at d=0.73 kpc (ATNFPC) (0.49-1.33 kpc, parallax \citep{yao2017new}) and PSR J1802-2124 (d=2.5 kpc) has V$_{trans}$=17.564 km/s at d=0.76 kpc (ATNFPC). For pulsars J1753-2240 (d=2.6 kpc) and J1759-2205 (d=2.9 kpc), the ATNFPC adopted distances are 3.23 and 3.26 kpc, respectively. Therefore, the adopted distances in ATNFPC for pulsars J1756-2251 and J1802-2124 are probably underestimated values. The actual speeds of these pulsars must then be not so small.

Pulsar J1803-2137 has V$_{trans}$=392.261 km/s at d=4.4 kpc (ATNFPC). The actual distance of this pulsar may be 1-2 kpc smaller than the value adopted in ATNFPC, making its speed more comparable to the average pulsar speed.

One of the young pulsars is J1801-2304 ($\ell$=6.837$\degr$, b= -0.066$\degr$, DM=1073.9 pc/cm$^3$, d=13.5 kpc, $\tau$ = 5.83$\times$10$^4$ yr, see Table~\ref{tab:3}) in the direction of the SNR W28 (G6.4-0.1) (d=2.5 kpc  \citep{guseinov2003observational}, d = 1.9 kpc (HI absorption) \citep{green2019,velazquez2002} and G7.0-0.1 (d=9.5 kpc \citep{guseinov2003observational}). As the DM value of the pulsar is too large, a physical connection to SNR W28 is unlikely, so it may simply be a chance projection.  A physical connection to SNR G7.0-0.1 considering the uncertainties in the distances is not ruled out but it is not very probable. Considering its position and the large DM, this young pulsar is probably related to a distant unobserved SNR beyond the Galactic central region.

Six of the pulsars in this group may be more luminous than the Crab pulsar at 1400 MHz (J1803-2137, J1812-2102, J1808-2057, J1759-2302, J1805-2032, J1801-2304) None of them are more luminous than the Crab pulsar at 400 MHz according to our adopted distances.

\subsection{10$\degr$ $<$ $\ell$ $<$ 14$\degr$; -2$\degr$ $<$ $b$ $<$ 2$\degr$}

There are 27 pulsars in this group, from which 6 are young ($\tau$$<$ 8$\times$10$^5$ yr) (see Table~\ref{tab:4}). The solid angle intersects the Sagittarius, Scutum, Norma, Perseus and Cygnus arms as well as the outer parts of the Galactic bulge region. Two of the young pulsars, J1809-1943 ($\ell$=10.727$\degr$, b=-0.158$\degr$, DM=178 pc/cm$^3$, d=3 kpc, $\tau$ = 3.1$\times$10$^4$ yr) and J1809-1917 ($\ell$=11.094$\degr$, b= 0.08$\degr$, DM=197.1 pc/cm$^3$, d=3.3 kpc, $\tau$ = 5.13$\times$10$^4$ yr), are located at the Scutum arm (see Table~\ref{tab:4}). The other young pulsars are probably located beyond the Norma arm.

There is a Sagittarius arm source \citep{reid2014trigonometric,reid2019,wu2014trigonometric} in this interval: G11.49-1.48 (d = 1.2 -- 1.3 kpc, arm offset = 120 -- 220 pc).
There are also several Scutum arm sources \citep{reid2014trigonometric,reid2019}: G11.91-0.61 (d = 3.05 -- 3.76 kpc \citep{sato2014trigonometric}); G12.68-0.18 (d = 2.25 -- 2.57 kpc \citep{immer2013}); G12.8-0.2 (d = 2.64 -- 3.27 kpc \citep{immer2013}); G12.88+0.48 (IRAS 18089-1732) (d = 2.27 -- 2.78 kpc \citep{immer2013}); G12.9-0.24 (d = 2.31 -- 2.61 kpc \citep{immer2013}); G12.9-0.26 (d = 2.34 -- 2.75 kpc \citep{immer2013}); G13.87+0.28 (d = 3.6 -- 4.35 kpc \citep{sato2014trigonometric}). Two sources located at the possible 3-kpc arm and a source at a 'connecting arm' are \citep{reid2014trigonometric,reid2019,sanna2014}: G10.62-0.38 (W31) (d = 4.52 -- 5.46 kpc); G12.02-0.03 (d = 8.77 -- 10.2 kpc) and G10.47+0.02 (d = 8 -- 9.2 kpc).

J1806-1920 is more luminous than the Crab pulsar by a factor of 1.4-5.2 times and J1809- is at least as luminous as Crab. J1811-1736 and J1814-1744 has a luminosity comparable to that of Crab, whereas J1801-1855 and J1811-1835 may be at most as luminous as  the Crab pulsar.

Distance for J1809-1943 from parallax measurements is given as 2.5${^{+0.4}_{-0.3}}$ kpc in \citet{ding2020magnetar}. Our adopted distance (3kpc$\pm$0.3 (see Table~\ref{tab:4} and Figure~\ref{fig:4})) for the pulsar is in agreement with the parallax distance.

\subsection{14$\degr$ $<$ $\ell$ $<$ 18$\degr$; -2$\degr$ $<$ $b$ $<$ 2$\degr$}

Only five of the 28 pulsars in this group are young ($\tau$$<$ $8\times$ 10$^5$ yr) (see Table~\ref{tab:5}). This interval intersects the Sagittarius (twice), Scutum (twice) and Norma arms as well as the Galactic bulge region. The young pulsars are located at or close to these arms and the Galactic bulge. It is possible that J1821-1419 may be located at a distant arm (Perseus?) beyond far-Sagittarus.

The Sagittarius arm sources \citep{reid2014trigonometric,reid2019,wu2014trigonometric} in this interval are: G14.33-0.64 (d = 1.01 -- 1.26 kpc, arm offset = 200 -- 440 pc \citealt{sato2010}); G14.63-0.57 (d = 1.77 -- 1.9 kpc, arm offset = -400 -- -260 pc); G15.03-0.67 (M 17) (d = 1.86 -- 2.11 kpc, arm offset = -570 -- -330 pc \citealt{xu2011}).
There is only one Scutum arm source \citet{reid2014trigonometric,reid2019,sato2014trigonometric}: G16.58-0.05 (d = 3.31 -- 3.9 kpc).

Although, the solid angle cuts through different parts of several arms, the number of young pulsars is relatively low, possibly due to the fact that the intersection depths through the arms are small.

Pulsars J1818-1422, J1818-1519 and J1820-1346 are more luminous than the Crab pulsar by a factor of 3.9-13.1, 2-7 and 1.7-5.9 times, respectively. J1819-1408 is probably at least as luminous as the Crab pulsar. Pulsars J1818-1541, J1817-1511, J1820-1529, J1825-1446, J1821-1419, J1822-1400, J1823-1347 may have luminosity comparable to the Crab pulsar's based on their adopted distances.

Pulsars J1818-1519 ($\ell$=15.55$\degr$, b= 0.192$\degr$, DM=845 pc/cm$^3$, d=10.5 kpc, $\tau$ = 3.63$\times$10$^6$ yr) and J1818-1422 ($\ell$=16.405$\degr$, b= 0.61$\degr$, DM=622 pc/cm$^3$, d=8.1 kpc, $\tau$ = 2.27$\times$10$^6$ yr) (see Table~\ref{tab:5}) are most probably more luminous than the Crab pulsar based on our adopted distances.

\subsection{18$\degr$ $<$ $\ell$ $<$ 22$\degr$; -2$\degr$ $<$ $b$ $<$ 2$\degr$}

In this group, 32 pulsars are included 9 of which are young ($\tau$$<$ $8\times$ 10$^5$ yr) (see Table~\ref{tab:6}). The youngest one, J1833-1034 ($\ell$=21.501$\degr$, b= -0.885$\degr$, DM=169.5 pc/cm$^3$, d=3.5 kpc, $\tau$ = 4.85$\times$10$^3$ yr) (see Table~\ref{tab:6})), is probably physically connected to the C-type SNR G21.5-0.9 (d(kinematic)$\sim$4.8 kpc \citealt{tian2008}, d = 4.3 kpc, \citealt{lee2019near} and references therein, d = 4.4$^{+0.2}_{-0.2}$ (HI absorption) \citealt{ranasinghe2018revised}, d>2.9 kpc (optical) \citealt{shan2018}). This pulsar is probably located at a slightly deviated part (z=-54 pc at 3.5 kpc (see Table~\ref{tab:6})) of the Scutum arm in a dense star formation region in between and possibly connected to the Sagittarius and the Norma arms. The other young pulsars of the group are located at or close to these three arms (J1822-1252 may be at Perseus or far-Sagittarius).

J1828-1101 is more luminous than the Crab pulsar up to a factor of 4.7. J1823-1115, J1824-1118, J1827-0958 and J1832-1021 are probably more luminous but may be comparable in the limit, whereas J1822-1252, J1823-1126, J1831-1223 and J1833-1055 are at most as luminous as the Crab pulsar.

\subsection{22$\degr$ $<$ $\ell$ $<$ 26$\degr$; -2$\degr$ $<$ $b$ $<$ 2$\degr$}

This solid angle intersects the Sagittarius and Scutum arms twice, possibly also intersecting the edge of the Norma arm at about $\ell$=22$\degr$ which is possibly separated from the Scutum at most by 1 kpc. As large parts of star formation regions are included within this interval, 16 young pulsars are present in the group out of a total of 41 pulsars (see Table~\ref{tab:7}). The young pulsars ($\tau$$<$ 8$\times$10$^5$ yr) are located mostly at or close to the Scutum arm, while several of them are probably related to the Norma or the Sagittarius arms.

There is one Scutum arm source and three Norma arm sources \citep{sato2014trigonometric,reid2014trigonometric,reid2019,brunthaler}: G25.7+0.04 (d = 7.87 -- 14.49 kpc) and G23.0-0.41 (d = 4.26 -- 4.98 kpc); G23.44-0.18 (d = 4.95 -- 7.25 kpc); G23.7-0.19 (d = 5.4 -- 7.3 kpc). High-mass Star Forming Region (HMSFR) source G23.65-0.12 (d = 2.84 -- 3.65 kpc \citealt{reid2014trigonometric,reid2019,bartkiewicz}) may be a Scutum arm source or in between the Scutum and the Sagittarius.

J1833-0827, a young radio pulsar ($\tau$ = 1.47$\times$10$^5$ yr), is probably more luminous than the Crab pulsar up to a factor of 3.5 based on our adopted distance. J1836-1008 ($\tau$ = 7.56$\times$10$^5$ yr), J1827-0750 ($\tau$ = 2.77$\times$10$^6$ yr) and J1840-0809 ($\tau$ = 6.44$\times$10$^6$ yr) (see Table~\ref{tab:7}) are possibly more luminous than the Crab pulsar. J1832-0827, J1840-0753, J1832-0644, J1835-0643, J1837-0653, J1839-0643 and J1840-0643 have luminosity possibly comparable to o that of Crab. J1834-0633 and J1834-0602 are probably less luminous than the Crab pulsar but may possibly be comparable in the limit.

\subsection{26$\degr$ $<$ $\ell$ $<$ 30$\degr$; -2$\degr$ $<$ $b$ $<$ 2$\degr$}

There are 50 pulsars in this interval. About 1/3 of these pulsars are young ($\tau$$<$ 8$\times$10$^5$ yr) (see Table~\ref{tab:8}) as this interval intersects the Scutum longitudinally including a large number of HII regions as well as molecular clouds. All but three of the young pulsars are at or close to the Scutum arm; J1849-0317 is definitely at the Sagittarius arm, J1844-0538 is at either Scutum or a distant part of Sagittarius, and J1838-0453 is either at Scutum, far-Sagittarius or even far-Perseus.

In this interval, the Scutum arm sources are \citep{reid2014trigonometric,reid2019}: G27.36-0.16 (d = 5.99 -- 12.05 kpc \citealt{sato2014trigonometric}); G28.86+0.06 (d = 6.54 -- 8.55 kpc); G29.86-0.04 (d = 5.52 -- 7.09 kpc \citealt{sato2014trigonometric}); G29.95-0.01 (W 43S) (d = 4.78 -- 5.84 kpc \citealt{sato2014trigonometric}).

Young and distant pulsar J1838-0453 ($\ell$=27.07$\degr$, b= 0.708$\degr$, DM=617.2 pc/cm$^3$, d=10.5 kpc, $\tau$ = 5.19$\times$10$^4$ yr) has z=130 pc (see Table~\ref{tab:8}), probably located at a deviated part of an SFR.

This group includes several possibly luminous pulsars as compared to the Crab pulsar: J1844-0538 may be up to about 5 times as luminous as the Crab pulsar. J1842-0612, J1838-0453 and J1848-0511 are probably less luminous than Crab pulsar but may be comparable in the limit. J1843-0355 can be as luminous, less luminous or up to 3 times more luminous than the Crab pulsar within distance uncertainties.

\subsection{30$\degr$ $<$ $\ell$ $<$ 34$\degr$; -2$\degr$ $<$ $b$ $<$ 2$\degr$}

Seven out of 31 pulsars in this group are young ($\tau$$<$ 8$\times$10$^5$ yr) (see Table~\ref{tab:9}) and probably located at or close to the Scutum arm.

The Scutum arm sources are \citep{reid2014trigonometric,reid2019,sato2014trigonometric,zhang2014}: G31.28+0.06 (d = 3.66 -- 5.12 kpc); G31.58+0.07 (W 43Main) (d = 4.27 -- 5.75 kpc); G32.04+0.05 (d = 4.97 -- 5.4 kpc). HMSFR source G33.64-0.22 (d = 5.88 -- 7.35 kpc \citealt{reid2014trigonometric,reid2019}) is probably a Scutum arm source (there is longitudinal intersection of the LOS through a large portion of the Scutum arm).

PSR J1844+00 is up to an order of magnitude more luminous than the Crab pulsar. Pulsars J1852+0031, J1850-0026 and J1848-0123 are possibly 1.4-2.8, 1.4-2.7 and 1.3-2.4 times as luminous as the Crab pulsar, respectively.

\subsection{34$\degr$ $<$ $\ell$ $<$ 38$\degr$; -2$\degr$ $<$ $b$ $<$ 2$\degr$}

Thirteen of the 47 pulsars in this group are young ($\tau$$<$ 8$\times$10$^5$ yr) (see Table~\ref{tab:10}). Young pulsar J1856+0113 ($\ell$=34.56$\degr$, b= -0.497$\degr$, DM=96.74 pc/cm$^3$, d=2.8 kpc, $\tau$ = 2.03$\times$10$^4$ yr) is physically connected to the C-type SNR W44 (G34.7-0.4, HI measurements indicates a distance of about 3 kpc \citep{radhakrishnan1972,green2019}, (d = 2.8 kpc \citealt{guseinov2003observational,lee2019near} and references therein, d = 3$^{+0.3}_{-0.3}$ kpc (HI absorption) \citealt{ranasinghe2018revised}, d=2.1$\pm$0.2 kpc (optical) \citealt{shan2018}). This pulsar-SNR pair is located at the Sagittarius arm which includes sources at similar distances. The other young pulsars of the group are probably located at or close to either the Sagittarius or the Scutum arms. Young and distant pulsar J1857+0210 ($\ell$=35.586$\degr$, b= -0.393$\degr$, DM=783 pc/cm$^3$, d=11 kpc, $\tau$ = 7.12 10$^5$ yr, see Table~\ref{tab:10}) may possibly be located close to the Perseus arm instead of the Sagittarius arm.

The Sagittarius arm sources are \citep{reid2014trigonometric,wu2014trigonometric}: G34.39+0.22 (d = 1.45 -- 1.69 kpc, arm offset = 160 -- 320 pc \citealt{kurayama2011}); G35.02+0.34 (d = 2.12 -- 2.56 kpc, arm offset = -90 -- -350 pc \citealt{reid2009}); G35.19-0.74 (d = 1.99 -- 2.43 kpc, arm offset = -20 -- -260 pc \citealt{zhang2009}); G35.2-1.73 (d = 2.85 -- 3.83 kpc, arm offset = -490 -- -930 pc \citealt{zhang2009}); G37.43+1.51 (d = 1.8 -- 1.96 kpc, arm offset = 70 -- 150 pc \citealt{reid2009}).

The Scutum and the Sagittarius arms are closer to each other in the approximate distance interval 4-8 kpc in these directions increasing the uncertainty of the young pulsar locations \citep{hou2009spiral,hou2014observed}.

Young pulsar J1857+0210 should have a luminosity comparable to that of the Crab pulsar if it is located beyond but close to the Perseus arm, while the younger pulsar J1857+0212 in this direction, probably located at the Sagittarius arm, has 1.2-2.4 times the Crab pulsar’s luminosity. Pulsars J1859+00 and J1903+0135 must be 2.9-8.2 and 2.3-4 times more luminous than the Crab pulsar, respectively. Pulsars J1859+00 and J1903+0135 must be 2.9-8.2 and 2.3-4 times more luminous than the Crab pulsar, respectively. J1901+0331 has luminosity 2.6-4.7 times the Crab pulsar's luminosity both at 400 MHz and 1400 MHz. J1855+0307, J1903+0327 and J1853+0505 may have luminosity comparable to or slightly less than that of the Crab pulsar. Luminous pulsar J1901+0331 has V$_{trans}$=1511.36 km/s at d=7 kpc (ATNFPC).

\subsection{38$\degr$ $<$ $\ell$ $<$ 42$\degr$; -2$\degr$ $<$ b $<$ 2$\degr$}

This crowded group has 55 members including some of the pulsars with the smallest and the largest $\tau$ values. Fourteen of these pulsars are young ($\tau$$<$ 8$\times$10$^5$ yr) which are probably located at or close to the Sagittarius arm. Pulsar J1901+0459 ($\ell$=38.49$\degr$, b=0.087$\degr$, DM=1108 pc/cm$^3$, d=13.5 kpc, $\tau$ = 8.86$\times$10$^5$ yr, see Table~\ref{tab:11}) may be located at or close to the Perseus arm or the Cygnus arm.

Young pulsar J1855+0527 ($\ell$=38.227$\degr$, b= 1.642$\degr$, DM=362 pc/cm$^3$, d=7 kpc, $\tau$ = 8.26$\times$10$^4$ yr, see Table~\ref{tab:11}) is probably located at a deviated part of an SFR as its distance from the Galactic plane is 201 pc.

For recycled millisecond (ms) pulsars J1905+0400 (P=3.8 ms, $\tau$=1.22$\times$10$^{10}$ yr), J1904+0412 (P=71 ms, $\tau$=1.02$\times$10$^{10}$ yr), J1904+0451 (P=6 ms, $\tau$=1.69$\times$10$^{10}$ yr) and J1914+0659 (P=18.5 ms, $\tau$=9.46$\times$10$^9$ yr) (ATNFPC), the real ages are not related to their characteristic ages as such radio pulsars are formed after the X-ray binary phase in close binary systems \citep{bisnovatyi,Bisnovatyi-Kogan}.

Although the number of pulsars is relatively large in this group, there is no known arm source in this interval.

J1901+0435 must be one order of magnitude more luminous than the Crab pulsar and J1906+0641 has luminosity 1.3-2.7 times larger than the Crab's. J1857+0526, J1907+0534, J1905+0600 and J1910+0534 may be as luminous as the Crab pulsar.

\subsection{42$\degr$ $<$ $\ell$ $<$ 46$\degr$; -2$\degr$ $<$ $b$ $<$ 2$\degr$}

The solid angle of this group cuts through the Sagittarius arm including a large number of HII regions and molecular clouds, also intersecting a small part of the Perseus arm and possibly the Cygnus arm. There are 11 young ($\tau$$<$ 8$\times$10$^5$ yr) pulsars (see Table~\ref{tab:12}) all of which are most probably located at or close to the Sagittarius arm in large distance intervals. Two of the 40 pulsars in this group are probably located close to the Perseus arm: J1905+0902 ($\ell$=42.555$\degr$, b=1.056$\degr$, DM=433.4 pc/cm$^3$, d=11 kpc, $\tau$ = 9.88$\times$10$^5$ yr) and J1913+1145 ($\ell$=45.92$\degr$, b=0.476$\degr$, DM=637 pc/cm$^3$, d=11.5 kpc, $\tau$ = 9.67$\times$10$^5$ yr, see Table~\ref{tab:12}).

There are Sagittarius arm sources in this interval \citep{reid2014trigonometric,reid2019,wu2014trigonometric}: G43.79-0.12 (OH 43.8-0.1) (d = 5.84 -- 6.21 kpc, arm offset = -470 -- -530 pc); G43.89-0.78 (d = 7.09 -- 9.9 kpc, arm offset = -770 -- 1210 pc); G45.07+0.13 (d = 7.35 -- 8.19 kpc, arm offset = -10 -- 290 pc); G45.45+0.05 (d = 7.35 -- 9.8 kpc, arm offset = -380 -- 1360 pc). There is only one Perseus arm source: G43.16+0.01 (W 49N) \citep{reid2014trigonometric,reid2019} (d(parallax) = 10.42 -- 11.9 kpc \citep{zhang2013parallaxes}, d(kinematic) = 10.43 -- 11.28 kpc \citep{choi2014trigonometric}).

Dense plasma regions of the Sagittarius arm in this longitude interval are possibly located at: $\ell$=42$\degr$, d=1.7-5.4 kpc, d=8-8.9 kpc; $\ell$=43$\degr$, d=1.7-3.6 kpc, 4.8-5.4 kpc, d=8-8.9 kpc; $\ell$=44$\degr$, d=1.7-2.7 kpc, 8-8.5 kpc; $\ell$=45$\degr$, d=1.7-2.2 kpc, d=4.8-4.9 kpc, d=6.6-8.5 kpc; $\ell$=46$\degr$, 1.7-2.2 kpc, 6-7.6 kpc \citep{anderson2014wise, hou2014observed}.

Young pulsar J1907+0918 ($\ell$=43.024$\degr$, b=0.73$\degr$, DM=357.9 pc/cm$^3$, d=8.5 kpc, $\tau$ = 3.8$\times$10$^4$ yr) with z=108 pc (see Table~\ref{tab:12}) is located at a deviated part of an SFR.

Pulsars J1916+0844, J1915+1009, J1920+1040 and J1913+1145 may have luminosity comparable to the Crab pulsar, while most of the others are definitely much less luminous than the Crab.

\subsection{46$\degr$ $<$ $\ell$ $<$ 50$\degr$; -2$\degr$ $<$ $b$ $<$ 2$\degr$}

This group includes 21 pulsars, three of which are young ($\tau$$<$ 8$\times$10$^5$ yr) and four of which with unknown $\tau$ (see Table~\ref{tab:13}). The solid angle intersects Sagittarius (longitudinal) and Perseus arms. There is no known HMSFR up to about 5 kpc in these directions.

There are 41 HII regions with measured distance within this solid angle \citep{anderson2014wise}: G046.495-00.241 and G049.589-00370 at about 3.8 kpc; 23 HII regions at about 5.3-5.8 kpc; G048.547-00.005 at about 9 kpc; 7 HII regions at about 10 kpc; G046.173+00.533 and G046.213+00.547 at about 11 kpc; G048.719+01.147, G049.408+00.332 and G049.420+00.320 at about 13 kpc; G046.948+00.371 at about 15 kpc; G047.094+00.492 and G047.100+00.479 at about 16 kpc. Only one of these HII regions coincides with the direction of a member of this group: HII region G046.792+00.284 in the LOS of young pulsar J1916+1225.

The Sagittarius arm sources in the interval are \citep{reid2014trigonometric,reid2019,wu2014trigonometric}: G49.19-0.34 (d = 4.98 -- 5.47 kpc, arm offset = -50 -- -90 pc); G49.48-0.36 (W 51 IRS2) (d = 3.75 -- 8.06 kpc, arm offset = -860 -- 740 pc); G49.48-0.38 (W 51M) (d = 5.12 -- 5.71 kpc, arm offset = -60 -- 0 pc). There is one Perseus arm source \citep{reid2014trigonometric,reid2019}: G48.6+0.02 (d(parallax) = 10.2 -- 11.36 kpc \citealt{zhang2013parallaxes}, d(kinematic) = 8.7 -- 9.63 kpc \citealt{choi2014trigonometric}).

\subsection{50$\degr$ $<$ $\ell$ $<$ 54$\degr$; -2$\degr$ $<$ $b$ $<$ 2$\degr$}

In this group, 9 of the 27 pulsars, 5 of which are young, have unknown $\tau$ (see Table~\ref{tab:14}). Relatively large number of pulsars with unmeasured $\dot{P}$ (unknown $\tau$) makes it more difficult to determine the locations and to make comparisons of their luminosity for the pulsars in this group.

Four of the 5 young ($\tau$$<$ 8$\times$10$^5$ yr) pulsars belong to the Sagittarius arm: J1926+1648 ($\ell$=51.859$\degr$, b=0.063$\degr$, DM=176.885 pc/cm$^3$, d=4 kpc, $\tau$ = 5.11 10$^5$ yr), J1928+1746 ($\ell$=52.931$\degr$, b=0.114$\degr$, DM=176.68 pc/cm$^3$, d=4 kpc, $\tau$ = 8.26$\times$10$^4$ yr), J1925+1720 ($\ell$=52.179$\degr$, b=0.59$\degr$, DM=223.3 pc/cm$^3$, d=5.3 kpc, $\tau$ = 1.15 10$^5$ yr), J1922+1733 ($\ell$=52.08$\degr$, b=1.23$\degr$, DM=238 pc/cm$^3$, d=5.4 kpc, $\tau$ = 2.8$\times$10$^5$ yr). Pulsar J1924+1631 ($\ell$=51.405$\degr$, b=0.318$\degr$, DM=518.5 pc/cm$^3$, d=10 kpc, $\tau$ = 1.28$\times$10$^5$ yr), which has the largest DM among the young pulsars of this group, most probably belongs to the Perseus arm (see Table~\ref{tab:14}).

J1926+1648 has V$_{trans}$ = 463.064 km/s at d=6 kpc (d(kinematic) = 4-9 kpc, \citep{yao2017new}) as displayed in ATNFPC. The transverse speed is more comparable to the average space velocity of pulsars (about 300 km/s) at our adopted distance of 4 kpc (see Table~\ref{tab:14} and Figure~\ref{fig:14}).

There is only one Sagittarius arm source \citep{reid2014trigonometric,reid2019}: G52.1+1.04 (IRAS 19213+1723) (d = 3.48 -- 4.65 kpc, arm offset = 50 -- 210 pc \citealt{wu2014trigonometric},d = 5.62--6.58 kpc \citealt{wu2019}).

All the pulsars with measured flux in this group are less luminous than the Crab pulsar.

\subsection{54$\degr$ $<$ $\ell$ $<$ 58$\degr$; -2$\degr$ $<$ $b$ $<$ 2$\degr$}

There are 19 pulsars in this group, four of which are young and four of them have unknown $\tau$ (see Table~\ref{tab:15}). Very young ($\tau$=2890 yr) pulsar J1930+1852 (DM=308 pc/cm$^3$) connected to SNR G54.1+0.3 (d=6$^{+2.2}_{-1.5}$ kpc \citep{guseinov2003observational}, d(optical)=6.3$^{+0.8}_{-0.7}$ kpc \citealt{shan2018,green2019}, d = 4.9 $^{+0.8}_{-0.8}$ (HI abosorption) \citep{ranasinghe2018revised,green2019}, d=8.2 kpc (association with CO) \citealt{lee2012,green2019}), type is C? (i.e. possibly composite type SNR) in Green's catalogue \citep{green2019}) is probably located at Sagittarius arm at 6.7$\pm$0.5 kpc. The other three young pulsars (all with $\tau$$<$6.3$\times$10$^4$ yr) must also belong to the Sagittarius. The four young pulsars have significantly different adopted distances (about 2.5, 4.5, 5.5 and 6.7 kpc) due to the fact that their LOSs intersect the Sagittarius arm longitudinally (see Table~\ref{tab:15}).

PSR J1939+2134 (d(parallax) = 1.5$^{+0.5}_{-0,3}$ \citep{reardon2016timing}, d(parallax) $>$ 3.23, \citep{matthews2016nanograv}), V$_{trans}$=6.755 km/s (d=3.5 kpc (ATNFPC)) has the Crab limits d$_{Crab}<$3.1 kpc (S400), d$_{Crab}$$<$2.1 kpc (S1400). PSR J1939+2134 was the first ms pulsar observed \citep{backer1982millisecond}, and it is currently the second fastest spinning pulsar known \citep{reardon2016timing} with V$_{radial}$=89.16 km/s \citep{matthews2016nanograv} which is isolated due to the disruption during the core-collapse SN). 

For the ms recycled pulsar J1939+2134, we adopted d=1.5-2.5 kpc (see Figure~\ref{fig:15} and Table~\ref{tab:15}) making it slightly less luminous than the Crab pulsar at 400 MHz and comparable to the Crab at 1400 MHz. In ATNFPC, d=3.5 kpc is given as the adopted distance in which case this old (recycled) ms pulsar must be more luminous than the Crab at both 400 MHz and 1400 MHz. As the luminosity of old ms pulsars are mostly much less than the Crab pulsar, d=3.5 kpc is probably an overestimated value.

Young pulsar J1932+2220 ($\ell$=57.356$\degr$, b= 1.554$\degr$, DM=219.2 pc/cm$^3$, d=5.5 kpc, $\tau$ = 3.98$\times$10$^4$ yr (ATNFPC)) has z=149 pc, is probably located at a deviated part of a SF from the Galactic plane (see Table~\ref{tab:15}).

\subsection{58$\degr$ $<$ $\ell$ $<$ 62$\degr$; -2$\degr$ $<$ $b$ $<$ 2$\degr$}

Six of the 14 pulsars are young in this group with one unknown $\tau$ (see Table~\ref{tab:16}).

Young pulsar J1939+2449 ($\ell$=60.174$\degr$, b= 1.361$\degr$, DM=142.88 pc/cm$^3$, d=6 kpc, $\tau$ = 5.6$\times$10$^5$ yr) is probably located close to the far end of the Sagittarius arm in this direction. J1940+2245 ($\ell$=58.629$\degr$, b= 0.127$\degr$, DM=222.4 pc/cm$^3$, d=8.2 kpc, $\tau$ = 3.23$\times$10$^5$ yr), J1940+2337 ($\ell$=59.397$\degr$, b= 0.529$\degr$, DM=252.1 pc/cm$^3$, d=8.8 kpc, $\tau$ = 1.13$\times$10$^5$ yr) and J1948+2333 ($\ell$=60.214$\degr$, b= -1.045$\degr$, DM=198.2 pc/cm$^3$, d=8.2 kpc, $\tau$ = 6.17$\times$10$^5$ yr) are at or close to the Perseus arm. Distant young pulsars J1934+2352 ($\ell$=58.966$\degr$, b= 1.814$\degr$, DM=355.5 pc/cm$^3$, d=14 kpc, $\tau$ = 2.16$\times$10$^4$ yr) and J1941+2525 ($\ell$=61.037$\degr$, b= 1.263$\degr$, DM=314.4 pc/cm$^3$, d=13 kpc, $\tau$ = 2.27$\times$10$^5$ yr) (see Table~\ref{tab:15}) may belong to the Outer or Outer+1 arm despite their large distances from the Galactic plane (about 440 pc and 290 pc, respectively) as the HMSFRs close to these directions have large positive offsets from the Galactic plane \citep{hou2014observed}.

Although, the pulsars in this solid angle interval have large distances with respect to their DM values as compared to the previous groups, all of them are probably less luminous than the Crab pulsar.

There is one Local arm source \citep{reid2014trigonometric,reid2019}: G59.78+0.06 (d = 2.07 -- 2.26 kpc \citealt{xu2009}).

\subsection{62$\degr$ $<$ $\ell$ $<$ 66$\degr$; -2$\degr$ $<$ $b$ $<$ 2$\degr$}

There are 9 pulsars in this group two of which are young (see Table~\ref{tab:17}). The two young pulsars of this group, J1948+2551 ($\ell$=62.207$\degr$, b= 0.131$\degr$, DM=289.27 pc/cm$^3$, d=8.7 kpc, $\tau$ = 3.45$\times$10$^5$ yr) and J1946+2611 ($\ell$=62.321$\degr$, b= 0.595$\degr$, DM=165 pc/cm$^3$, d=6.3 kpc, $\tau$ = 3.14$\times$10$^5$ yr), are probably located at the Perseus arm and close to the Sagittarius arm, respectively, though J1946+2611 may be close to the Perseus arm.

Proper motions of four of the pulsars in this solid angle interval were measured. For three of these pulsars, the distances we have adopted are in agreement with the adopted distances displayed in ATNFPC. J1955+2908 is an old ms pulsar (P=6.133 ms, \.{P}=2.97$\times$10$^{-20}$, $\tau$ = 3.27$\times$10$^9$ yr) which is most probably located at a smaller distance than the one adopted in ATNFPC (d=6.3 kpc) considering its DM as compared to the other members of this group and the Crab limit on its distance (d$_{Crab}<$7.2 kpc) together with its large $\tau$ value. The transverse speeds at our adopted distances are reasonable (100-300 km/s).

All the pulsars in this group are probably less luminous than the Crab pulsar. Pulsar J1948+2551's luminosity may be close to the Crab's, considering the uncertainty in its luminosity.

\subsection{66$\degr$ $<$ $\ell$ $<$ 70$\degr$; -2$\degr$ $<$ $b$ $<$ 2$\degr$}

At least four of the 11 pulsars in this interval are young ($\tau$ is unknown for 5 pulsars (see Table~\ref{tab:18})) that they must be located at or close to the Perseus and the Sagittarius arms based on their DM values. Pulsar J2002+30 of unknown $\tau$  may also be young considering its small latitude (b= -0.187$\degr$) and the possible distance interval (6-8 kpc) making it a member of the Perseus arm.

There is a Local arm source \citep{reid2014trigonometric,reid2019}: G69.54-0.97 (ON 1) (d = 2.39 -- 2.54 kpc) which is not related to the pulsars in this group.

Pulsar J2004+3137, which has the largest DM in this group, is more luminous than the Crab pulsar up to a factor of about 2.2. This pulsar has a high transverse speed (about 400 km/s at d=7.8 kpc) as compared to the average space velocity of pulsars. Pulsar J2002+3217 is possibly as luminous as the Crab within the distance uncertainties.

\subsection{70$\degr$ $<$ $\ell$ $<$ 74$\degr$; -2$\degr$ $<$ $b$ $<$ 2$\degr$}

There are 5 pulsars in this interval. The only young pulsar of this group, J2004+3429 ($\ell$=71.425$\degr$, b= 1.571$\degr$, DM=351 pc/cm$^3$, d=12.5 kpc, $\tau$ = 1.85$\times$10$^4$ yr) (see Table~\ref{tab:19}), must be located in a HMSFR, probably within the Outer arm or the possible Outer+1 arm (see e.g. \citealt{hou2014observed} for the possible structures of distant arms in these directions). The distance of this very young pulsar from the Galactic plane is 288 pc < z < 398 pc, based on the large latitude of this pulsar together with its large distance. This does not prevent this pulsar to locate in a HMSFR within a Galactic arm as there are several giant molecular clouds (GMCs) at about 11.5-13.5 kpc with large positive offsets from the Galactic plane \citep{hou2014observed}.

All the members of this group are less luminous than Crab pulsar, despite the fact that their adopted distances are large.

\subsection{74$\degr$ $<$ $\ell$ $<$ 78$\degr$; -2$\degr$ $<$ $b$ $<$ 2$\degr$}

There are four pulsars in the group, two of them are very young and one of them is young (see Table~\ref{tab:20}).

The Local arm sources are \citep{reid2014trigonometric,reid2019}: G74.03-1.71 (d = 1.55 -- 1.63 kpc \citet{xu2013nature}); G75.76+0.33 (d = 3.26 -- 3.8 kpc \citet{xu2013nature}); G75.78+0.34 (ON 2N) (d = 3.41 -- 4.02 kpc); G76.38-0.61 (d = 1.22 -- 1.39 kpc \citet{xu2013nature}). All these sources are foreground objects with respect to the pulsars in this interval. There is one Outer arm source: G75.29+1.32 (d = 8.8 -- 9.7 kpc \citealt{sanna2012}).

In the ATNFPC, d(kinematic, HI measurements)=1.8 kpc is given as the adopted distance for the very young pulsar J2021+3651 ($\ell$=75.222$\degr$, b= 0.111$\degr$, DM=367.5 pc/cm$^3$, d=11 kpc, $\tau$ = 1.72$\times$10$^4$ yr), which is a significantly underestimated value considering the large DM of this pulsar in this direction. PWN G75.2+0.1 is powered by J2021+3651 \citep{hessels2004observations}. \citet{abdo2013} give d=6-12 kpc for this pulsar. \citet{kirichenko2015optical} give a distance of 1.8 $^{+1.7}_{-1.4}$ kpc for J2021+3651 from X-ray and extinction-distance relation. Comparing the DM of this pulsar with the others in this interval and the neighbouring ones, also considering the possible location and structure of the Outer arm in this direction, we have adopted d=10-12 kpc for J2021+3651 (see Figure~\ref{fig:20} and Table~\ref{tab:20}).

PSR J2022+3842 ($\ell$=76.888$\degr$, b= 0.96$\degr$, DM=429.1 pc/cm$^3$, d=13 kpc, $\tau$ = 8.94$\times$10$^3$ yr, z=218 pc) is another very young pulsar in this interval which has the largest DM among the four pulsars. This pulsar is probably connected to the C-type SNR G76.9+1.0 (d=10$\pm$3 \citealt{arumugasamy2014xmm,arzoumanian2011discovery}, d=12.6 kpc \citealt{guseinov2003observational}). We have adopted d=11-15 kpc (see Figure~\ref{fig:20} and Table~\ref{tab:20}) taking into account the large latitude and the large DM of the pulsar together with the large positive offsets of the GMCs from the Galactic plane close to this direction at large distances \citep{hou2014observed}.

For the young pulsar J2030+3641 ($\ell$=76.123$\degr$, b= -1.438$\degr$, DM=246 pc/cm$^3$, d=9.5 kpc, $\tau$ = 4.88$\times$10$^5$ yr, z=-238 pc), \citet{abdo2013} give d=3$\pm$1 kpc. As the latitude of this pulsar is relatively large and its DM is not so small to locate it at 3 kpc, we have adopted d=8.7-11 kpc (see Figure~\ref{fig:20} and Table~\ref{tab:20}). Since this pulsar is below the Galactic plane, its transverse speed must be high at such large distances. Note that the adopted distance of J2030+3641 is 7 kpc in ATNFPC which leads to a similar conclusion.

All the three young pulsars of the group are possibly located at the Outer arm, though it may also be possible that J2030+3641 is closer to the Perseus arm and J2022+3842 is a member of the possible Outer+1 arm instead of the Outer arm.

Based on the Crab limit on their distances, three of the four pulsars should be less luminous than the Crab pulsar despite their large distances.

\subsection{78$\degr$ $<$ $\ell$ $<$ 90$\degr$; -2$\degr$ $<$ $b$ $<$ 2$\degr$}

78$\degr$ $<$ $\ell$ $<$ 82$\degr$; -2$\degr$ $<$ $b$ $<$ 2$\degr$

There is only one pulsar in this interval, namely J2032+4127 ($\ell$=80.224$\degr$, b= 1.028$\degr$, DM=114.67 pc/cm$^3$, d=1.4-1.7 kpc, $\tau$ = 2.01$\times$10$^5$ yr) (see Table~\ref{tab:21}). This young pulsar is a member of the OB-association Cygnus OB2 at d=1.4-1.7 kpc \citep{lyne2015binary}.

There are several Local arm sources in this interval \citep{reid2014trigonometric,reid2019}: G78.88+0.7 (AFGL 2591) (d = 3.23 -- 3.45 kpc \citealt{rygl2012}) ; G79.73+0.99 (IRAS 20290+4052) (d = 1.25 -- 1.48 kpc \citealt{rygl2012}); G79.87+1.17 (d = 1.55 -- 1.69 kpc \cite{xu2013nature}); G80.79-1.92 (NML Cyg) (d = 1.5 -- 1.75 kpc \citealt{zhang2012x}); G80.86+0.38 (DR 20) (d = 1.38 -- 1.55 kpc \citealt{rygl2012}); G81.75+0.59 (DR 21) (d = 1.43 -- 1.58 kpc \citealt{rygl2012}); G81.87+0.78 (W 75N) (d = 1.23 -- 1.37 kpc \citealt{rygl2012}). Therefore, young pulsar J2032+4127 must be a member of the Local arm. Note that the DM-based distance in YMW17 model is 4.62 kpc, though a distance of 1.33 kpc is adopted for this pulsar in ATNFPC.

82$\degr$ $<$ $\ell$ $<$ 86$\degr$; -2$\degr$ $<$ $b$ $<$ 2$\degr$

There is no radio pulsar observed in this solid angle interval which is a surprising fact since the LOSs pass through both the Local arm, the Perseus arm and the Outer arm including several SFRs. The reason may partly be the large positive offsets of the HMSFRs from the Galactic plane and also the lack of pulsar surveys for these directions.

86$\degr$ $<$ $\ell$ $<$ 90$\degr$; -2$\degr$ $<$ $b$ $<$ 2$\degr$

There are two pulsars observed in this interval; J2113+4644 ($\ell$=89.003$\degr$, b= -1.266$\degr$, DM=141.26 pc/cm$^3$, d=1.9-2.6 kpc, $\tau$ = 2.25$\times$10$^7$ yr) which has the Crab limits on its distance d$_{Crab}<$3.1 kpc (S400) and d$_{Crab}<$1.7 kpc (S1400) and J2053+4650 ($\ell$=86.861$\degr$, b= 1.302$\degr$, DM=98.0828 pc/cm$^3$, $\tau$ = 1.16$\times$10$^9$ yr) (see Table~\ref{tab:21}). Since J2113+4644 is possibly a middle aged pulsar (10$^6$ yr < $\tau$ < 10$^7$ yr) with a negative latitude in this direction, its distance may be larger with respect to its DM. The kinematic distance from HI absorption measurements is d=4$\pm$1 kpc which is in accordance with the YMW17 model but the parallax measurement \citep{deller2019microarcsecond} was used for the adopted distance (d=2.2 kpc) in ATNFPC. This pulsar is probably located at 2.5-3 kpc with the luminosity on the same order of magnitude as Crab's at both S400 and S1400.

\section{Conclusions}

Constructing distance vs DM relations is essential to determine the 3D distributions of both radio pulsars and the plasma in the Galaxy. In order to make such relations as reliable as possible, we have considered some basic criteria as presented above. Galactic distributions of the clouds and the plasma are still not well known as seen in the models on the Galactic structure. These models include significant differences with respect to each other in the locations, depths and widths of the arms even though the amount of observational data has greatly increased in recent years. This is the main factor for large uncertainties in the distances of many pulsars, especially the distant ones. Using the fact that young pulsars are located in or close to star formation regions helps to decrease these uncertainties by making comparisons. Consequently, these young pulsars are used as calibrators to improve the distance values of older ones in the same solid angle intervals.

The very large values of DM for some of the pulsars (in particular the magnetar J1745-2900) and smaller but still large DM for some other pulsars in the Galactic central directions suggest the existence of very high density plasma with possible `gaps` for the region around the Galactic center. There is no pulsar with measured DM in the interval 600-800 pc/cm$^3$ in the central directions. The number of radio pulsars detected in these directions being surprisingly small strengthens the probability of the presence of large density plasma in the central region.

In Taylor-Cordes \citep{taylor1993pulsar} (553 pulsars) and Cordes-Lazio \citep{ne2001, ne20012} (1143 pulsars) models, there are biases leading to underestimation of the distances for some nearby pulsars and overestimation of the distances for some distant ones. This is true especially in the Taylor-Cordes model which is based on the arm structure introduced in Georgelin and Georgelin (1976). A significant fraction (134 out of 553) of the pulsars seem to be located outside the Galaxy (d$>$30 kpc) according to the Taylor-Cordes model. Pulsars discussed in section ~\ref{sec:3.1} J1745-3040, J1740-3015 and J1752-2806 have underestimated distance values in ATNFPC (see Table ~\ref{tab:1}). 
These distance values lead to an overestimation of the average free electron density for the region up to 0.4 kpc from the Sun in the central directions. The characteristic ages of these three pulsars are 5.46 $\times$ $10^5$ yr, 2.06 $\times$ $10^4$ yr, and 1.1 $\times$ $10^6$ yr, respectively. Although they are young pulsars, with the ATNFPC adopted distances, these pulsars do not seem to be part of any Galactic arm, while with our adopted distances, J1740-3015 (d=2.8 kpc) is most probably located within the Scutum arm and the other two, J1752-2806 (d=1.3 kpc) and J1745-3040 (d=1.7 kpc) which have also similar RM values, are in or close to the Sagittarius arm. Our estimations on the distances of these pulsars based on comparisons with the other nearby and/or young pulsars in this group should give a more reliable average free electron density for the central directions up to about 3 kpc as well as better luminosity and space velocity values for these pulsars.
 
YMW17 electron distribution model (which is currently used as the default distance -- DM model in ATNFPC) includes a large number of calibrator pulsars based on measurements of parallax, kinematic measurements, connections of pulsars to other sources and clusters. ATNFPC adopted distances are significantly different for some pulsars as compared to YMW17, especially the ones based on kinematic measurements. As the minimum and maximum limits on pulsar distances from kinematic measurements differ significantly in many cases, we do not use kinematic distances for the calibration of distance -- DM relations in our model.

\begin{table}
	\centering
	\caption{Average free electron density in the solid angle interval 358$\degr$ $<$ $\ell$ $<$ 78$\degr$; -2$\degr$ $<$ $b$ $<$ 2$\degr$.
		n is the average free electron density in the line of sight, calculated from the inverse of the slopes of Figures~\ref{fig:1}--\ref{fig:20} in 1/cm$^3$.
	In the solid angle interval 358$\degr$ $<$ $\ell$ $<$ 2$\degr$; -2$\degr$ $<$ $b$ $<$ 2$\degr$$^{(1)}$ radio pulsars whose DM values are from 18 pc/cm$^3$ to 584 pc/cm$^3$ (see Figure~\ref{fig:1} and Table ~\ref{tab:1}) are located.   
In the solid angle interval 358$\degr$ $<$ $\ell$ $<$ 2$\degr$; -2$\degr$ $<$ $b$ $<$ 2$\degr$$^{(2)}$ radio pulsars whose DM values are from 835 pc/cm$^3$ to 1133 pc/cm$^3$ (see Figure~\ref{fig:1} and Table ~\ref{tab:1}) are located.   
In the solid angle interval 358$\degr$ $<$ $\ell$ $<$ 2$\degr$; -2$\degr$ $<$ $b$ $<$ 2$\degr$$^{(3)}$ radio pulsars whose DM values are 1168 pc/cm$^3$ and 1456 pc/cm$^3$ (see Figure~\ref{fig:1} and Table ~\ref{tab:1}) are located.}

	\label{tab:22}
	\resizebox{\columnwidth}{!}{\begin{tabular}{lcc} 
			\hline
			Solid Angle Interval & n & err  \\
			& 10$^{5}$/cm$^3$ & 10$^{5}$/cm$^3$ \\
			\hline
			358$\degr$ $<$ $\ell$ $<$2$\degr$; -2$\degr$ $<$ $b$ $<$ 2$\degr$$^{(1)}$	&	7.35	&	$\pm$0.27	\\
			358$\degr$ $<$ $\ell$ $<$2$\degr$; -2$\degr$ $<$ $b$ $<$ 2$\degr$$^{(2)}$	&	9.90	&	$\pm$4.80	\\
			358$\degr$ $<$ $\ell$ $<$2$\degr$; -2$\degr$ $<$ $b$ $<$ 2$\degr$$^{(3)}$	&	5.88 &	$\pm$16.90	\\
			2$\degr$ $<$ $\ell$ $<$ 6$\degr$; -2$\degr$ $<$ $b$ $<$ 2$\degr$ & 8.00	&	$\pm$0.70	\\
			6$\degr$ $<$ $\ell$ $<$ 10$\degr$; -2$\degr$ $<$ $b$ $<$ 2$\degr$ & 8.20	&	$\pm$0.94	\\
			10$\degr$ $<$ $\ell$ $<$ 14$\degr$; -2$\degr$ $<$ $b$ $<$ 2$\degr$ & 8.26	&	$\pm$0.96	\\
			14$\degr$ $<$ $\ell$ $<$ 18$\degr$; -2$\degr$ $<$ $b$ $<$ 2$\degr$ & 8.00	&	$\pm$0.58	\\
			18$\degr$ $<$ $\ell$ $<$ 22$\degr$; -2$\degr$ $<$ $b$ $<$ 2$\degr$ & 7.87	&	$\pm$0.62	\\
			22$\degr$ $<$ $\ell$ $<$ 26$\degr$; -2$\degr$ $<$ $b$ $<$ 2$\degr$ & 7.52	&	$\pm$0.73	\\
			26$\degr$ $<$ $\ell$ $<$ 30$\degr$; -2$\degr$ $<$ $b$ $<$ 2$\degr$ & 7.25	&	$\pm$0.37	\\
			30$\degr$ $<$ $\ell$ $<$ 34$\degr$; -2$\degr$ $<$ $b$ $<$ 2$\degr$ & 9.43	&	$\pm$0.36	\\
			34$\degr$ $<$ $\ell$ $<$ 38$\degr$; -2$\degr$ $<$ $b$ $<$ 2$\degr$ & 8.85	&	$\pm$0.47	\\
			38$\degr$ $<$ $\ell$ $<$ 42$\degr$; -2$\degr$ $<$ $b$ $<$ 2$\degr$ & 6.33	&	$\pm$0.16	\\
			42$\degr$ $<$ $\ell$ $<$ 46$\degr$; -2$\degr$ $<$ $b$ $<$ 2$\degr$ & 5.62	&	$\pm$0.19	\\
			46$\degr$ $<$ $\ell$ $<$ 50$\degr$; -2$\degr$ $<$ $b$ $<$ 2$\degr$ & 4.65	&	$\pm$0.26	\\
			50$\degr$ $<$ $\ell$ $<$ 54$\degr$; -2$\degr$ $<$ $b$ $<$ 2$\degr$ & 5.15	&	$\pm$0.24	\\
			54$\degr$ $<$ $\ell$ $<$ 58$\degr$; -2$\degr$ $<$ $b$ $<$ 2$\degr$ & 5.13	&	$\pm$0.34	\\
			58$\degr$ $<$ $\ell$ $<$ 62$\degr$; -2$\degr$ $<$ $b$ $<$ 2$\degr$ & 2.74	&	$\pm$0.19	\\
			62$\degr$ $<$ $\ell$ $<$ 66$\degr$; -2$\degr$ $<$ $b$ $<$ 2$\degr$ & 3.12	&	$\pm$0.16	\\
			66$\degr$ $<$ $\ell$ $<$ 70$\degr$; -2$\degr$ $<$ $b$ $<$ 2$\degr$ & 3.18	&	$\pm$0.31	\\
			70$\degr$ $<$ $\ell$ $<$ 74$\degr$; -2$\degr$ $<$ $b$ $<$ 2$\degr$ & 4.98	&	$\pm$2.62	\\
			74$\degr$ $<$ $\ell$ $<$ 78$\degr$; -2$\degr$ $<$ $b$ $<$ 2$\degr$ & 4.74	&	$\pm$1.48	\\
			\hline
	\end{tabular}}
\end{table}

Although, there are more than 8000 plasma regions observed up to date (see e.g. WISE catalogue 2021 \citealt{anderson2014wise}), distribution of the HII regions in the Galaxy is highly uncertain and the density distributions within HII regions are unknown. Because of the lack of information on the contribution of each HII to DM and the large uncertainty in their distances, it is possible only to determine the average free electron density in different directions (see  Table~\ref{tab:22}). In order to get the average densities of plasma, all the available data directly or indirectly related to the locations of radio pulsars, HII regions, SNRs and molecular clouds are taken into consideration in this work. Distance -- DM relations for pulsars in solid angle intervals relate the distribution of pulsars to the distribution of plasma in each interval. Making comparisons between pulsars in each solid angle interval improves such relations. As the number of known radio pulsars and their observational data increased significantly in the last decade, it is now possible to get more reliable distance -- DM relations in small solid angle intervals following the works by \citet{guseinov2002trustworthy,guseinov2007}.

\FloatBarrier

\section*{Acknowledgements}

This work was supported by TUBITAK through project no: 115F028.

%%%%%%%%%%%%%%%%%%%%%%%%%%%%%%%%%%%%%%%%%%%%%%%%%%
\section*{Data Availability}

The data underlying this article are available in the article.

%The inclusion of a Data Availability Statement is a requirement for articles published in MNRAS. Data Availability Statements provide a standardised format for readers to understand the availability of data underlying the research results described in the article. The statement may refer to original data generated in the course of the study or to third-party data analysed in the article. The statement should describe and provide means of access, where possible, by linking to the data or providing the required accession numbers for the relevant databases or DOIs.



\bibliographystyle{mnras}
\bibliography{mnras} % if your bibtex file is called example.bib

%\begin{thebibliography}{99}

%\end{thebibliography}

%%%%%%%%%%%%%%%%%%%%%%%%%%%%%%%%%%%%%%%%%%%%%%%%%%

%%%%%%%%%%%%%%%%% APPENDICES %%%%%%%%%%%%%%%%%%%%%

\FloatBarrier
\appendix

\section{Figures and Tables}
\label{app}

Figures and tables for distance-DM relations in the solid angle interval 2$\degr$ $<$ $\ell$  $<$ 90$\degr$; -2$\degr$ $<$ $b$ $<$ 2$\degr$ are given in this section.

\begin{table}
	\centering
	\caption{Radio pulsars in the solid angle interval 2$\degr$ $<$ $\ell$  $<$ 6$\degr$; -2$\degr$ $<$ $b$ $<$ 2$\degr$ with measured DM. See Table~\ref{tab:1} for column descriptions.}
	
	\label{tab:2}
	\begin{adjustbox}{width=0.48\textwidth}
		
\begin{tabular}{lccccccc|ccccr} 
			
			\hline
			
			\multicolumn{8}{c|}{Data from ATNFPC} & \multicolumn{5}{c}{Adopted and Calculated Data in this Study} \\ \hline
			
			PSR & $\ell$ & b &  DM & err & RM & err & $\tau$ & RM/DM & d & err & d$_{Crab}$ & z \\
			
			& degree & degree & pc/cm$^3$ & pc/cm$^3$ & rad/m$^2$ & rad/m$^2$ & Myr & & kpc & kpc   & kpc & pc\\
			
			\hline
			
			J1755-2725	&	2.432	&	-1.135	&	115	&	$\pm$5	&	-	&	-	&	302	&	-	&	2.2	&	+0.4, -0.3	&	11	&	-44	\\
			
			J1750-2536	&	3.413	&	0.778	&	178.4	&	$\pm$0.01	&	-	&	-	&	-	&	-	&	3	&	 +0.4, -0.3	&	12	&	41	\\
			
			J1757-2421	&	5.281	&	0.054	&	179.454	&	$\pm$0.011	&	16	&	$\pm$5.0	&	0.285	&	0.09	&	3	&	$\pm$0.3	&	10.5$^{\dagger}$, 2.8	&	3	\\
			
			J1748-2444	&	3.955	&	1.564	&	207.33	&	$\pm$0.09	&	-	&	-	&	63	&	-	&	3.4	&	$\pm$1.3	&	13	&	93	\\
			
			J1758-2540	&	4.263	&	-0.807	&	218.2	&	$\pm$1.3	&	-207	&	$\pm$9.3	&	21.5	&	-0.95	&	3.4	&	$\pm$0.4	&	9	&	-48	\\
\hline
			
		\end{tabular}
		
	\end{adjustbox}
\end{table}

\begin{table}
	\centering
	\caption{Radio pulsars in the solid angle interval 6$\degr$ $<$ $\ell$  $<$ 10$\degr$; -2$\degr$ $<$ $b$ $<$ 2$\degr$ with measured DM. See Table~\ref{tab:1} for column descriptions.}
	\label{tab:3}
	\begin{adjustbox}{width=0.48\textwidth}
		
		\begin{tabular}{lccccccc|ccccr} 
			
			\hline
			
			\multicolumn{8}{c|}{Data from ATNFPC} & \multicolumn{5}{c}{Adopted and Calculated Data in this Study} \\ \hline
			
			PSR & $\ell$ & b &  DM & err & RM & err & $\tau$ & RM/DM & d & err & d$_{Crab}$ & z \\
			
			& degree & degree & pc/cm$^3$ & pc/cm$^3$ & rad/m$^2$ & rad/m$^2$ & Myr & & kpc & kpc   & kpc & pc\\
			
			\hline
			
			J1756-2251	&	6.499	&	0.948	&	121.196	&	$\pm$0.005	&	10	&	$\pm$11.0	&	443	&	0.08	&	2.1	&	$\pm$1.0	&	10	&	35	\\
			
			J1802-2124	&	8.382	&	0.611	&	149.6258	&	$\pm$0.0006	&	286	&	$\pm$14.0	&	2760	&	1.91	&	2.5	&	$\pm$1.0	&	9	&	27	\\
			
			J1753-2240	&	6.299	&	1.663	&	158.6	&	$\pm$0.4	&	-	&	-	&	1550	&	-	&	2.6	&	$\pm$1.0	&	19	&	75	\\
			
			J1759-2205	&	7.472	&	0.81	&	177.157	&	$\pm$0.005	&	6	&	$\pm$7.0	&	0.673	&	0.03	&	2.9	&	$\pm$0.3	&	10.5$^{\dagger}$, 6.6	&	41	\\
			
			J1803-2137	&	8.395	&	0.146	&	233.99	&	$\pm$0.05	&	-36.1	&	$\pm$0.2	&	0.0158	&	-0.15	&	3.3	&	$\pm$1.0	&	9.8$^{\dagger}$, 2.4	&	8	\\
\hline
			
		\end{tabular}
		
	\end{adjustbox}
\end{table}

\begin{table}
	\centering
	\caption{Radio pulsars in the solid angle interval 10$\degr$ $<$ $\ell$  $<$ 14$\degr$; -2$\degr$ $<$ $b$ $<$ 2$\degr$ with measured DM. See Table~\ref{tab:1} for column descriptions.}
	\label{tab:4}
	\begin{adjustbox}{width=0.48\textwidth}
		
		\begin{tabular}{lccccccc|ccccr} 
			
			\hline
			
			\multicolumn{8}{c|}{Data from ATNFPC} & \multicolumn{5}{c}{Adopted and Calculated Data in this Study} \\ \hline
			
			PSR & $\ell$ & b &  DM & err & RM & err & $\tau$ & RM/DM & d & err & d$_{Crab}$ & z \\
			
			& degree & degree & pc/cm$^3$ & pc/cm$^3$ & rad/m$^2$ & rad/m$^2$ & Myr & & kpc & kpc   & kpc & pc\\
			
			\hline
			
			J1809-1943	&	10.727	&	-0.158	&	178	&	$\pm$5	&	76	&	$\pm$1.0	&	0.031	&	0.43	&	3	&	$\pm$0.3	&	9	&	-8	\\
			
			J1809-1917	&	11.094	&	0.08	&	197.1	&	$\pm$0.4	&	41	&	$\pm$17.0	&	0.0514	&	0.21	&	3.3	&	$\pm$0.4	&	4	&	5	\\
			
			J1810-2005	&	10.545	&	-0.563	&	241	&	$\pm$0.3	&	-15	&	$\pm$14.0	&	3530	&	-0.06	&	4	&	$\pm$0.5	&	6	&	-39	\\
			
			J1812-1718	&	13.109	&	0.538	&	255.1	&	$\pm$1.8	&	70.6	&	$\pm8.9$	&	1	&	0.28	&	4.2	&	$\pm$0.5	&	7	&	39	\\
			
			J1801-1909	&	10.295	&	1.781	&	264	&	$\pm$9	&	-	&	-	&	25	&	-	&	5	&	$\pm$1.0	&	11	&	155	\\
\hline
			
		\end{tabular}
		
	\end{adjustbox}
\end{table}

\begin{table}
	\centering
	\caption{Radio pulsars in the solid angle interval 14$\degr$ $<$ $\ell$  $<$ 18$\degr$; -2$\degr$ $<$ $b$ $<$ 2$\degr$ with measured DM. See Table~\ref{tab:1} for column descriptions.}
	\label{tab:5}
	\begin{adjustbox}{width=0.48\textwidth}
		
		\begin{tabular}{lccccccc|ccccr} 
			
			\hline
			
			\multicolumn{8}{c|}{Data from ATNFPC} & \multicolumn{5}{c}{Adopted and Calculated Data in this Study} \\ \hline
			
			PSR & $\ell$ & b &  DM & err & RM & err & $\tau$ & RM/DM & d & err & d$_{Crab}$ & z \\
			
			& degree & degree & pc/cm$^3$ & pc/cm$^3$ & rad/m$^2$ & rad/m$^2$ & Myr & & kpc & kpc   & kpc & pc\\
			
			\hline
			
			J1819-1318	&	17.497	&	0.829	&	35.1	&	$\pm$1.5	&	-	&	-	&	40.2	&	-	&	0.8	&	$\pm$0.2	&	19	&	12	\\
			
			J1819-17	&	14.146	&	-0.904	&	67	&	$\pm$23.0	&	-	&	-	&	-	&	-	&	1.5	&	$\pm$0.4	&	20	&	-24	\\
			
			J1826-1419	&	17.398	&	-1.142	&	160	&	$\pm$1.0	&	-	&	-	&	1.39	&	-	&	3.3	&	$\pm$1	&	-	&	-66	\\
			
			J1819-1458	&	16.02	&	0.081	&	196	&	$\pm$0.4	&	-	&	-	&	0.12	&	-	&	3.5	&	$\pm$0.5	&	-	&	5	\\
			
			J1818-1556	&	15.087	&	-0.227	&	230	&	$\pm$4.0	&	-	&	-	&	21.4	&	-	&	4.2	&	$\pm$1.3	&	11	&	-17	\\
\hline
			
		\end{tabular}
		
	\end{adjustbox}
\end{table}

\begin{table}
	\centering
	\caption{Radio pulsars in the solid angle interval 18$\degr$ $<$ $\ell$  $<$ 22$\degr$; -2$\degr$ $<$ $b$ $<$ 2$\degr$ with measured DM. See Table~\ref{tab:1} for column descriptions.}
	\label{tab:6}
	\begin{adjustbox}{width=0.48\textwidth}
		
		\begin{tabular}{lccccccc|ccccr} 
			
			\hline
			
			\multicolumn{8}{c|}{Data from ATNFPC} & \multicolumn{5}{c}{Adopted and Calculated Data in this Study} \\ \hline
			
			PSR & $\ell$ & b &  DM & err & RM & err & $\tau$ & RM/DM & d & err & d$_{Crab}$ & z \\
			
			& degree & degree & pc/cm$^3$ & pc/cm$^3$ & rad/m$^2$ & rad/m$^2$ & Myr & & kpc & kpc   & kpc & pc\\
			
			\hline
			
			J1825-0935	&	21.449	&	1.324	&	19.3833	&	$\pm$0.0009	&	65.2	&	$\pm$0.2	&	0.233	&	3.36	&	1	&	$\pm$0.2	&	7.8$^{\dagger}$, 2.3	&	23	\\
			
			J1835-1020	&	21.98	&	-1.302	&	113.7	&	$\pm$0.9	&	142.8	&	$\pm$3.2	&	0.81	&	1.26	&	2.6	&	$\pm$0.8	&	5	&	-59	\\
			
			J1825-1108	&	20.047	&	0.641	&	121	&	$\pm$19	&	-	&	-	&	16.1	&	-	&	2	&	$\pm$0.6	&	21	&	22	\\
			
			J1835-1106	&	21.222	&	-1.513	&	132.679	&	$\pm$0.003	&	42.9	&	$\pm$2.1	&	0.128	&	0.32	&	2.8	&	$\pm$0.4	&	8.6$^{\dagger}$, 4.7	&	-74	\\
			
			J1830-1059	&	20.812	&	-0.478	&	159.7	&	$\pm$0.01	&	47	&	$\pm$5.0	&	0.107	&	0.29	&	2.9	&	$\pm$0.5	&	32.4$^{\dagger}$, 6.1	&	-24	\\
\hline
			
		\end{tabular}
		
	\end{adjustbox}
\end{table}

\begin{table}
	\centering
	\caption{Radio pulsars in the solid angle interval 22$\degr$ $<$ $\ell$  $<$ 26$\degr$; -2$\degr$ $<$ $b$ $<$ 2$\degr$ with measured DM. See Table~\ref{tab:1} for column descriptions.}
	\label{tab:7}
	\begin{adjustbox}{width=0.48\textwidth}
		
		\begin{tabular}{lccccccc|ccccr} 
			
			\hline
			
			\multicolumn{8}{c|}{Data from ATNFPC} & \multicolumn{5}{c}{Adopted and Calculated Data in this Study} \\ \hline
			
			PSR & $\ell$ & b &  DM & err & RM & err & $\tau$ & RM/DM & d & err & d$_{Crab}$ & z \\
			
			& degree & degree & pc/cm$^3$ & pc/cm$^3$ & rad/m$^2$ & rad/m$^2$ & Myr & & kpc & kpc   & kpc & pc\\
			
			\hline
			
			J1832-0836	&	23.109	&	0.257	&	28.19422	&	$\pm$0.00009	&	25	&	$\pm$17.0	&	5210	&	0.89	&	1	&	$\pm$0.5	&	7	&	4	\\
			
			J1839-0627	&	25.807	&	-0.266	&	88.5	&	$\pm$0.7	&	-	&	-	&	58.3	&	-	&	2.1	&	$\pm$0.6	&	14	&	-10	\\
			
			J1842-0800	&	24.829	&	-1.766	&	188.6	&	$\pm$0.01	&	-	&	-	&	103	&	-	&	3.5	&	$\pm$1.1	&	24	&	-108	\\
			
			J1835-0946	&	22.451	&	-0.999	&	193.3	&	$\pm$0.5	&	-	&	-	&	139	&	-	&	3.1	&	$\pm$0.9	&	18	&	-54	\\
			
			J1843-0806	&	24.806	&	-1.934	&	215.8	&	$\pm$0.9	&	167	&	$\pm$15.0	&	0.49	&	0.77	&	4	&	$\pm$1.2	&	12	&	-135	\\
\hline
			
		\end{tabular}
		
	\end{adjustbox}
\end{table}

\begin{table}
	\centering
	\caption{Radio pulsars in the solid angle interval 26$\degr$ $<$ $\ell$  $<$ 30$\degr$; -2$\degr$ $<$ $b$ $<$ 2$\degr$ with measured DM. See Table~\ref{tab:1} for column descriptions.}
	\label{tab:8}
	\begin{adjustbox}{width=0.48\textwidth}
		
		\begin{tabular}{lccccccc|ccccr} 
			
			\hline
			
			\multicolumn{8}{c|}{Data from ATNFPC} & \multicolumn{5}{c}{Adopted and Calculated Data in this Study} \\ \hline
			
			PSR & $\ell$ & b &  DM & err & RM & err & $\tau$ & RM/DM & d & err & d$_{Crab}$ & z \\
			
			& degree & degree & pc/cm$^3$ & pc/cm$^3$ & rad/m$^2$ & rad/m$^2$ & Myr & & kpc & kpc   & kpc & pc\\
			
			\hline
			
			J1841-04	&	27.49	&	0.09	&	29	&	$\pm$3	&	-	&	-	&	-	&	-	&	0.7	&	$\pm$0.2	&	-	&	1	\\
			
			J1849-0317	&	29.834	&	-1.173	&	42.9	&	$\pm$2.8	&	-31.9	&	$\pm$3.8	&	0.481	&	-0.74	&	1.5	&	$\pm$0.5	&	9	&	-31	\\
			
			J1834-0426	&	27.042	&	1.749	&	79.308	&	$\pm$0.008	&	100	&	$\pm$5.0	&	63.9	&	1.26	&	2.3	&	$\pm$0.5	&	5.3$^{\dagger}$, 3.3	&	70	\\
			
			J1844-0433	&	28.096	&	-0.548	&	123.158	&	$\pm$0.02	&	7	&	$\pm$13.0	&	4.01	&	0.06	&	2.4	&	+1.0,-0.5	&	16.5$^{\dagger}$, 7.1	&	-23	\\
			
			J1847-0402	&	28.876	&	-0.939	&	141.979	&	$\pm$0.005	&	117	&	$\pm$8.0	&	0.183	&	0.82	&	3.5	&	+0.5,-0.3	&	5.4$^{\dagger}$, 3.4	&	-57	\\
\hline
			
		\end{tabular}
		
	\end{adjustbox}
\end{table}

\begin{table}
	\centering
	\caption{Radio pulsars in the solid angle interval 30$\degr$ $<$ $\ell$  $<$ 34$\degr$; -2$\degr$ $<$ $b$ $<$ 2$\degr$ with measured DM. See Table~\ref{tab:1} for column descriptions.}
	\label{tab:9}
	\begin{adjustbox}{width=0.48\textwidth}
		
		\begin{tabular}{lccccccc|ccccr} 
			
			\hline
			
			\multicolumn{8}{c|}{Data from ATNFPC} & \multicolumn{5}{c}{Adopted and Calculated Data in this Study} \\ \hline
			
			PSR & $\ell$ & b &  DM & err & RM & err & $\tau$ & RM/DM & d & err & d$_{Crab}$ & z \\
			
			& degree & degree & pc/cm$^3$ & pc/cm$^3$ & rad/m$^2$ & rad/m$^2$ & Myr & & kpc & kpc   & kpc & pc\\
			
			\hline
			
			J1851-0053	&	32.1	&	-0.318	&	24	&	$\pm$4	&	-7	&	$\pm$16.0	&	25.6	&	-0.29	&	0.55	&	$\pm$0.15	&	8	&	-3	\\
			
			J1848-0023	&	32.267	&	0.447	&	30.6	&	$\pm$0.1	&	-	&	-	&	5.29	&	-	&	0.7	&	$\pm$0.2	&	10	&	5	\\
			
			J1843-0000	&	32.013	&	1.768	&	101.5	&	$\pm$0.8	&	-52.9	&	$\pm$3.8	&	1.79	&	-0.52	&	3	&	+1.0, -0.5	&	4	&	93	\\
			
			J1846+0051	&	33.162	&	1.44	&	140	&	$\pm$3	&	-60.1	&	$\pm$8.1	&	0.613	&	-0.43	&	4	&	+1.0, -0.5	&	13	&	101	\\
			
			J1848-0123	&	31.339	&	0.039	&	159.531	&	$\pm$0.008	&	580	&	$\pm$30.0	&	1.99	&	3.64	&	3.5	&	$\pm$0.5	&	5.3$^{\dagger}$, 2	&	2	\\
\hline
			
		\end{tabular}
		
	\end{adjustbox}
\end{table}

\begin{table}
	\centering
	\caption{Radio pulsars in the solid angle interval 34$\degr$ $<$ $\ell$  $<$ 38$\degr$; -2$\degr$ $<$ $b$ $<$ 2$\degr$ with measured DM. See Table~\ref{tab:1} for column descriptions.}
	\label{tab:10}
	\begin{adjustbox}{width=0.48\textwidth}
		
		\begin{tabular}{lccccccc|ccccr} 
			
			\hline
			
			\multicolumn{8}{c|}{Data from ATNFPC} & \multicolumn{5}{c}{Adopted and Calculated Data in this Study} \\ \hline
			
			PSR & $\ell$ & b &  DM & err & RM & err & $\tau$ & RM/DM & d & err & d$_{Crab}$ & z \\
			
			& degree & degree & pc/cm$^3$ & pc/cm$^3$ & rad/m$^2$ & rad/m$^2$ & Myr & & kpc & kpc   & kpc & pc\\
			
			\hline
			
			J1857+0057	&	34.417	&	-0.805	&	82.39	&	$\pm$0.11	&	104	&	$\pm$19.0	&	104	&	1.26	&	2.4	&	+0.5, -0.3	&	21$^{\dagger}$, 7.8	&	-34	\\
			
			J1856+0113	&	34.56	&	-0.497	&	96.74	&	$\pm$0.12	&	-140	&	$\pm$30.0	&	0.0203	&	-1.45	&	2.8	&	$\pm$0.3	&	25.4$^{\dagger}$, 17.2	&	-24	\\
			
			J1901+0156	&	35.818	&	-1.367	&	105.394	&	$\pm$0.007	&	-122	&	$\pm$9.0	&	1.94	&	-1.16	&	3.2	&	+0.5, -0.3	&	12.7$^{\dagger}$, 12	&	-76	\\
			
			J1850+0124	&	34.025	&	0.959	&	118.89	&	$\pm$0.05	&	-	&	-	&	5180	&	-	&	3	&	$\pm$0.5	&	17	&	50	\\
			
			J1853+0056	&	34.017	&	-0.036	&	180.9	&	$\pm$1.2	&	-	&	-	&	0.204	&	-	&	3.8	&	$\pm$0.3	&	16	&	-2	\\
\hline
			
		\end{tabular}
		
	\end{adjustbox}
\end{table}

\begin{table}
	\centering
	\caption{Radio pulsars in the solid angle interval 38$\degr$ $<$ $\ell$  $<$ 42$\degr$; -2$\degr$ $<$ $b$ $<$ 2$\degr$ with measured DM. See Table~\ref{tab:1} for column descriptions.}
	\label{tab:11}
	\begin{adjustbox}{width=0.48\textwidth}
		
		\begin{tabular}{lccccccc|ccccr} 
			
			\hline
			
			\multicolumn{8}{c|}{Data from ATNFPC} & \multicolumn{5}{c}{Adopted and Calculated Data in this Study} \\ \hline
			
			PSR & $\ell$ & b &  DM & err & RM & err & $\tau$ & RM/DM & d & err & d$_{Crab}$ & z \\
			
			& degree & degree & pc/cm$^3$ & pc/cm$^3$ & rad/m$^2$ & rad/m$^2$ & Myr & & kpc & kpc   & kpc & pc\\
			
			\hline
			
			J1908+0734	&	41.585	&	-0.27	&	11.104	&	$\pm$0.011	&	-19	&	$\pm$18.0	&	4.08	&	-1.71	&	0.7	&	$\pm$0.2	&	25.1$^{\dagger}$, 10.2	&	-3	\\
			
			J1905+0400	&	38.095	&	-1.289	&	25.6923	&	$\pm$0.0012	&	-	&	-	&	12200	&	-	&	1.1	&	$\pm$0.2	&	33	&	-25	\\
			
			J1909+0641	&	40.941	&	-0.94	&	36.7	&	$\pm$0.2	&	-	&	-	&	3.65	&	-	&	1.3	&	$\pm$0.2	&	22	&	-21	\\
			
			J1907+0602	&	40.182	&	-0.894	&	82.1	&	$\pm$1.1	&	-	&	-	&	0.0195	&	-	&	2.2	&	+0.5, -0.4	&	128	&	-34	\\
			
			J1901+0621	&	39.687	&	0.756	&	94	&	$\pm$7	&	-229	&	$\pm$13.0	&	732	&	-2.44	&	2.5	&	$\pm$0.5	&	11	&	33	\\
\hline
			
		\end{tabular}
		
	\end{adjustbox}
\end{table}

\begin{table}
	\centering
	\caption{Radio pulsars in the solid angle interval 42$\degr$ $<$ $\ell$  $<$ 46$\degr$; -2$\degr$ $<$ $b$ $<$ 2$\degr$ with measured DM. See Table~\ref{tab:1} for column descriptions.}
	\label{tab:12}
	\begin{adjustbox}{width=0.48\textwidth}
		
		\begin{tabular}{lccccccc|ccccr} 
			\hline
			
			\multicolumn{8}{c|}{Data from ATNFPC} & \multicolumn{5}{c}{Adopted and Calculated Data in this Study} \\ \hline
			
			PSR & $\ell$ & b &  DM & err & RM & err & $\tau$ & RM/DM & d & err & d$_{Crab}$ & z \\
			
			& degree & degree & pc/cm$^3$ & pc/cm$^3$ & rad/m$^2$ & rad/m$^2$ & Myr & & kpc & kpc   & kpc & pc\\
			
			\hline
			
			J1917+0834	&	43.575	&	-1.892	&	29.18	&	$\pm$0.06	&	-	&	-	&	1.93	&	-	&	1.2	&	+0.4, -0.2	&	70.7$^{\dagger}$, 13.4	&	-40	\\
			
			J1915+0738	&	42.466	&	-1.803	&	39	&	$\pm$0.08	&	-6.8	&	$\pm$1.5	&	7.39	&	-0.17	&	1.7	&	$\pm$0.3	&	34$^{\dagger}$, 12.8	&	-53	\\
			
			J1916+0951	&	44.556	&	-1.019	&	60.953	&	$\pm$0.006	&	97	&	$\pm$1.0	&	1.7	&	1.59	&	2	&	$\pm$0.3	&	10.5$^{\dagger}$, 5.9	&	-36	\\
			
			J1913+0904	&	43.502	&	-0.684	&	95.3	&	$\pm$0.6	&	-	&	-	&	0.147	&	-	&	3	&	$\pm$0.5	&	12	&	-36	\\
			
			J1914+1122	&	45.623	&	0.199	&	100	&	$\pm$10	&	360.9	&	$\pm$7.2	&	14.5	&	3.61	&	2.9	&	+0.3, -0.5	&	44.7$^{\dagger}$, 10.1	&	10	\\
\hline
			
		\end{tabular}
		
	\end{adjustbox}
\end{table}

\begin{table}
	\centering
	\caption{Radio pulsars in the solid angle interval 46$\degr$ $<$ $\ell$  $<$ 50$\degr$; -2$\degr$ $<$ $b$ $<$ 2$\degr$ with measured DM. See Table~\ref{tab:1} for column descriptions.}
	\label{tab:13}
	\begin{adjustbox}{width=0.48\textwidth}
		
		\begin{tabular}{lccccccc|ccccr} 
			
			\hline
			
			\multicolumn{8}{c|}{Data from ATNFPC} & \multicolumn{5}{c}{Adopted and Calculated Data in this Study} \\ \hline
			
			PSR & $\ell$ & b &  DM & err & RM & err & $\tau$ & RM/DM & d & err & d$_{Crab}$ & z \\
			
			& degree & degree & pc/cm$^3$ & pc/cm$^3$ & rad/m$^2$ & rad/m$^2$ & Myr & & kpc & kpc   & kpc & pc\\
			
			\hline
			
			J1918+1541	&	49.894	&	1.36	&	13	&	$\pm$2	&	-	&	-	&	2.31	&	-	1&	1	&	$\pm$0.2	&	52$^{\dagger}$	&	24	\\
			
			J1918+1444	&	49.096	&	0.866	&	27.202	&	$\pm$0.017	&	-41.7	&	$\pm$3.5	&	0.0881	&	-1.53	&	1.7	&	$\pm$0.3	&	37.1$^{\dagger}$, 7.5	&	26	\\
			
			J1911+1347	&	47.518	&	1.809	&	30.9883	&	$\pm$0.0009	&	-	&	-	&	4330	&	-	&	2	&	$\pm$0.3	&	26	&	63	\\
			
			J1910+1256	&	46.564	&	1.795	&	38.0684	&	$\pm$0.0005	&	78	&	$\pm$19.0	&	8160	&	2.05	&	2.5	&	+0.3, -0.5	&	11	&	78	\\
			
			J1921+1419	&	49.058	&	0.021	&	91.64	&	$\pm$0.04	&	164.8	&	$\pm$3.1	&	1.75	&	1.8	&	3.4	&	+0.4, -0.5	&	26.2$^{\dagger}$, 9.1	&	1	\\
\hline
			
		\end{tabular}
		
	\end{adjustbox}
\end{table}

\begin{table}
	\centering
	\caption{Radio pulsars in the solid angle interval 50$\degr$ $<$ $\ell$  $<$ 54$\degr$; -2$\degr$ $<$ $b$ $<$ 2$\degr$ with measured DM. See Table~\ref{tab:1} for column descriptions.}
	\label{tab:14}
	\begin{adjustbox}{width=0.48\textwidth}
		
		\begin{tabular}{lccccccc|ccccr} 
			
			\hline
			
			\multicolumn{8}{c|}{Data from ATNFPC} & \multicolumn{5}{c}{Adopted and Calculated Data in this Study} \\ \hline
			
			PSR & $\ell$ & b &  DM & err & RM & err & $\tau$ & RM/DM & d & err & d$_{Crab}$ & z \\
			
			& degree & degree & pc/cm$^3$ & pc/cm$^3$ & rad/m$^2$ & rad/m$^2$ & Myr & & kpc & kpc   & kpc & pc\\
			
			\hline
			
			J1929+16	&	51.747	&	-0.692	&	12	&	$\pm$9	&	-	&	-	&	-	&	-	&	0.8	&	$\pm$0.2	&	-	&	-10	\\
			
			J1926+1613	&	51.252	&	-0.06	&	32.9	&	$\pm$0.01	&	-	&	-	&	-	&	-	&	1.25	&	$\pm$0.25	&	-	&	-1	\\
			
			J1932+1756	&	53.464	&	-0.52	&	53.189	&	$\pm$0.009	&	-	&	-	&	1530	&	-	&	1.75	&	+0.65, -0.45	&	35	&	-16	\\
			
			J1935+1726	&	53.368	&	-1.374	&	61.6	&	$\pm$0.1	&	-	&	-	&	-	&	-	&	2	&	+0.7, -0.5	&	9	&	-48	\\
			
			J1927+1856	&	53.811	&	0.941	&	99	&	$\pm$1	&	74.4	&	$\pm$5.8	&	2.11	&	0.75	&	2.8	&	+0.7, -0.5	&	31.6$^{\dagger}$, 10.1	&	46	\\
\hline
			
		\end{tabular}
		
	\end{adjustbox}
\end{table}

\begin{table}
	\centering
	\caption{Radio pulsars in the solid angle interval 54$\degr$ $<$ $\ell$  $<$ 58$\degr$; -2$\degr$ $<$ $b$ $<$ 2$\degr$ with measured DM. See Table~\ref{tab:1} for column descriptions.}
	\label{tab:15}
	\begin{adjustbox}{width=0.48\textwidth}
		
		\begin{tabular}{lccccccc|ccccr} 
			
			\hline
			
			\multicolumn{8}{c|}{Data from ATNFPC} & \multicolumn{5}{c}{Adopted and Calculated Data in this Study} \\ \hline
			
			PSR & $\ell$ & b &  DM & err & RM & err & $\tau$ & RM/DM & d & err & d$_{Crab}$ & z \\
			
			& degree & degree & pc/cm$^3$ & pc/cm$^3$ & rad/m$^2$ & rad/m$^2$ & Myr & & kpc & kpc   & kpc & pc\\
			
			\hline
			
			J1929+2121	&	56.117	&	1.751	&	66	&	$\pm$12	&	-	&	-	&	5.36	&	-	&	2	&	+0.8, -0.4	&	16	&	61	\\
			
			J1939+2134	&	57.509	&	-0.29	&	71.0237	&	$\pm$0.0013	&	7.73	&	$\pm$0.1	&	235	&	0.11	&	2.1	&	+0.4, -0.6	&	3$^{\dagger}$, 2	&	-11	\\
			
			J1938+2213	&	57.903	&	0.308	&	91	&	$\pm$3	&	-	&	-	&	0.062	&	-	&	2.5	&	$\pm$0.5	&	47$^{\dagger}$, 9.7	&	13	\\
			
			J1946+2052	&	57.657	&	-1.975	&	93.965	&	$\pm$0.003	&	-	&	-	&	292	&	-	&	4	&	+1.0, -0.5	&	30	&	-138	\\
			
			J1934+19	&	54.609	&	-0.397	&	97.6	&	$\pm$0.01	&	-	&	-	&	-	&	-	&	2.6	&	$\pm$0.5	&	24	&	-18	\\
\hline
			
		\end{tabular}
		
	\end{adjustbox}
\end{table}

\begin{table}
	\centering
	\caption{Radio pulsars in the solid angle interval 58$\degr$ $<$ $\ell$  $<$ 62$\degr$; -2$\degr$ $<$ $b$ $<$ 2$\degr$ with measured DM. See Table~\ref{tab:1} for column descriptions.}
	\label{tab:16}
	\begin{adjustbox}{width=0.48\textwidth}
		
		\begin{tabular}{lccccccc|ccccr} 
			
			\hline
			
			\multicolumn{8}{c|}{Data from ATNFPC} & \multicolumn{5}{c}{Adopted and Calculated Data in this Study} \\ \hline
			
			PSR & $\ell$ & b &  DM & err & RM & err & $\tau$ & RM/DM & d & err & d$_{Crab}$ & z \\
			
			& degree & degree & pc/cm$^3$ & pc/cm$^3$ & rad/m$^2$ & rad/m$^2$ & Myr & & kpc & kpc   & kpc & pc\\
			
			\hline
			
			J1954+2407	&	61.367	&	-1.872	&	80.5	&	$\pm$0.4	&	-	&	-	&	2.9	&	-	&	3	&	+1.0, -0.5	&	24	&	-98	\\
			
			J1946+24	&	60.306	&	-0.376	&	96	&	$\pm$0.1	&	-	&	-	&	-	&	-	&	3	&	$\pm$0.5	&	-	&	-20	\\
			
			J1946+2244	&	59.301	&	-1.07	&	140	&	$\pm$20	&	2	&	$\pm$20.0	&	23.8	&	0.01	&	5.9	&	+0.6, -0.4	&	27$^{\dagger}$, 18	&	-110	\\
			
			J1950+2414	&	61.098	&	-1.169	&	142.0842	&	$\pm$0.0031	&	-	&	-	&	3630	&	-	&	5.9	&	+0.7, -0.5	&	22	&	-120	\\
			
			J1939+2449	&	60.174	&	1.361	&	142.88	&	$\pm$0.07	&	76.2	&	$\pm$1.1	&	0.56	&	0.53	&	6	&	+1.0, -0.4	&	27$^{\dagger}$	&	143	\\
			
			J1943+2210	&	58.439	&	-0.73	&	174.089	&	$\pm$0.003	&	-	&	-	&	9170	&	-	&	7	&	+0.7, -0.5	&	37	&	-89	\\
			
			J1944+2236	&	58.9	&	-0.663	&	185.45	&	$\pm$0.12	&	-	&	-	&	7680	&	-	&	7.4	&	+0.7, -0.5	&	24	&	-86	\\
			
			J1949+2306	&	59.931	&	-1.423	&	196.3	&	$\pm$0.5	&	-	&	-	&	170	&	-	&	8.2	&	+1.2, -0.5	&	24	&	-204	\\
			
			J1948+2333	&	60.214	&	-1.045	&	198.2	&	$\pm$0.8	&	-	&	-	&	0.617	&	-	&	8.2	&	+1.0, -0.4	&	14	&	-150	\\
			
			J1940+2245	&	58.629	&	0.127	&	222.4	&	$\pm$1.3	&	-	&	-	&	0.323	&	-	&	8.2	&	$\pm$0.4	&	19	&	18	\\
			
			J1946+2535	&	61.809	&	0.283	&	248.81	&	$\pm$0.04	&	-	&	-	&	1.45	&	-	&	8.5	&	$\pm$0.5	&	11	&	42	\\
			
			J1940+2337	&	59.397	&	0.529	&	252.1	&	$\pm$0.3	&	-	&	-	&	0.113	&	-	&	8.8	&	$\pm$0.6	&	28	&	81	\\
			
			J1941+2525	&	61.037	&	1.263	&	314.4	&	$\pm$0.4	&	-	&	-	&	0.227	&	-	&	13	&	+2.0,-1.2	&	15	&	287	\\
			
			J1934+2352	&	58.966	&	1.814	&	355.5	&	$\pm$0.2	&	-	&	-	&	0.0216	&	-	&	14	&	+3.0, -2.0	&	30	&	443	\\
			
			\hline
			
		\end{tabular}
	\end{adjustbox}
\end{table}

	\begin{table}
		\centering
		\caption{Radio pulsars in the solid angle interval 62$\degr$ $<$ $\ell$  $<$ 66$\degr$; -2$\degr$ $<$ $b$ $<$ 2$\degr$ with measured DM. See Table~\ref{tab:1} for column descriptions.}
		\label{tab:17}
		\begin{adjustbox}{width=0.48\textwidth}
			
			\begin{tabular}{lccccccc|ccccr} 
				
				\hline
				
				\multicolumn{8}{c|}{Data from ATNFPC} & \multicolumn{5}{c}{Adopted and Calculated Data in this Study} \\ \hline

				PSR & $\ell$ & b &  DM & err & RM & err & $\tau$ & RM/DM & d & err & d$_{Crab}$ & z \\
				
				& degree & degree & pc/cm$^3$ & pc/cm$^3$ & rad/m$^2$ & rad/m$^2$ & Myr & & kpc & kpc   & kpc & pc\\
				
				\hline
				
				J1954+2923	&	65.924	&	0.772	&	7.932	&	$\pm$0.007	&	-18	&	$\pm$3.0	&	3950	&	-2.27	&	0.5	&	$\pm$0.2	&	18.3$^{\dagger}$, 2.6	&	7	\\
				
				J1957+2516	&	62.769	&	-1.973	&	44.137	&	$\pm$0.003	&	-	&	-	&	2290	&	-	&	2	&	+1.0, -0.5	&	53	&	-69	\\
				
				J1955+2908	&	65.839	&	0.443	&	104.5159	&	$\pm$0.0006	&	22.4	&	$\pm$8.8	&	3270	&	0.21	&	4.2	&	$\pm$0.5	&	12.1$^{\dagger}$, 7.1	&	32	\\
				
				J1957+2831	&	65.524	&	-0.225	&	138.99	&	$\pm$0.08	&	-43.7	&	$\pm$1.8	&	1.57	&	-0.31	&	5.3	&	$\pm$0.5	&	7	&	-21	\\
				
				J1946+2611	&	62.321	&	0.595	&	165	&	$\pm$3	&	-88.7	&	$\pm$1.7	&	0.314	&	-0.54	&	6.3	&	$\pm$0.5	&	38.3$^{\dagger}$, 9	&	65	\\
				
				J1953+2732	&	64.205	&	0.059	&	195.4	&	$\pm$0.9	&	-	&	-	&	11.9	&	-	&	6.5	&	+0.4, -0.8	&	23	&	7	\\
				
				J1955+2527	&	62.74	&	-1.571	&	209.971	&	$\pm$0.003	&	-	&	-	&	8470	&	-	&	8	&	+2.0, -1.0	&	14	&	-219	\\
				
				J1948+2551	&	62.207	&	0.131	&	289.27	&	$\pm$0.05	&	-	&	-	&	0.345	&	-	&	8.7	&	$\pm$0.6	&	9	&	20	\\
				
				J1952+2630	&	63.254	&	-0.376	&	315.338	&	$\pm$0.002	&	-	&	-	&	76.9	&	-	&	10	&	$\pm$1.0	&	26	&	-66	\\
				
				\hline
				
			\end{tabular}
			
		\end{adjustbox}
	\end{table}

\begin{table}
	\centering
	\caption{Radio pulsars in the solid angle interval 66$\degr$ $<$ $\ell$  $<$ 70$\degr$; -2$\degr$ $<$ $b$ $<$ 2$\degr$ with measured DM. See Table~\ref{tab:1} for column descriptions.}
	\label{tab:18}
	\begin{adjustbox}{width=0.48\textwidth}
		
		\begin{tabular}{lccccccc|ccccr} 
			
			\hline
			
			\multicolumn{8}{c|}{Data from ATNFPC} & \multicolumn{5}{c}{Adopted and Calculated Data in this Study} \\ \hline

			PSR & $\ell$ & b &  DM & err & RM & err & $\tau$ & RM/DM & d & err & d$_{Crab}$ & z \\
			
			& degree & degree & pc/cm$^3$ & pc/cm$^3$ & rad/m$^2$ & rad/m$^2$ & Myr & & kpc & kpc   & kpc & pc\\
			
			\hline
			
			J1953+30	&	66.57	&	1.306	&	43.61	&	$\pm$0.11	&	-	&	-	&	-	&	-	&	2	&	+1.0, -0.5	&	-	&	46	\\
			
			J2006+3102	&	68.667	&	-0.53	&	107.16	&	$\pm$0.01	&	-	&	-	&	0.104	&	-	&	4.3	&	+0.6, -0.4	&	14	&	-40	\\
			
			J2000+29	&	66.232	&	-0.478	&	132.5	&	$\pm$1.4	&	-	&	-	&	-	&	-	&	5	&	+0.7, -0.5	&	-	&	-42	\\
			
			J2010+3051	&	68.967	&	-1.326	&	133.756	&	$\pm$0.002	&	-64	&	$\pm$15.0	&	15700	&	-0.48	&	6.2	&	+1.0, -0.5	&	24	&	-143	\\
			
			J2002+3217	&	69.261	&	0.879	&	142.21	&	$\pm$0.03	&	-90.2	&	$\pm$1.3	&	0.105	&	-0.63	&	6.4	&	+0.8, -0.4	&	20$^{\dagger}$, 6.8	&	98	\\
			
			J2013+3058	&	69.485	&	-1.886	&	148.7	&	$\pm$0.1	&	-	&	-	&	28.8	&	-	&	7	&	+1.0, -0.5	&	29	&	-221	\\
			
			J1952+30	&	66.184	&	1.529	&	188.6	&	$\pm$0.6	&	-	&	-	&	-	&	-	&	7.5	&	+1.0, -0.5	&	-	&	200	\\
			
			J2007+3120	&	69.042	&	-0.535	&	191.5	&	$\pm$0.2	&	-	&	-	&	0.618	&	-	&	7	&	+0.7, -0.5	&	21	&	-65	\\
			
			J2002+30	&	67.917	&	-0.187	&	196	&	$\pm$38	&	-	&	-	&	-	&	-	&	7	&	$\pm$1.0	&	-	&	-23	\\
			
			J1958+30	&	66.856	&	0.417	&	199.3	&	$\pm$0.4	&	-	&	-	&	-	&	-	&	7.2	&	$\pm$1.0	&	-	&	52	\\
			
			J2004+3137	&	69.011	&	0.021	&	234.82	&	$\pm$0.008	&	30	&	$\pm$6.0	&	0.449	&	0.13	&	7.8	&	$\pm$0.5	&	12.3$^{\dagger}$, 5.6	&	3	\\
			
			\hline
			
		\end{tabular}
		
	\end{adjustbox}
\end{table}

\begin{table}
	\centering
	\caption{Radio pulsars in the solid angle interval 70$\degr$ $<$ $\ell$  $<$ 74$\degr$; -2$\degr$ $<$ $b$ $<$ 2$\degr$ with measured DM. See Table~\ref{tab:1} for column descriptions.}
	\label{tab:19}
	\begin{adjustbox}{width=0.48\textwidth}
		
		\begin{tabular}{lccccccc|ccccr} 
			
			\hline
			
			\multicolumn{8}{c|}{Data from ATNFPC} & \multicolumn{5}{c}{Adopted and Calculated Data in this Study} \\ \hline

			PSR & $\ell$ & b &  DM & err & RM & err & $\tau$ & RM/DM & d & err & d$_{Crab}$ & z \\
			
			& degree & degree & pc/cm$^3$ & pc/cm$^3$ & rad/m$^2$ & rad/m$^2$ & Myr & & kpc & kpc   & kpc & pc\\
			
			\hline
			
			J2018+3431	&	73.044	&	-0.841	&	222.35	&	$\pm$0.07	&	-	&	-	&	3.34	&	-	&	8.5	&	+1.2, -0.8	&	15	&	-125	\\
			
			J2009+3326	&	71.103	&	0.124	&	263.6	&	$\pm$0.8	&	-	&	-	&	15.5	&	-	&	8.2	&	$\pm$0.8	&	19	&	18	\\
			
			J2011+3331	&	71.32	&	-0.049	&	298.58	&	$\pm$0.06	&	-	&	-	&	8.27	&	-	&	8.6	&	$\pm$0.9	&	12	&	-7	\\
			
			J2004+3429	&	71.425	&	1.571	&	351	&	$\pm$0.1	&	-	&	-	&	0.0185	&	-	&	12.5	&	$\pm$2.0	&	23	&	343	\\
			
			J2010+3230	&	70.391	&	-0.495	&	371.8	&	$\pm$0.5	&	-	&	-	&	6.32	&	-	&	12	&	$\pm$2.0	&	22	&	-104	\\
			
			\hline
			
		\end{tabular}
		
	\end{adjustbox}
\end{table}

\begin{table}
	\centering
	\caption{Radio pulsars in the solid angle interval 74$\degr$ $<$ $\ell$  $<$ 78$\degr$; -2$\degr$ $<$ $b$ $<$ 2$\degr$ with measured DM. See Table~\ref{tab:1} for column descriptions.}
	
	\label{tab:20}
	\begin{adjustbox}{width=0.48\textwidth}
		
		\begin{tabular}{lccccccc|ccccr} 
			
			\hline
			
			\multicolumn{8}{c|}{Data from ATNFPC} & \multicolumn{5}{c}{Adopted and Calculated Data in this Study} \\ \hline

			PSR & $\ell$ & b &  DM & err & RM & err & $\tau$ & RM/DM & d & err & d$_{Crab}$ & z \\
			
			& degree & degree & pc/cm$^3$ & pc/cm$^3$ & rad/m$^2$ & rad/m$^2$ & Myr & & kpc & kpc   & kpc & pc\\
			
			\hline
			
			J2029+3744	&	76.898	&	-0.727	&	190.66	&	$\pm$0.03	&	-6	&	$\pm$1.8	&	1.56	&	-0.03	&	7	&	$\pm$1.0	&	11$^{\dagger}$, 9.7	&	-89	\\
			
			J2030+3641	&	76.123	&	-1.438	&	246	&	$\pm$0.7	&	514	&	$\pm$1.0	&	0.488	&	2.09	&	9.5	&	+1.5, -0.8	&	19	&	-238	\\
			
			J2021+3651	&	75.222	&	0.111	&	367.5	&	$\pm$1	&	524	&	$\pm$4.0	&	0.0172	&	1.43	&	11	&	$\pm$1.0	&	24	&	21	\\
			
			J2022+3842	&	76.888	&	0.96	&	429.1	&	$\pm$0.5	&	270	&	$\pm$0.1	&	0.00894	&	0.63	&	13	&	$\pm$2.0	&	-	&	218	\\
			
			\hline
			
		\end{tabular}
		
	\end{adjustbox}
\end{table}

\begin{table}
	\centering
	\caption{Radio pulsars in the solid angle interval 78$\degr$ $<$ $\ell$  $<$ 90$\degr$; -2$\degr$ $<$ $b$ $<$ 2$\degr$ with measured DM. In the first column J--Name of the pulsar is given. The second and third columns are galactic latitude ($\ell$) and longitude (b) of the pulsar in degrees. The fourth and fifth columns are dispersion measure (DM) values and their error in pc/cm$^3$. The sixth and seventh columns are rotation measure (RM) values and their error in rad/m$^2$. The eighth column is the characteristic ages ($\tau$) of the pulsars in Myr. The ninth column is the ratio of rotation measure to dispersion measure (RM/DM). The tenth and eleventh columns are the adopted distances (d) and their error in kpc.  The twelfth column is the Crab limit for the distance (d$_{Crab}$) in kpc. The thirteenth column is the distance to the Galactic plane (z) calculated from the adopted distance in pc. The symbol $^{\dagger}$ indicates the Crab (S400) luminosity limit. The symbol $^{\star}$ indicates distance and distance error from parallax (Deller et al. (2019)).}
	
	\label{tab:21}
	\begin{adjustbox}{width=0.48\textwidth}
		
		\begin{tabular}{lccccccc|ccccr} 
			
			\hline
			
			\multicolumn{8}{c|}{Data from ATNFPC} & \multicolumn{5}{c}{Adopted and Calculated Data in this Study} \\ \hline

			PSR & $\ell$ & b &  DM & err & RM & err & $\tau$ & RM/DM & d & err & d$_{Crab}$ & z \\
			
			& degree & degree & pc/cm$^3$ & pc/cm$^3$ & rad/m$^2$ & rad/m$^2$ & Myr & & kpc & kpc   & kpc & pc\\
			
			\hline
			
			J2053+4650	&	86.861	&	1.302	&	98.0828	&	$\pm$0.0006	&	-174	&	$\pm$11.0	&	1160	&	-1.77	&	-	&	-	&	-	&	39	\\
			
			J2032+4127	&	80.224	&	1.028	&	114.67	&	$\pm$0.04	&	215	&	$\pm$1.0	&	0.201	&	1.87	&	1.5	&	+0.2, -0.1	&	-	&	27	\\
			
			J2113+4644	&	89.003	&	-1.266	&	141.26	&	$\pm$0.09	&	-218.7	&	$\pm$0.1	&	22.5	&	-1.55	&	2.2$^{\star}$	&	+0.36, -0.32$^{\star}$	&	3.1$^{\dagger}$, 1.7	&	-49	\\
			
			\hline
			
		\end{tabular}
		
	\end{adjustbox}
\end{table}

\FloatBarrier

\begin{figure}
	
	\includegraphics[width=\columnwidth]{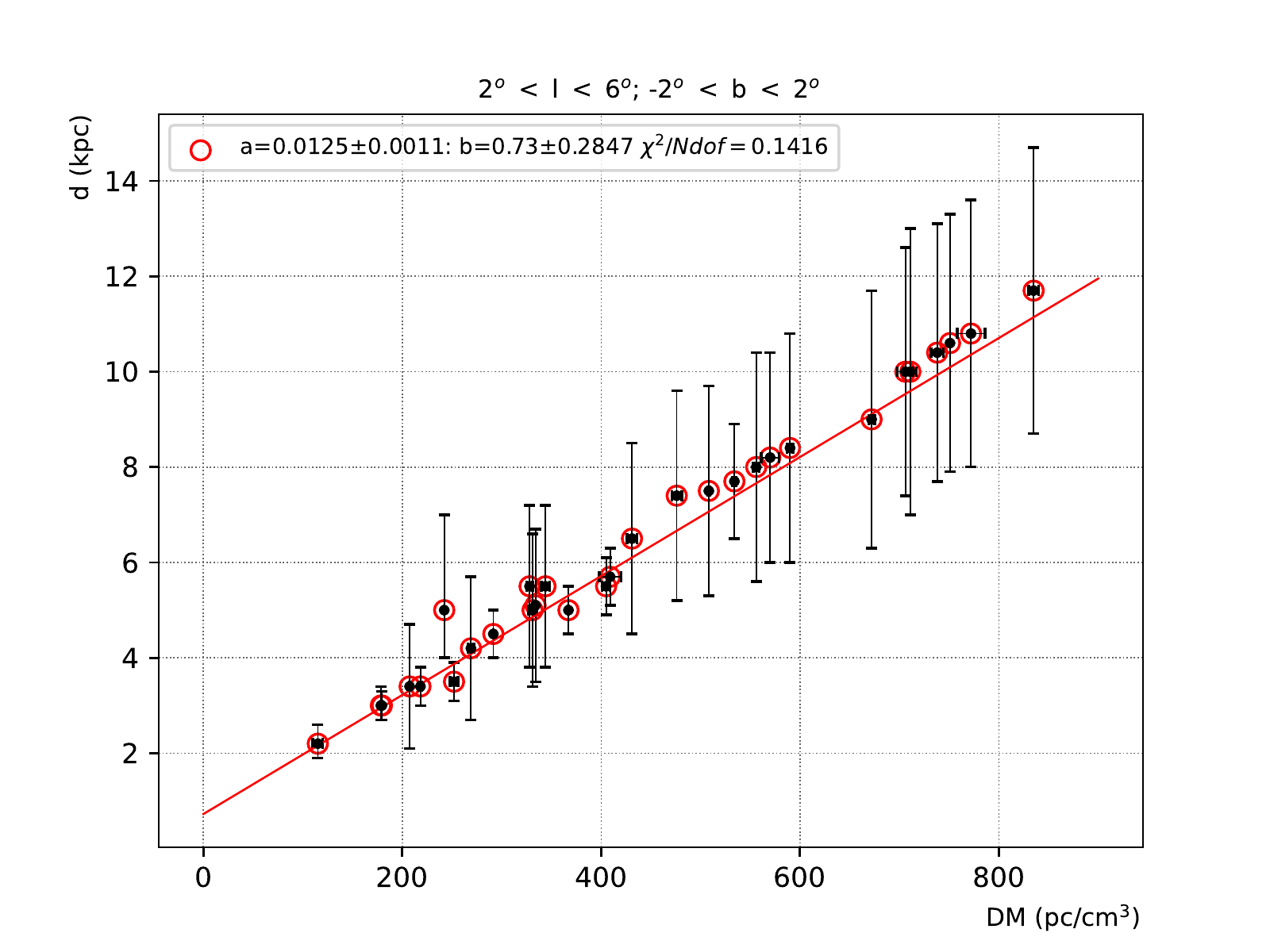}
	\caption{Distance-DM relation for radio pulsars in the solid angle interval 2$\degr$ $<$ $\ell$  $<$ 6$\degr$; -2$\degr$ $<$ $b$ $<$ 2$\degr$. The data are fit to a linear function (shown in red) with a slope a, and an intercept b.}
	\label{fig:2}
\end{figure}

\begin{figure}
	
	\includegraphics[width=\columnwidth]{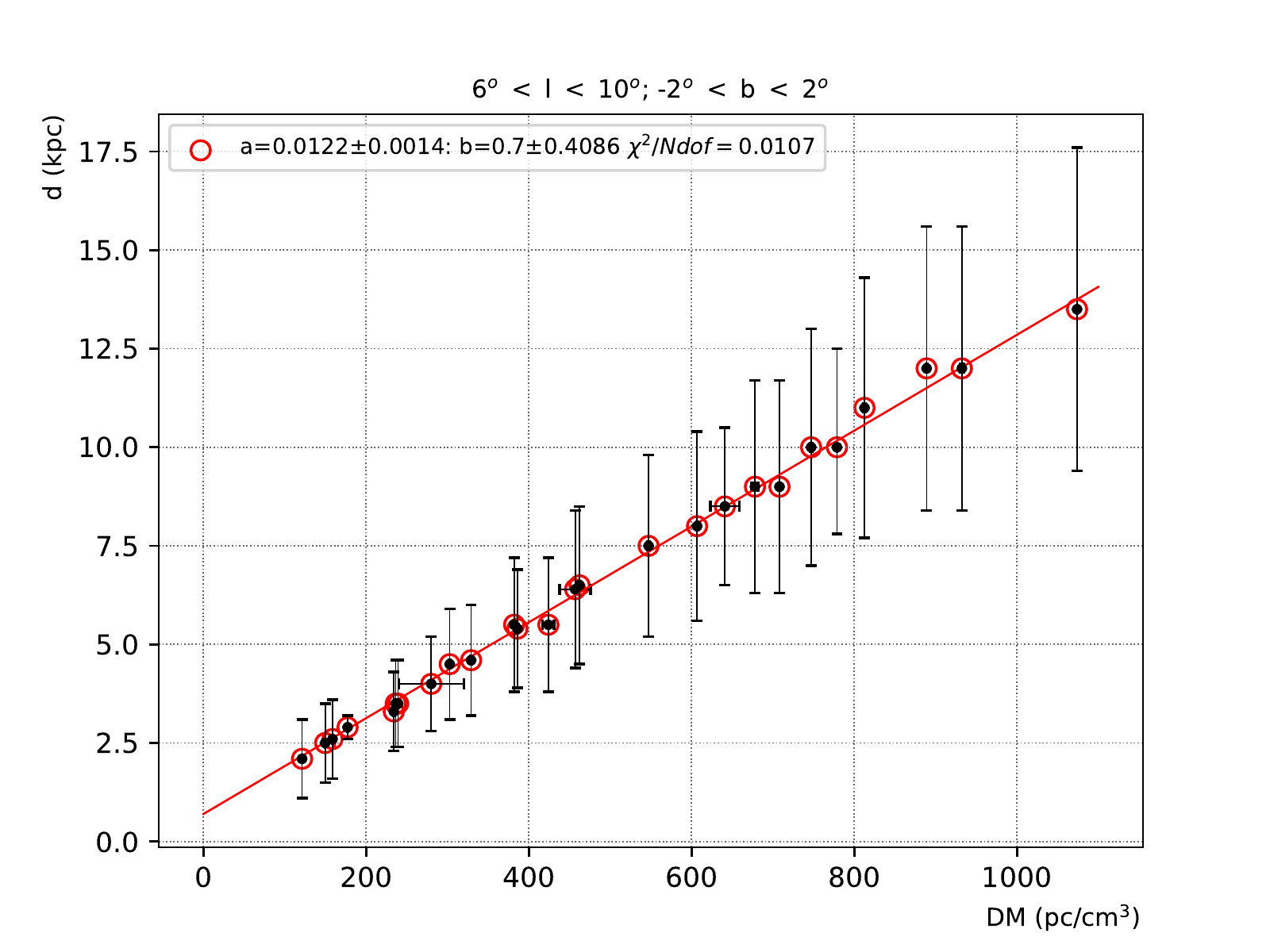}
	\caption{Distance-DM relation for radio pulsars in the solid angle interval 6$\degr$ $<$ $\ell$ $<$ 10$\degr$; -2$\degr$ $<$ $b$ $<$ 2$\degr$. The data are fit to a linear function (shown in red) with a slope a, and an intercept b.}
	\label{fig:3}
\end{figure}

\begin{figure}
	
	\includegraphics[width=\columnwidth]{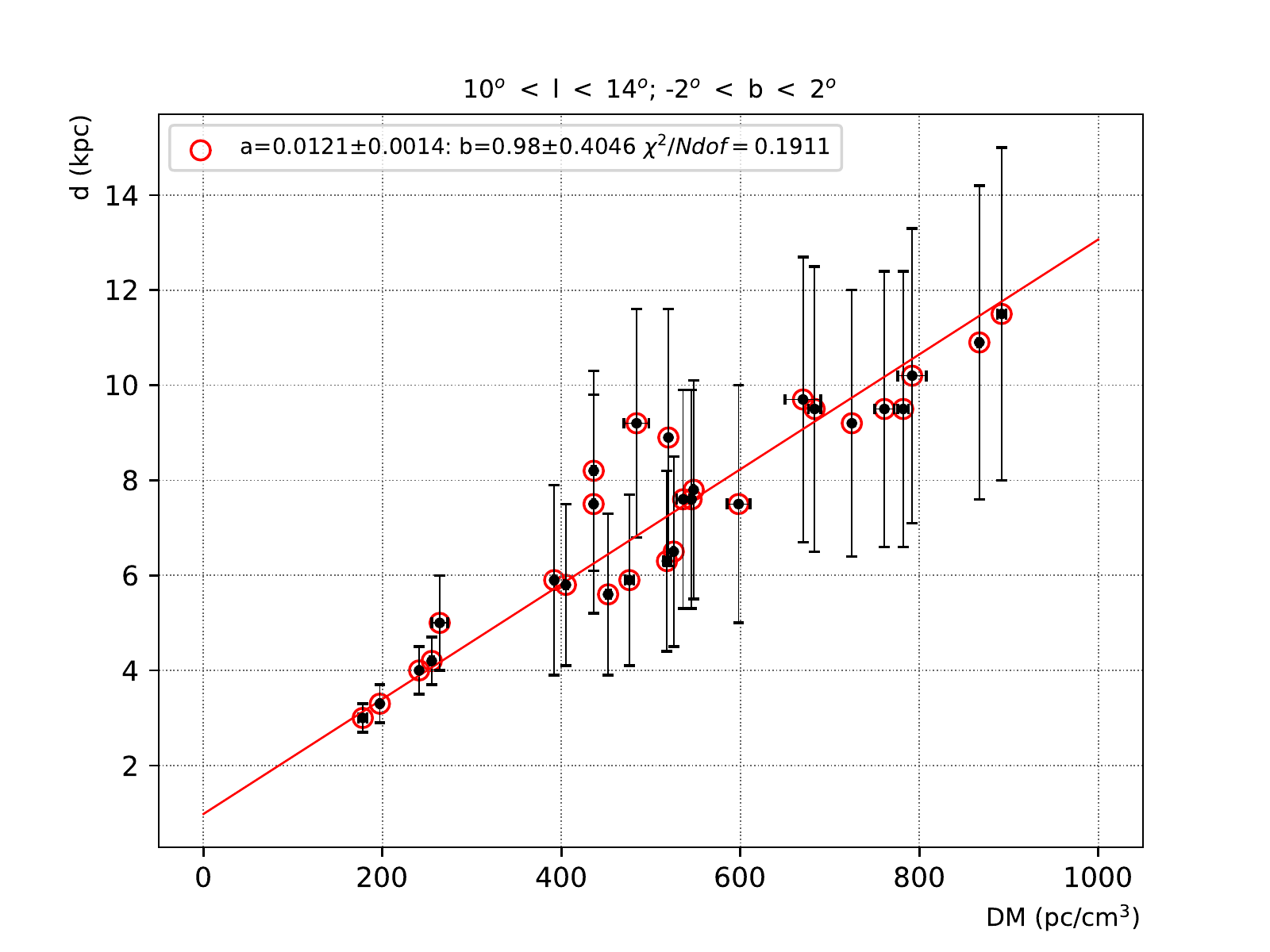}
	\caption{Distance-DM relation for radio pulsars in the solid angle interval 10$\degr$ $<$ $\ell$ $<$ 14$\degr$; -2$\degr$ $<$ $b$ $<$ 2$\degr$. The data are fit to a linear function (shown in red) with a slope a, and an intercept b.}
	\label{fig:4}
\end{figure}

\begin{figure}
	
	\includegraphics[width=\columnwidth]{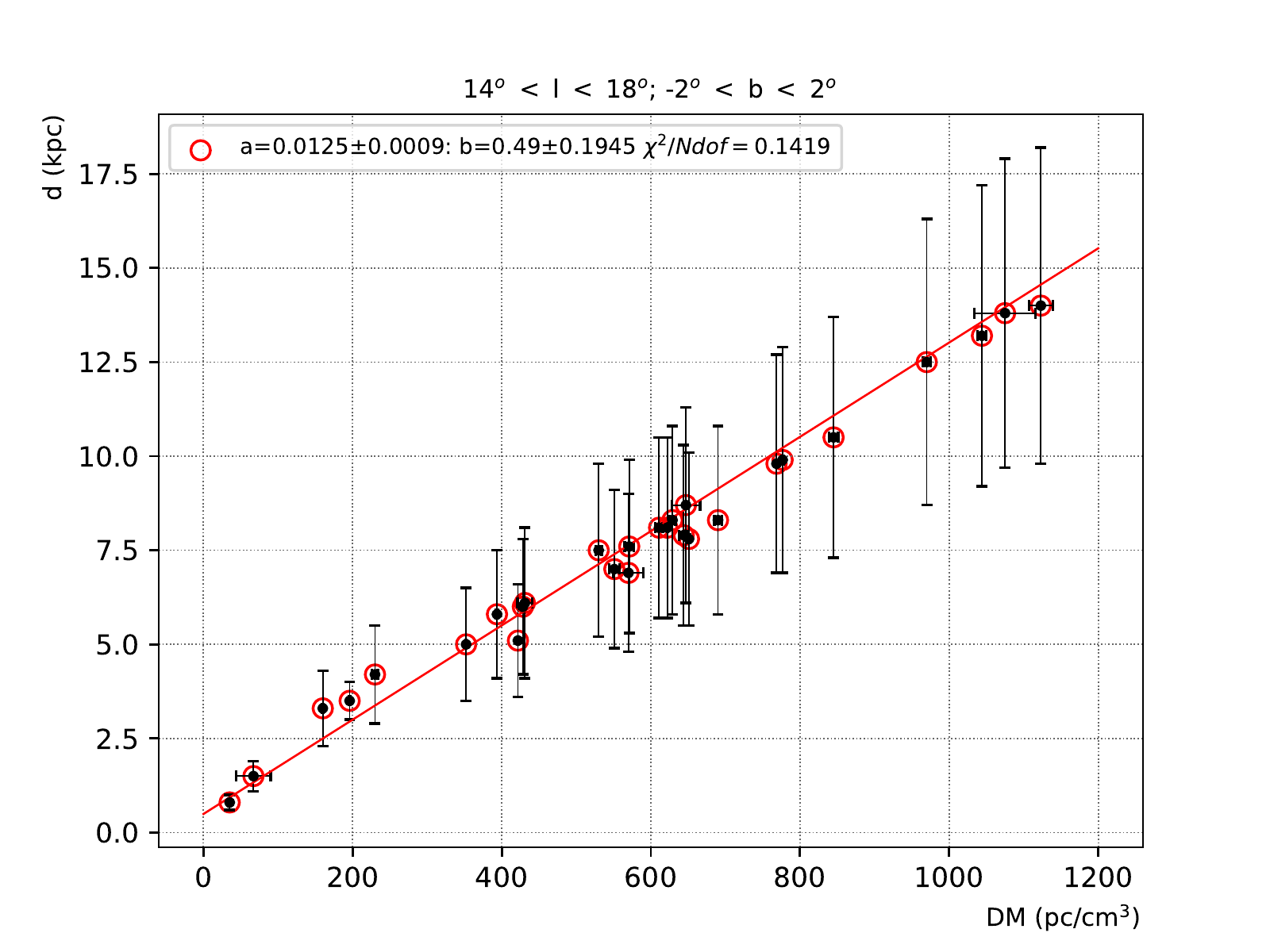}
	\caption{Distance-DM relation for radio pulsars in the solid angle interval 14$\degr$ $<$ $\ell$ $<$ 18$\degr$; -2$\degr$ $<$ $b$ $<$ 2$\degr$. The data are fit to a linear function (shown in red) with a slope a, and an intercept b.}
	\label{fig:5}
\end{figure}

\begin{figure}
	
	\includegraphics[width=\columnwidth]{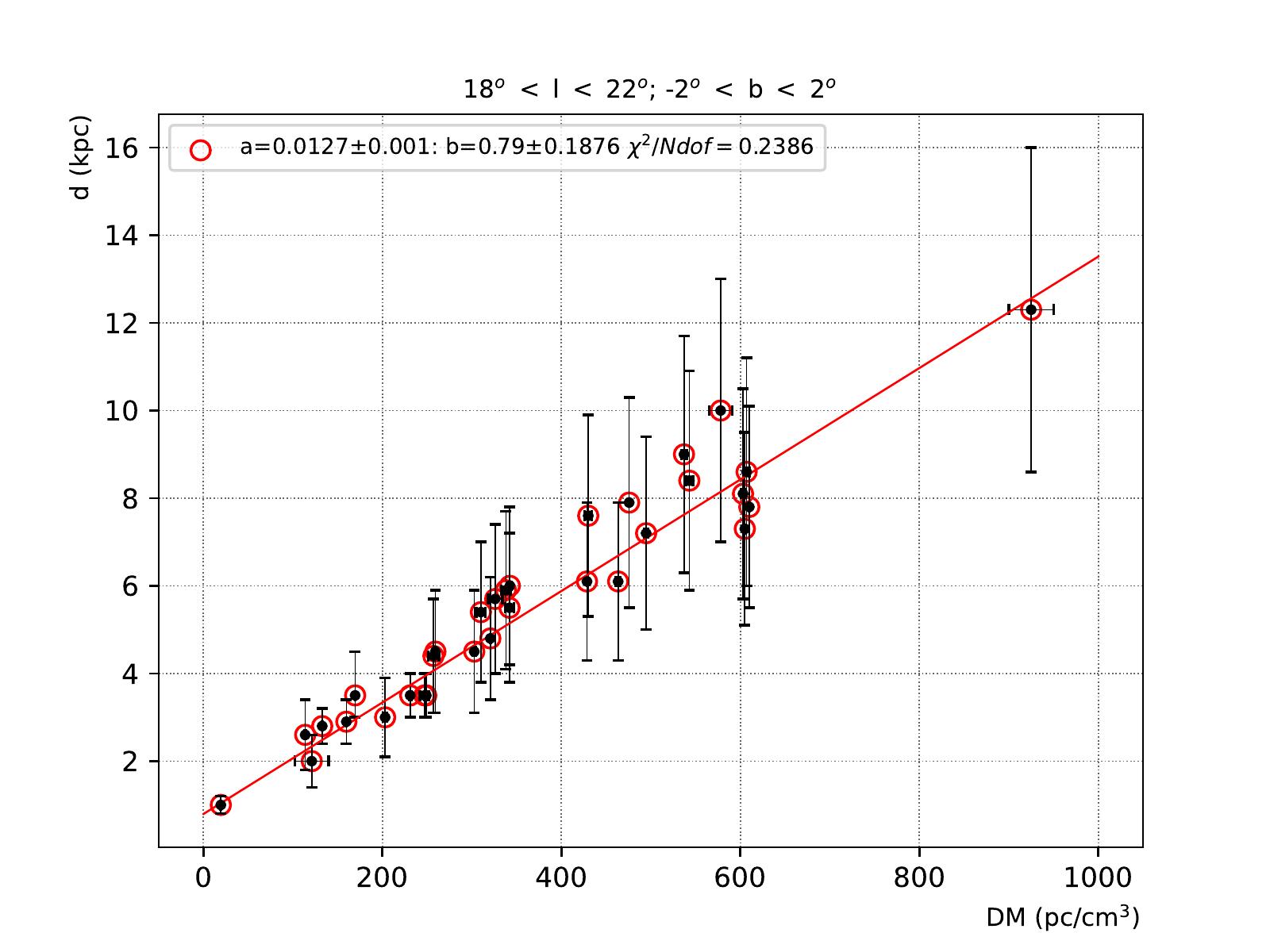}
	\caption{Distance-DM relation for radio pulsars in the solid angle interval 18$\degr$ $<$ $\ell$ $<$ 22$\degr$; -2$\degr$ $<$ $b$ $<$ 2$\degr$. The data are fit to a linear function (shown in red) with a slope a, and an intercept b.}
	\label{fig:6}
\end{figure}

\begin{figure}
	
	\includegraphics[width=\columnwidth]{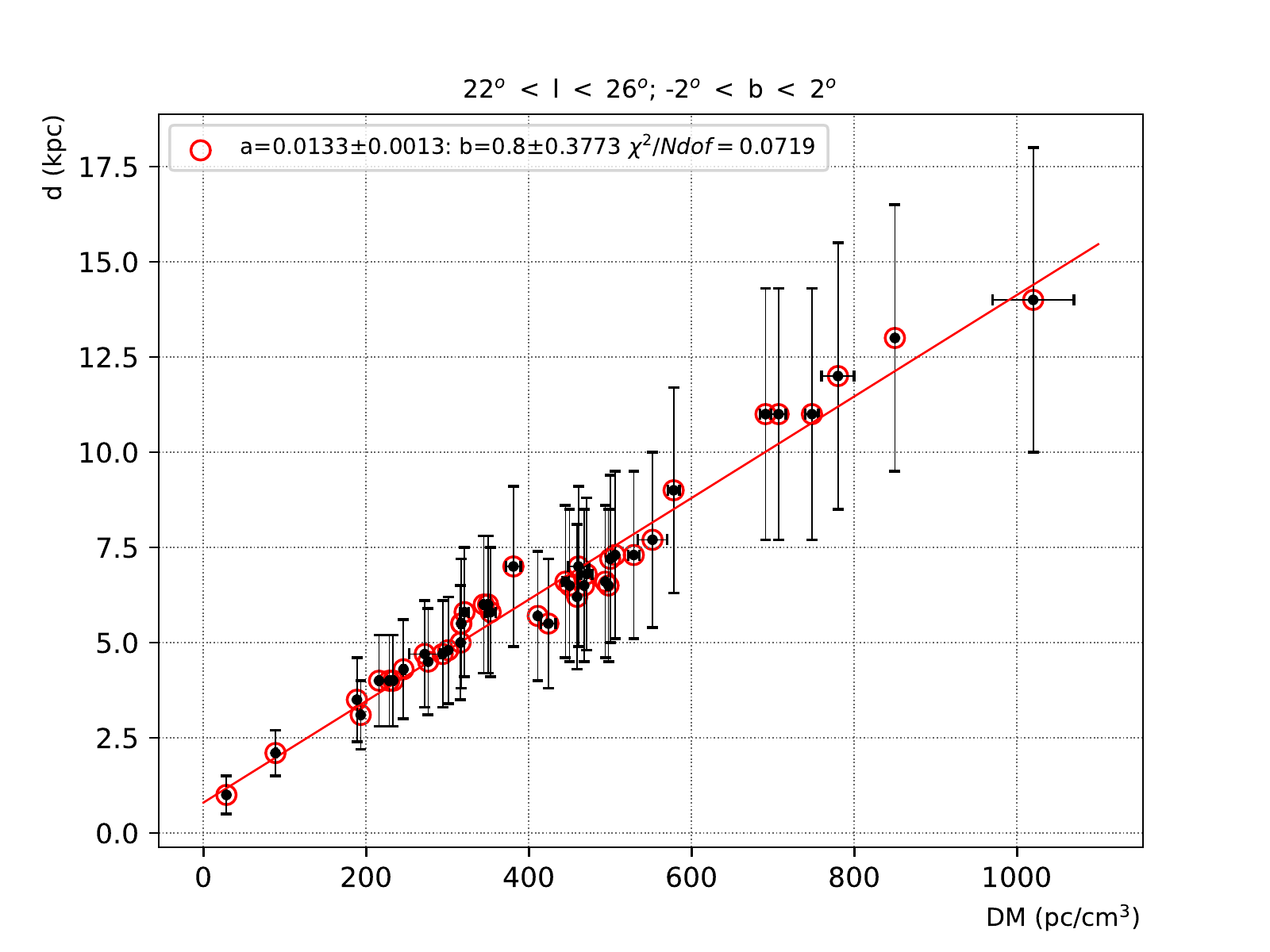}
	\caption{Distance-DM relation for radio pulsars in the solid angle interval 22$\degr$ $<$ $\ell$ $<$ 26$\degr$; -2$\degr$ $<$ $b$ $<$ 2$\degr$. The data are fit to a linear function (shown in red) with a slope a, and an intercept b.}
	\label{fig:7}
\end{figure}

\begin{figure}
	
	\includegraphics[width=\columnwidth]{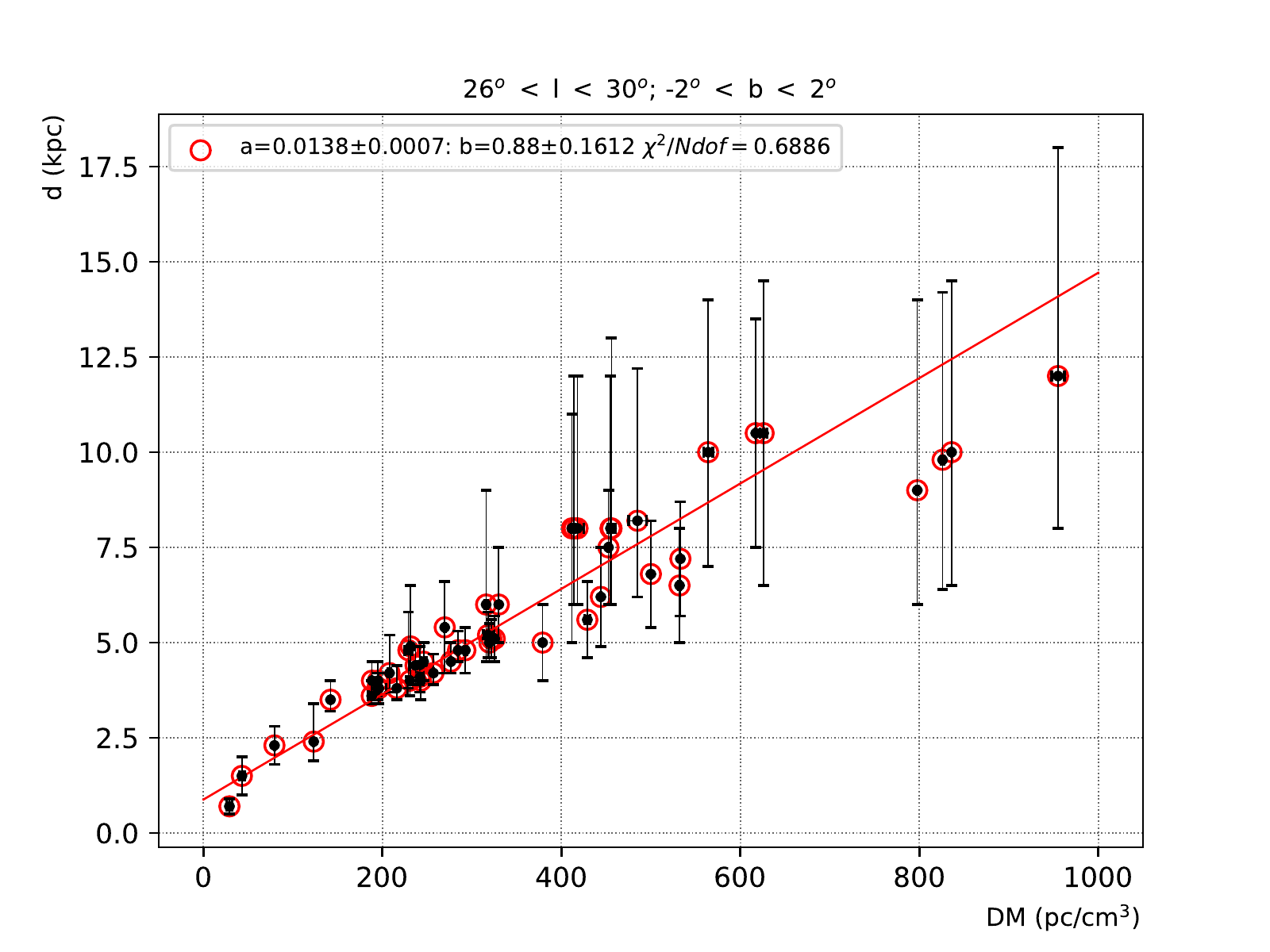}
	\caption{Distance-DM relation for radio pulsars in the solid angle interval 26$\degr$ $<$ $\ell$ $<$ 30$\degr$; -2$\degr$ $<$ $b$ $<$ 2$\degr$. The data are fit to a linear function (shown in red) with a slope a, and an intercept b.}
	\label{fig:8}
\end{figure}

\begin{figure}
	
	\includegraphics[width=\columnwidth]{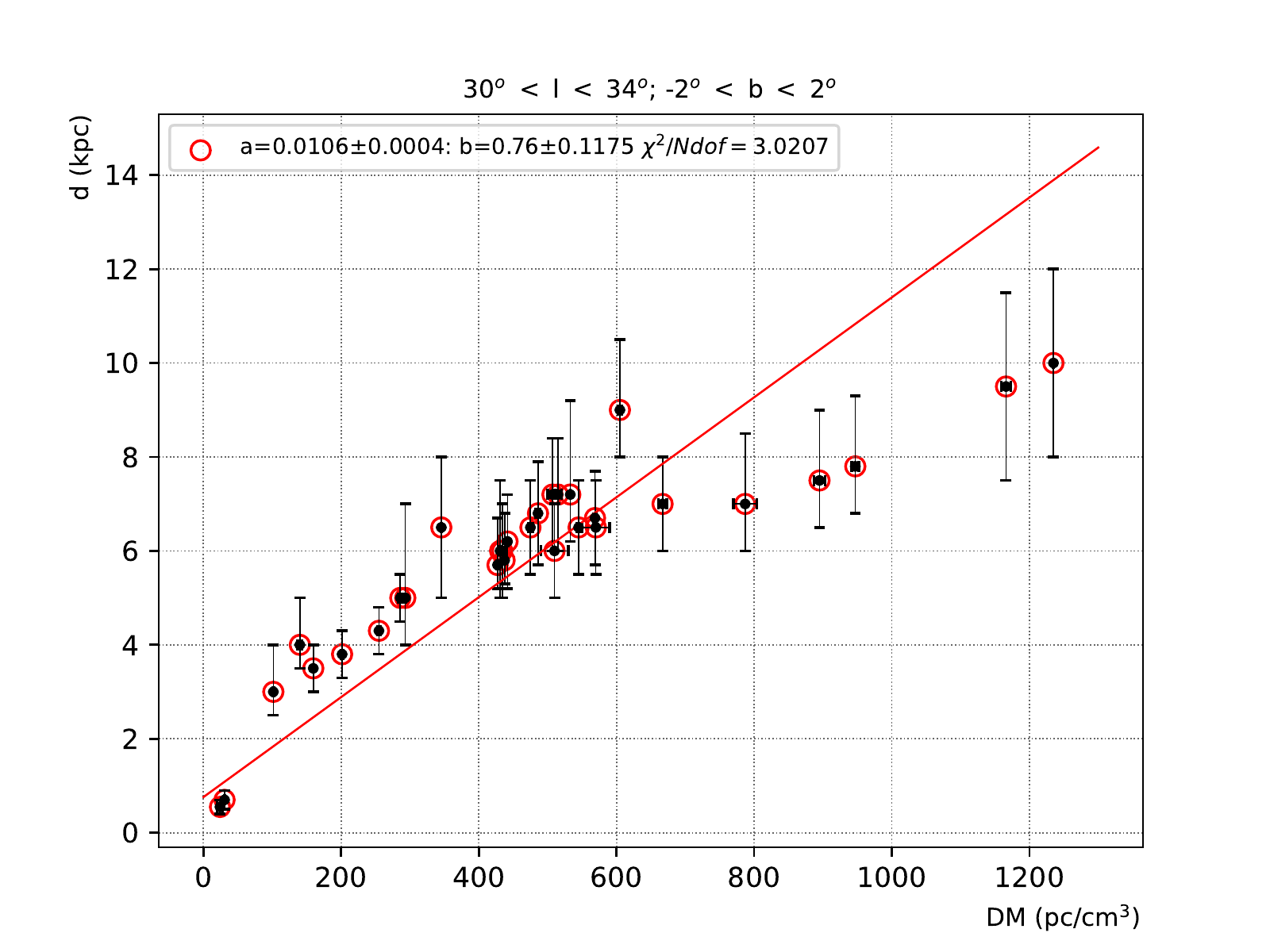}
	\caption{Distance-DM relation for radio pulsars in the solid angle interval 30$\degr$ $<$ $\ell$ $<$ 34$\degr$; -2$\degr$ $<$ $b$ $<$ 2$\degr$. The data are fit to a linear function (shown in red) with a slope a, and an intercept b.}
	\label{fig:9}
\end{figure}

\begin{figure}
	
	\includegraphics[width=\columnwidth]{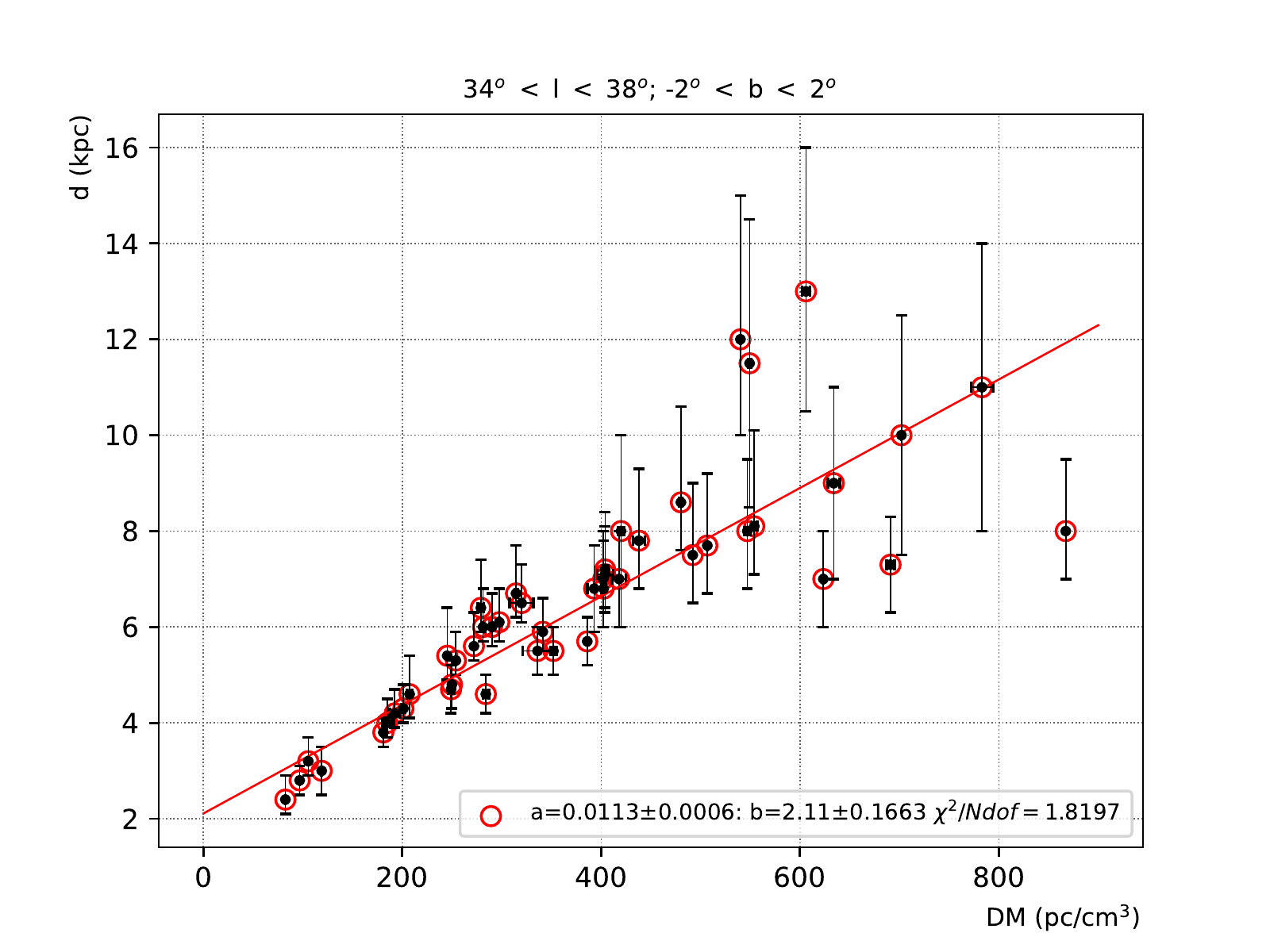}
	\caption{Distance-DM relation for radio pulsars in the solid angle interval 34$\degr$ $<$ $\ell$ $<$ 38$\degr$; -2$\degr$ $<$ $b$ $<$ 2$\degr$. The data are fit to a linear function (shown in red) with a slope a, and an intercept b.}
	\label{fig:10}
\end{figure}

\begin{figure}
	
	\includegraphics[width=\columnwidth]{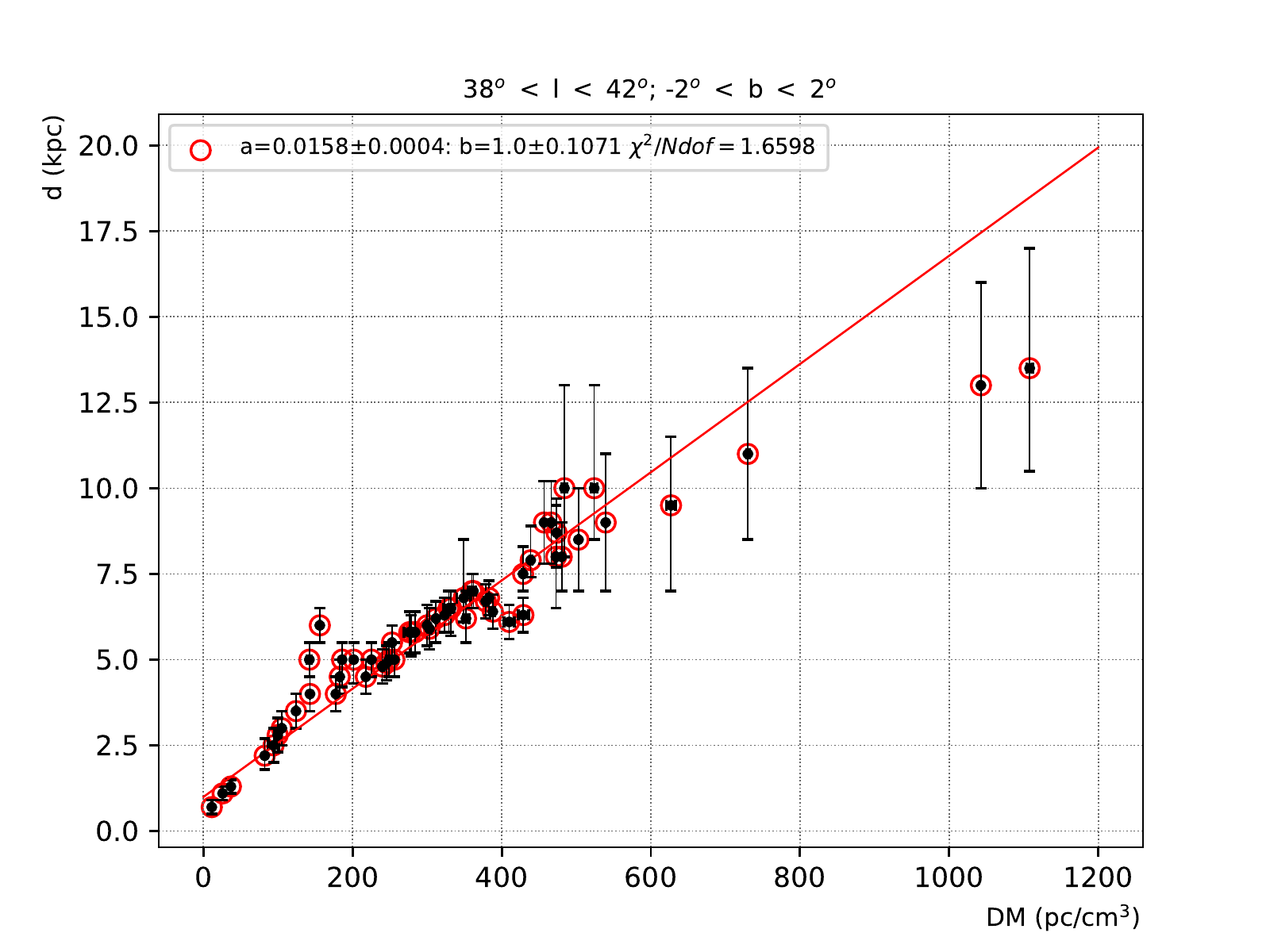}
	\caption{Distance-DM relation for radio pulsars in the solid angle interval 38$\degr$ $<$ $\ell$ $<$ 42$\degr$; -2$\degr$ $<$ $b$ $<$ 2$\degr$. The data are fit to a linear function (shown in red) with a slope a, and an intercept b.}
	\label{fig:11}
\end{figure}

\begin{figure}
	
	\includegraphics[width=\columnwidth]{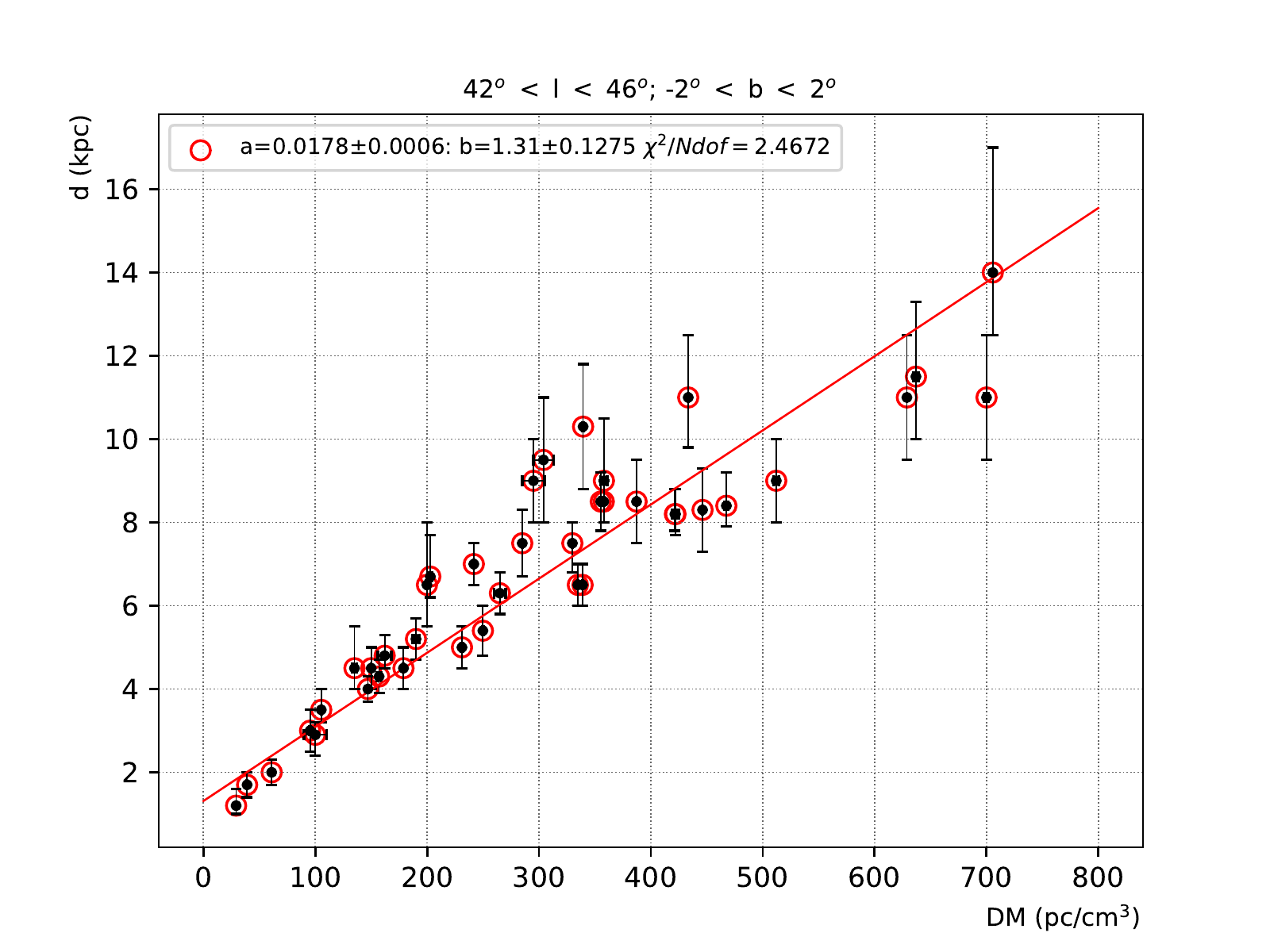}
	\caption{Distance-DM relation for radio pulsars in the solid angle interval 42$\degr$ $<$ $\ell$ $<$ 46$\degr$; -2$\degr$ $<$ $b$ $<$ 2$\degr$. The data are fit to a linear function (shown in red) with a slope a, and an intercept b.}
	\label{fig:12}
\end{figure}

\begin{figure}
	
	\includegraphics[width=\columnwidth]{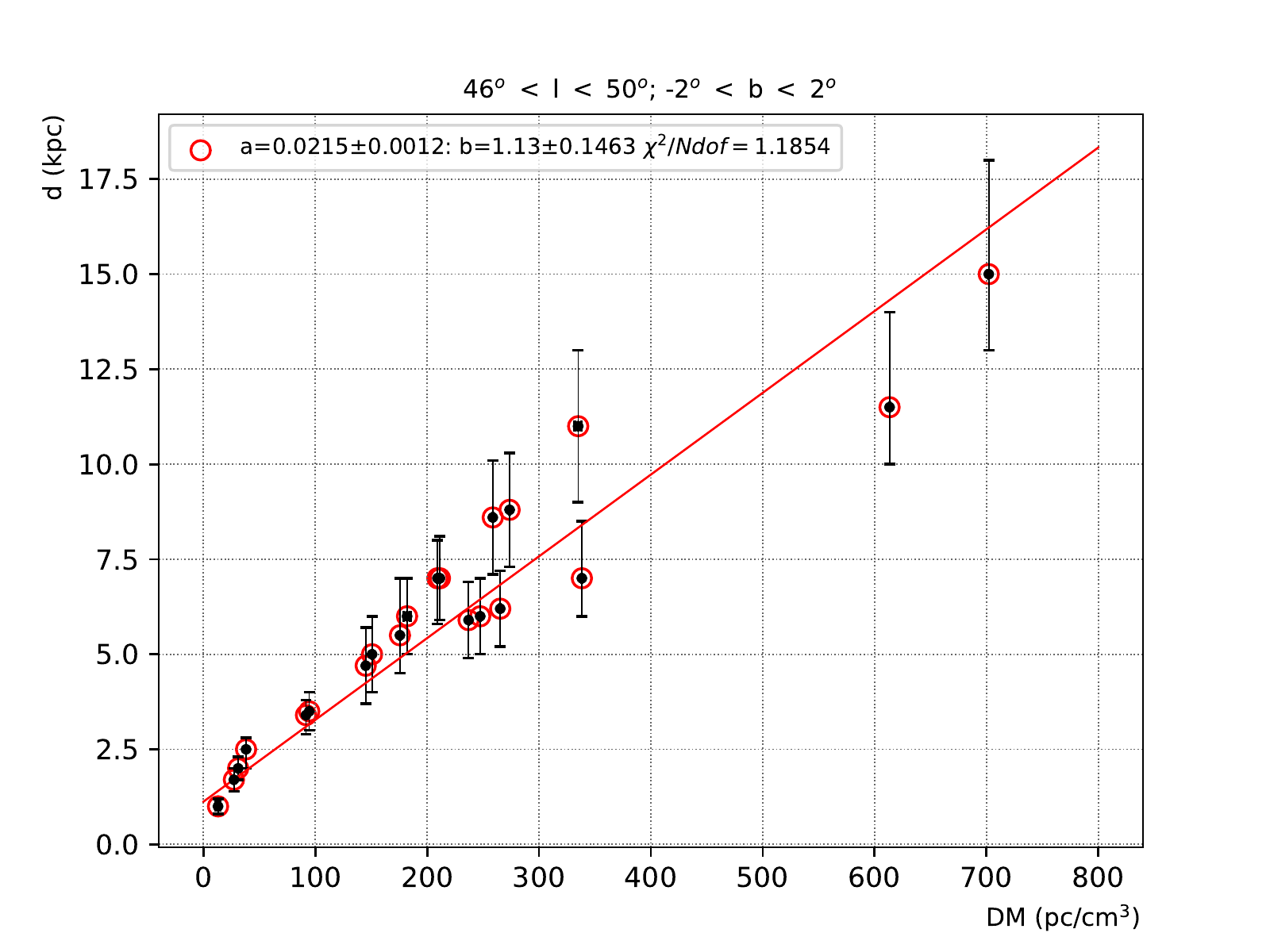}
	\caption{Distance-DM relation for radio pulsars in the solid angle interval 46$\degr$ $<$ $\ell$ $<$ 50$\degr$; -2$\degr$ $<$ $b$ $<$ 2$\degr$. The data are fit to a linear function (shown in red) with a slope a, and an intercept b. }
	\label{fig:13}
\end{figure}

\begin{figure}
	
	\includegraphics[width=\columnwidth]{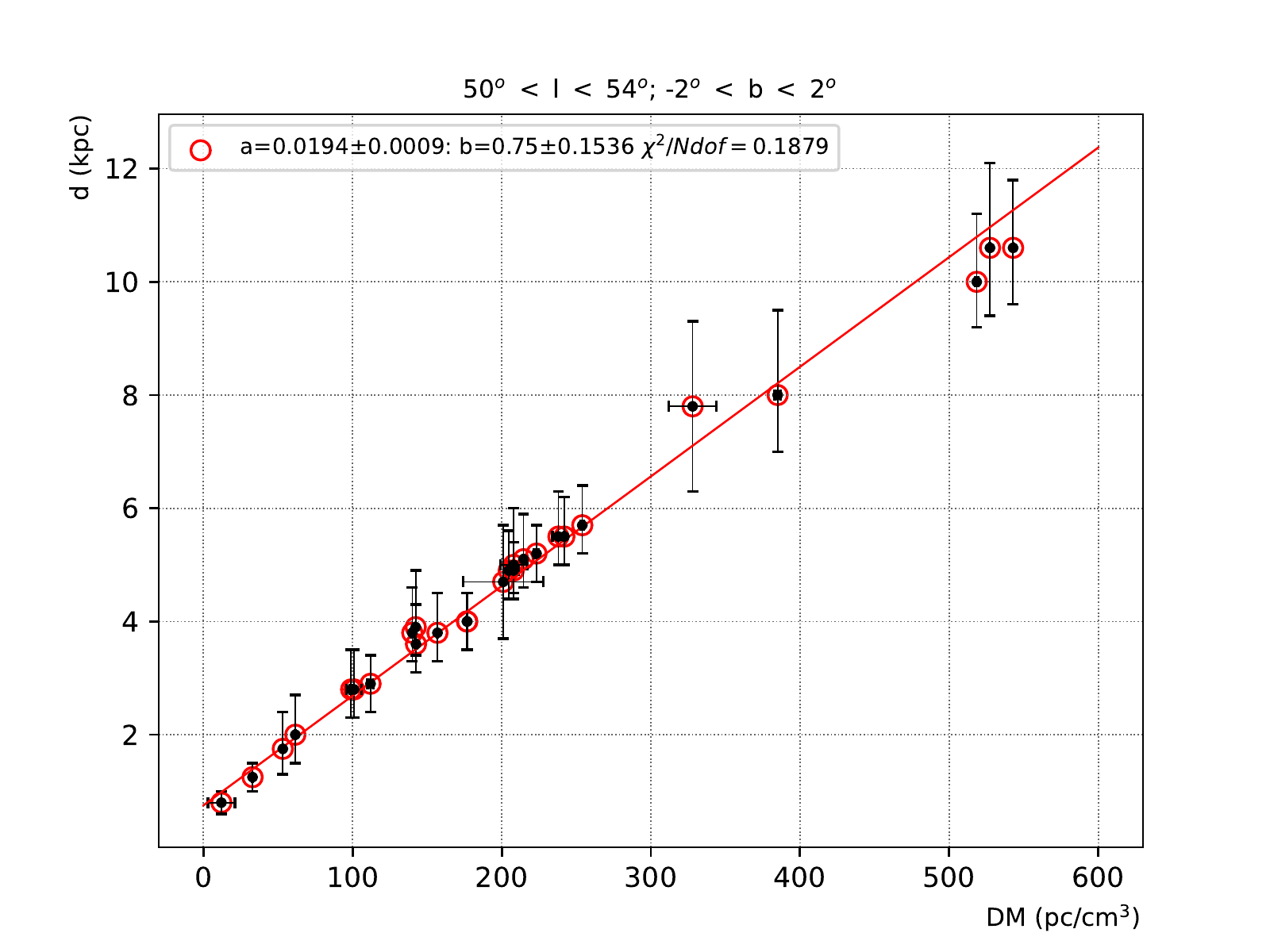}
	\caption{Distance-DM relation for radio pulsars in the solid angle interval 50$\degr$ $<$ $\ell$ $<$ 54$\degr$; -2$\degr$ $<$ b $<$ 2$\degr$. The data are fit to a linear function (shown in red) with a slope a, and an intercept b. }
	\label{fig:14}
\end{figure}

\begin{figure}
	
	\includegraphics[width=\columnwidth]{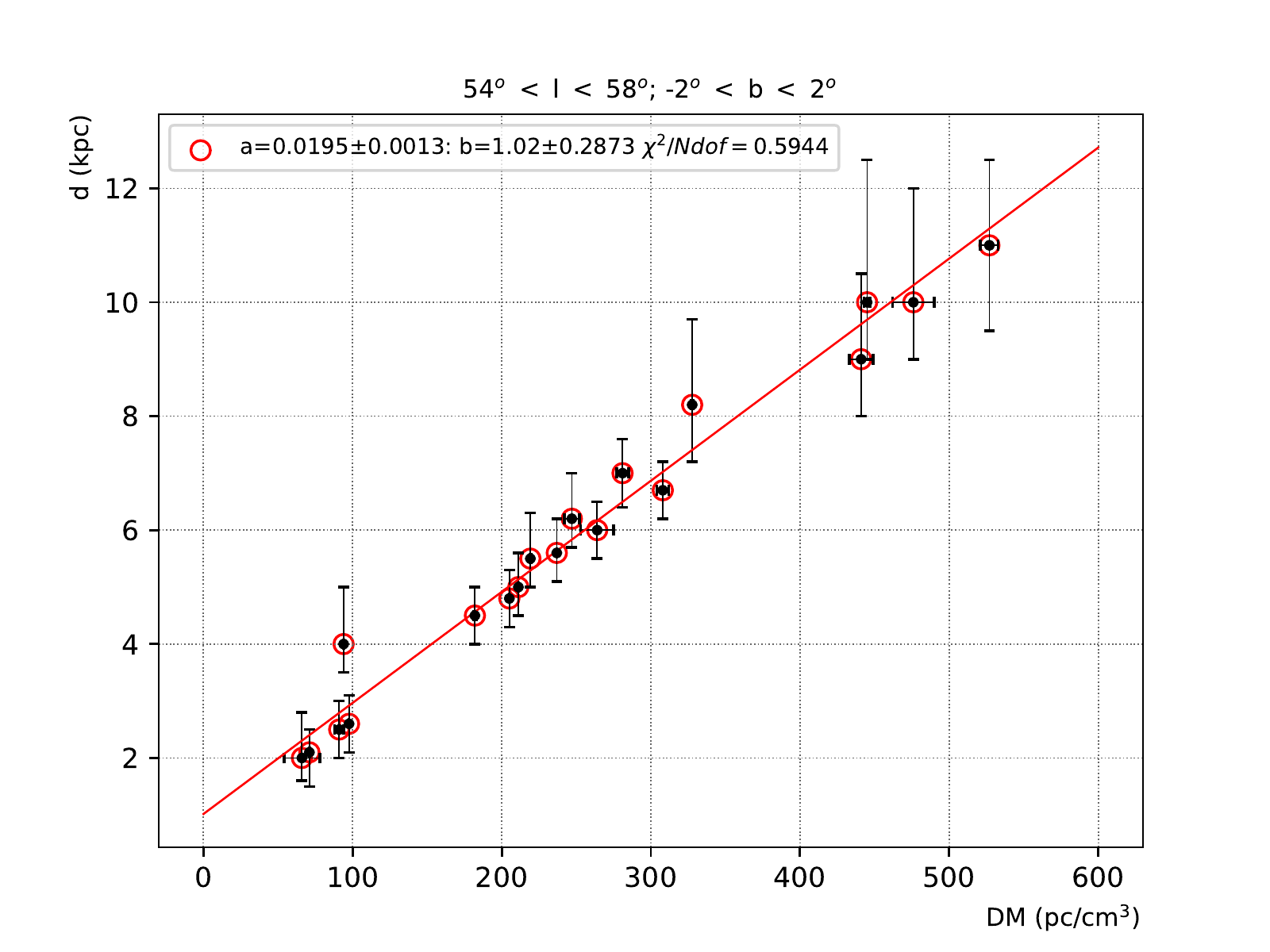}
	\caption{Distance-DM relation for radio pulsars in the solid angle interval 54$\degr$ $<$ $\ell$ $<$ 58$\degr$; -2$\degr$ $<$ $b$ $<$ 2$\degr$. The data are fit to a linear function (shown in red) with a slope a, and an intercept b.}
	\label{fig:15}
\end{figure}

\begin{figure}
	
	\includegraphics[width=\columnwidth]{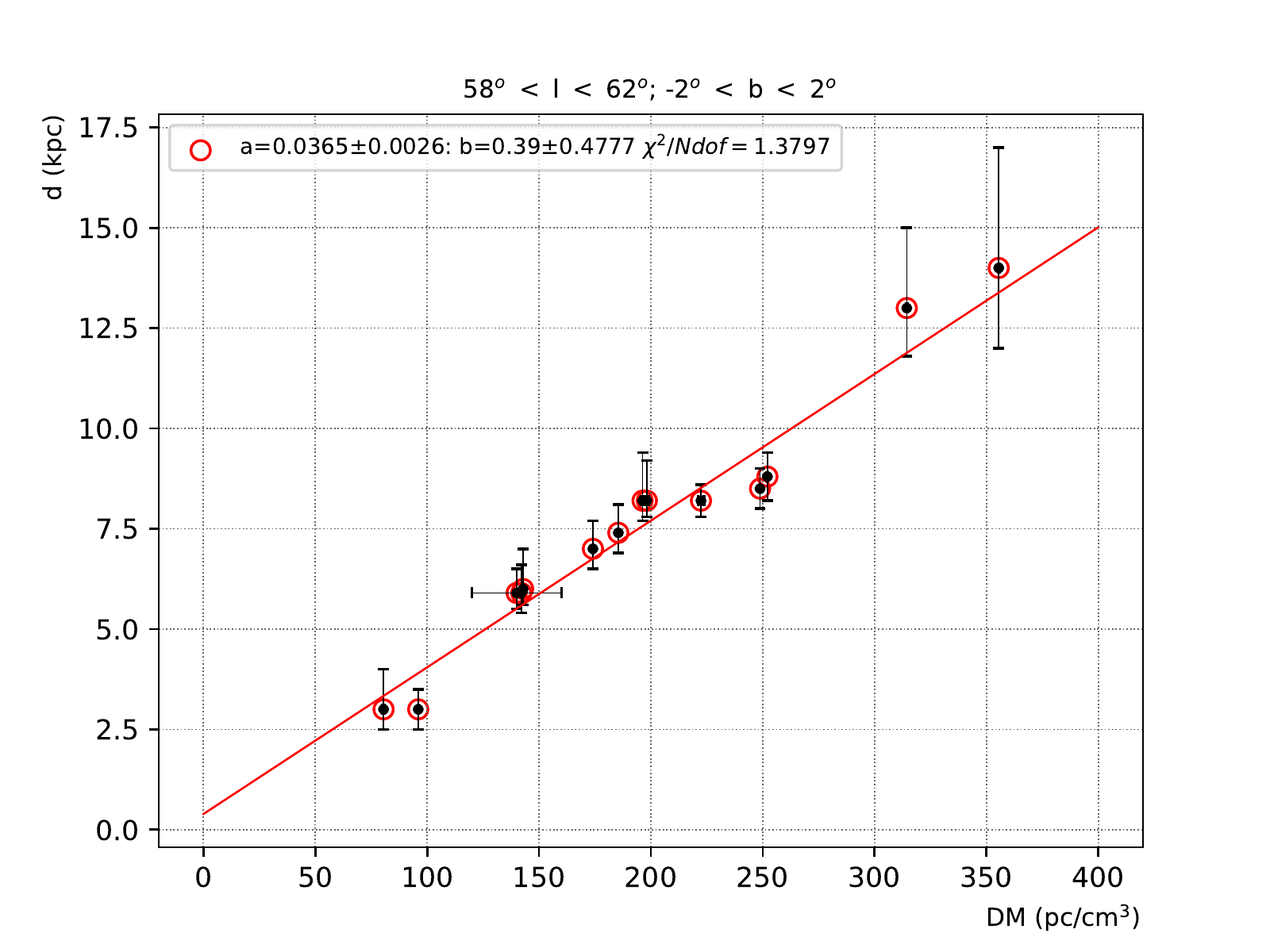}
	\caption{Distance-DM relation for radio pulsars in the solid angle interval 58$\degr$ $<$ $\ell$ $<$ 62$\degr$; -2$\degr$ $<$ $b$ $<$ 2$\degr$. The data are fit to a linear function (shown in red) with a slope a, and an intercept b.}
	\label{fig:16}
\end{figure}

\begin{figure}
	
	\includegraphics[width=\columnwidth]{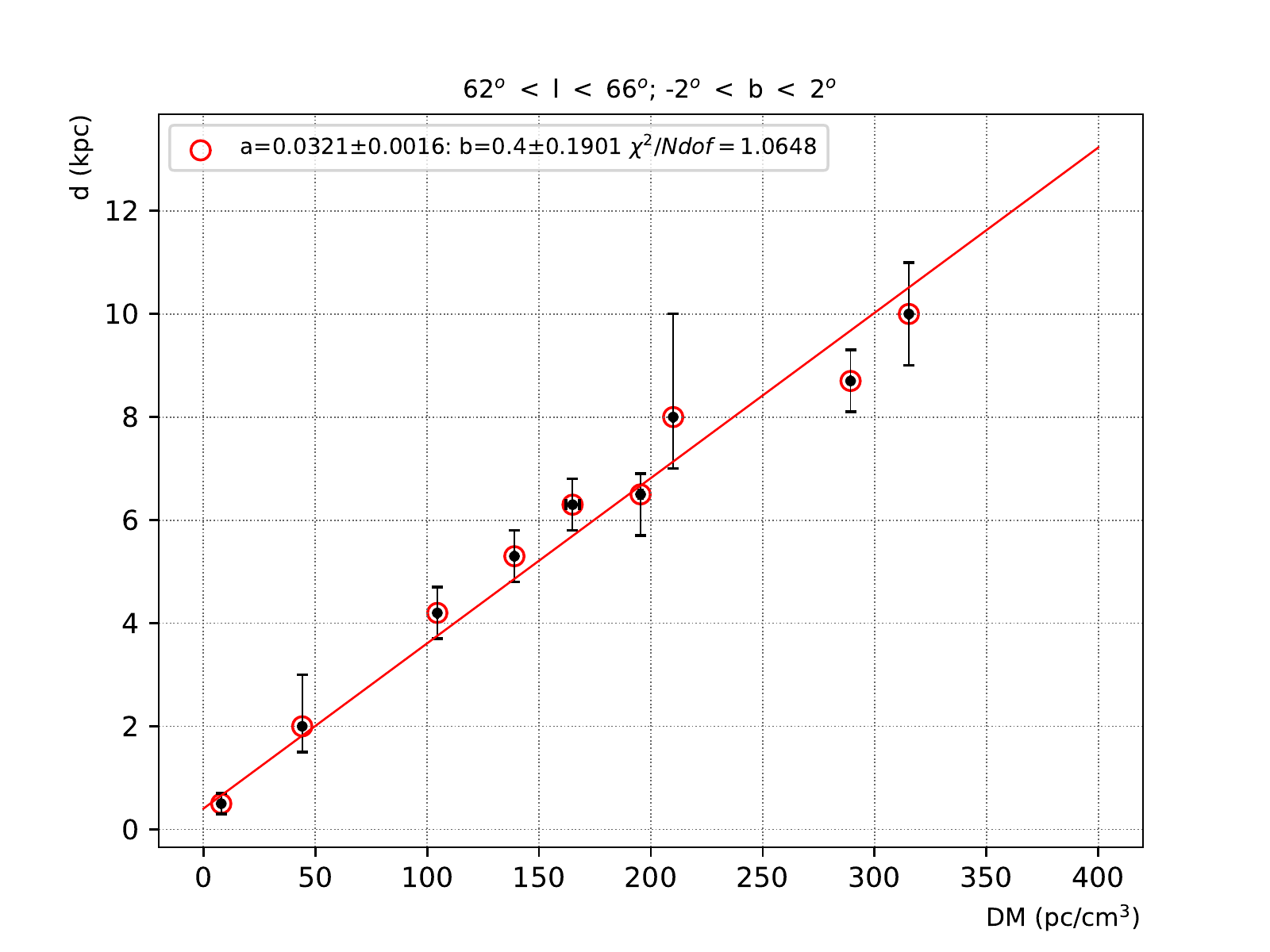}
	\caption{Distance-DM relation for radio pulsars in the solid angle interval 62$\degr$ $<$ $\ell$ $<$ 66$\degr$; -2$\degr$ $<$ $b$ $<$ 2$\degr$. The data are fit to a linear function (shown in red) with a slope a, and an intercept b.}
	\label{fig:17}
\end{figure}

\begin{figure}
	
	\includegraphics[width=\columnwidth]{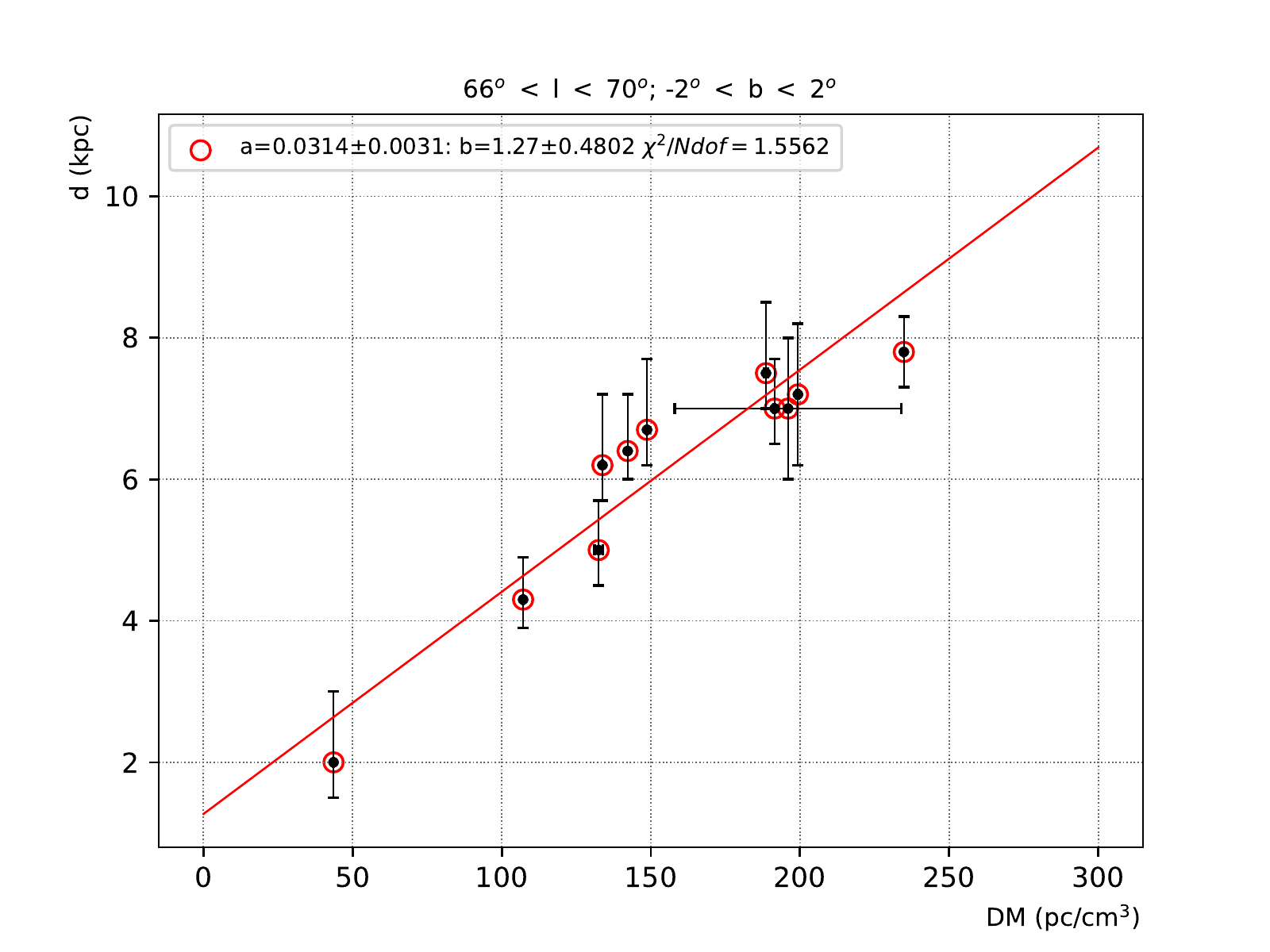}
	\caption{Distance-DM relation for radio pulsars in the solid angle interval 66$\degr$ $<$ $\ell$ $<$ 70$\degr$; -2$\degr$ $<$ $b$ $<$ 2$\degr$. The data are fit to a linear function (shown in red) with a slope a, and an intercept b.}
	\label{fig:18}
\end{figure}

\begin{figure}
	
	\includegraphics[width=\columnwidth]{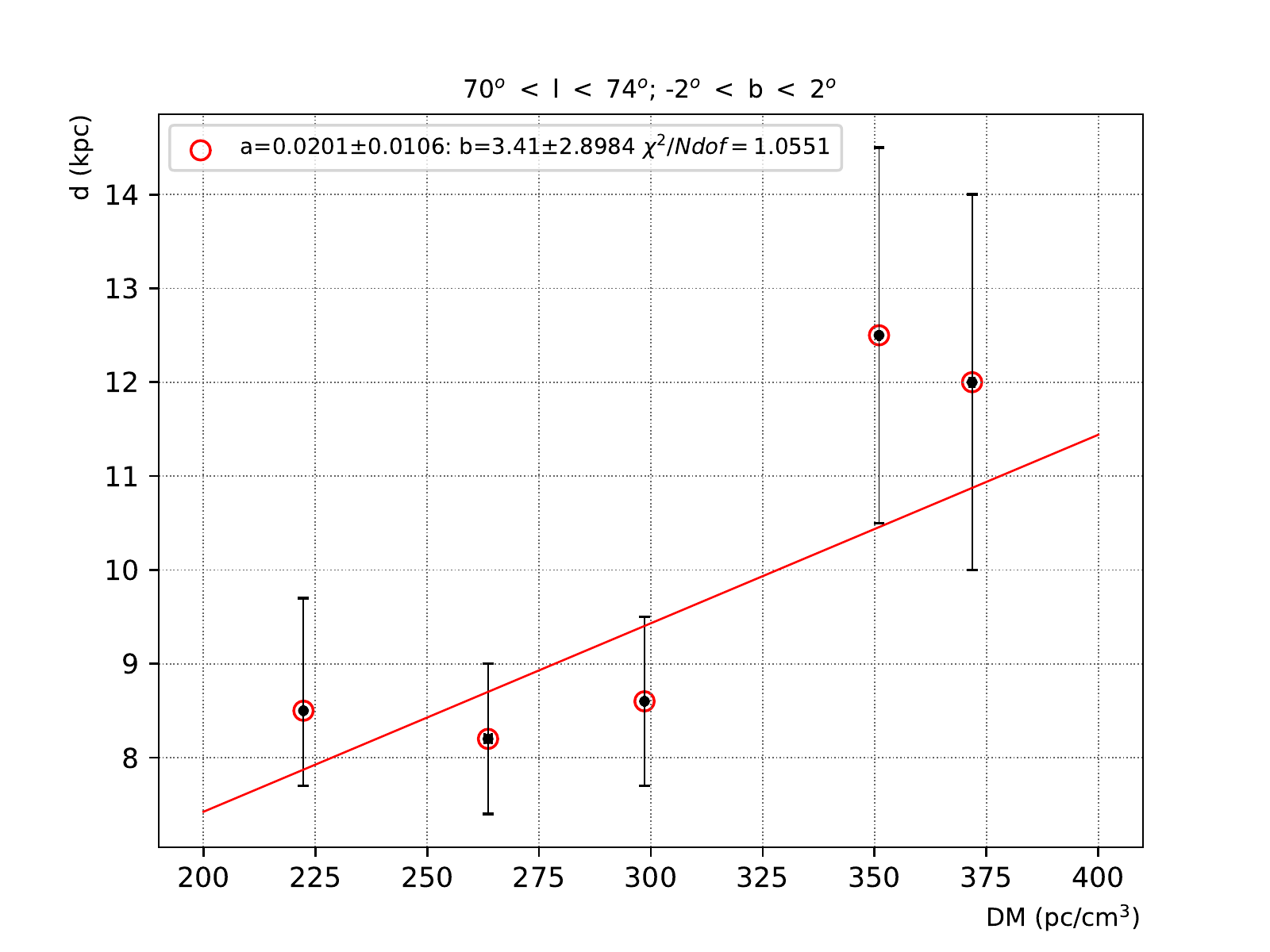}
	\caption{Distance-DM relation for radio pulsars in the solid angle interval 70$\degr$ $<$ $\ell$ $<$ 74$\degr$; -2$\degr$ $<$ $b$ $<$ 2$\degr$. The data are fit to a linear function (shown in red) with a slope a, and an intercept b.}
	\label{fig:19}
\end{figure}

\begin{figure}
	
	\includegraphics[width=\columnwidth]{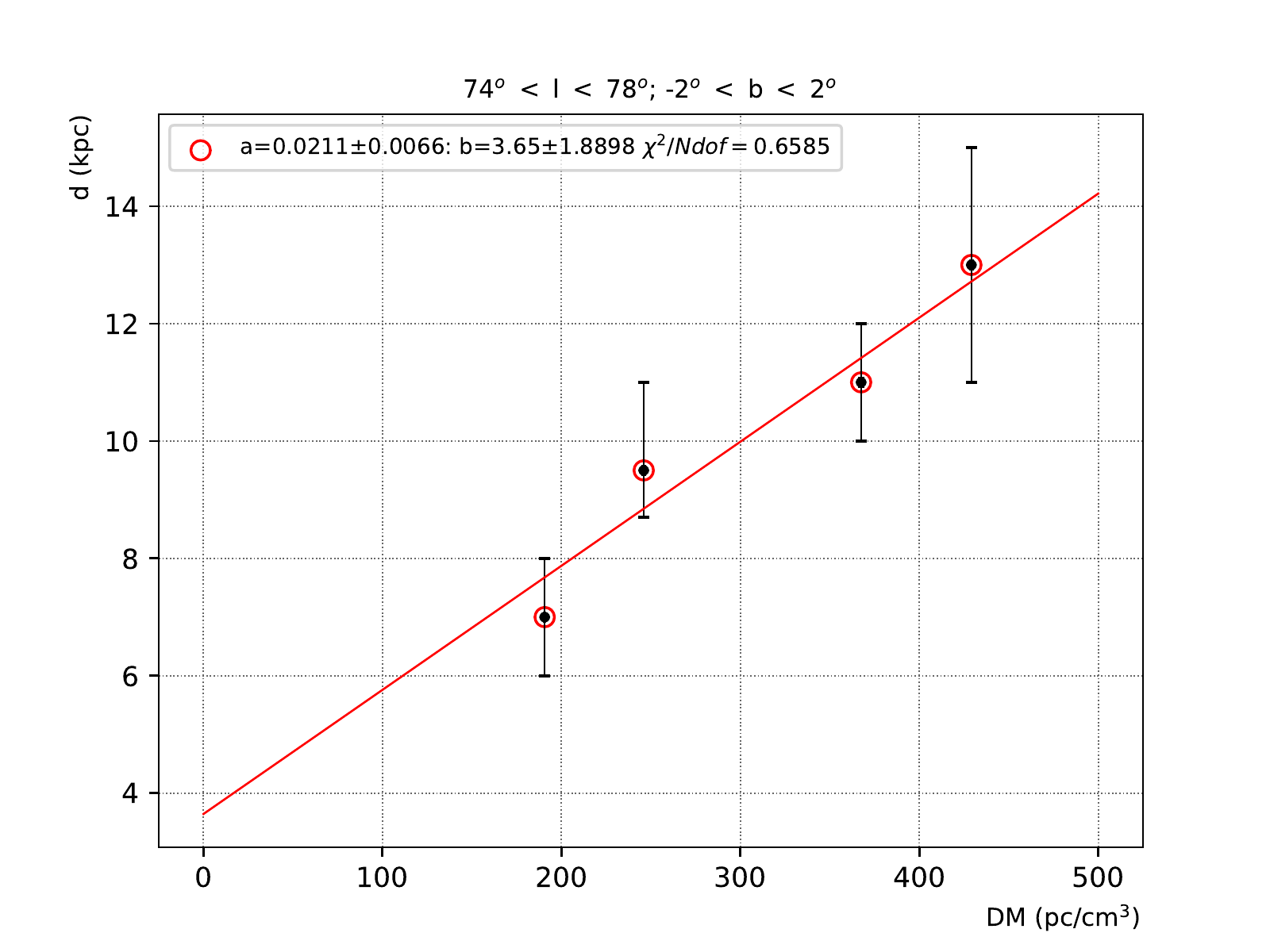}
	\caption{Distance-DM relation for radio pulsars in the solid angle interval 74$\degr$ $<$ $\ell$ $<$ 78$\degr$; -2$\degr$ $<$ $b$ $<$ 2$\degr$. The data are fit to a linear function (shown in red) with a slope a, and an intercept b.}
	\label{fig:20}
\end{figure}
\FloatBarrier


%%%%%%%%%%%%%%%%%%%%%%%%%%%%%%%%%%%%%%%%%%%%%%%%%%

\let\clearpage\relax

% Don't change these lines
\bsp	% typesetting comment
\label{lastpage}
\end{document}